\documentstyle[rmp,naps]{reviewtex}
\makeindex
\renewcommand{\vec}[1]{\relax\ifmmode\mathchoice
{\mbox{\boldmath$\relax\displaystyle#1$}}
{\mbox{\boldmath$\relax\textstyle#1$}}
{\mbox{\boldmath$\relax\scriptstyle#1$}}
{\mbox{\boldmath$\relax\scriptscriptstyle#1$}}\else
\hbox{\boldmath$\relax\textstyle#1$}\fi}
\newcommand{\tit}[1]{}
\newcommand{\disstit}[1]{}
\newcommand{\pag}[2]{pp.~#1.}
\newcommand{\art}[4]{{\em #1} {\bf #2}, #3.} 
\newcommand{\etal}{\mbox{\it et al.}}
\newcommand{\bib}[1]{\\[-3mm]\bibitem{#1}}

\begin{document}
\tighten
\onecolumn
\twocolumn[\hsize\textwidth\columnwidth\hsize\csname
@twocolumnfalse\endcsname

\title{\mbox{ }\\[-2.5cm]Traffic and Related Self-Driven Many-Particle Systems}
\author{\normalsize Dirk Helbing$^{1,2,3,}\protect\thanks{}$\\[2mm]
\normalsize $^1$Institute for Economics and Traffic, Dresden University of Technology,\\ 
\normalsize Andreas-Schubert-Str. 23, D-01062 Dresden, Germany\\[2mm]
\normalsize $^2$II. Institute of Theoretical
Physics,  University of Stuttgart,\\ 
\normalsize Pfaffenwaldring 57/III, D-70550 Stuttgart, Germany\\[2mm]
\normalsize $^3$Collegium Budapest~-- Institute for Advanced Study,\\
\normalsize Szenth\'{a}roms\'{a}g utca 2,
H-1014 Budapest, Hungary} 
\maketitle              
\begin{abstract}
Since the subject of traffic dynamics has captured the interest of
physicists, many surprising effects have been revealed and explained.
Some of the questions now understood are the following:
Why are vehicles sometimes stopped by so-called ``phantom traffic jams'',
although they all like to drive fast? 
What are the mechanisms behind stop-and-go traffic?
Why are there several different kinds of congestion, and how are they
related?
Why do most traffic jams occur considerably before the road capacity
is reached?
Can a temporary reduction of the traffic volume cause a lasting
traffic jam?
Under which conditions can speed limits speed up traffic?
Why do pedestrians moving in opposite directions normally organize in lanes,
while similar systems are ``freezing by heating''?
Why do panicking pedestrians produce dangerous deadlocks?
All these questions have been answered by applying and extending 
methods from statistical physics and non-linear
dynamics to self-driven many-particle systems. This review article
on traffic introduces (i) empirically data, facts, and observations,
(ii) the main approaches to pedestrian and vehicle traffic,
(iii) microscopic (particle-based), mesoscopic (gas-kinetic), and macroscopic
(fluid-dynamic) models. Attention is also paid to the formulation of a
micro-macro link, to aspects of universality, and to other unifying concepts
like a general modelling framework for self-driven many-particle systems, including
spin systems. Subjects like the
optimization of traffic flows and relations to biological or socio-economic
systems such as bacterial colonies, flocks of birds, panics, and stock market
dynamics are touched as well.
\end{abstract}
]
\footnotetext{$^*$ Electronic address: helbing@trafficforum.org}
\tableofcontents
\section{Introduction} \label{I}

\subsection{Motivation: History of traffic modelling and traffic impact on society}\label{IA}

The interest in the subject of traffic dynamics is surprisingly old. 
In 1935, Greenshields has carried out  early studies of vehicular traffic, and
already in the 1950ies, a considerable publication activity is documented by journals on
operations research and engineering. These papers have already addressed
topics like the fundamental diagram between traffic flow and vehicle density
or the instability of traffic flow, which are still relevant. The reason
becomes understandable through the following quotation from H. Greenberg
in 1959:
\par
{\em ``The volume of vehicular traffic in the past several years has rapidly
  outstripped the capacities of the nation's highways. It has become
  increasingly necessary to understand the dynamics of traffic flow and obtain
  a mathematical description of the process.''}
\par
More than fourty years later, the situation has deteriorated a lot. Some
cities like Los Angeles and San Francisco suffer from heavy traffic congestion around the
clock. In Europe, as well, the time that drivers spend standing in traffic jams amounts 
to several days each year. During holiday seasons, jams may grow up
to more than 100 kilometers in size. Vehicle emissions like SO$_2$, NO$_x$,
CO, CO$_2$, dust particles, smog, and noise have reached or even exceeded a level 
comparable to those by industrial production and those by private households, being harmful to
the environment and human health. On average, every driver is involved in one
accident during his lifetime. In Germany alone, the financial damage by traffic due to accidents and
environmental impacts is estimated 100 billion dollars each year. The economic
loss due to congested traffic is of the same order of magnitude. However, the
traffic volume is still growing because of increased demands for mobility
and modern logistics, including e-commerce. 
\par
Without any doubt, an efficient transportation system
is essential for the functioning and success of modern, industrialized
societies. But the days, when freeways were free ways, are over. Facing the
increasing problems of individual traffic, the following questions came up:
Is it still affordable and publicly acceptable to expand the 
infrastructure? Will drivers still buy cars in view of streets that
effectively turn into parking lots? 
It is for these reasons that automobile companies started to worry 
about their own future and decided to spend considerable amounts of
money for research on traffic dynamics and the question how the available infrastructure
could be used more efficiently by new technologies (telematics).
\par
At this time, physicists were motivated to think about what physics could
contribute to the field of traffic dynamics. Although, among mathematicians,
physicists, and chemists, there were already some early pioneers like
Whitham, Prigogine, Montroll, and K\"uhne, the main activities started in 1992
and 1993 with papers by Biham {\em et al.} (1992), Nagel and Schreckenberg
(1992), as well as Kerner and Konh\"auser (1993). These papers
initiated an avalanche of publications in various international physics journals. 
Meanwhile, it is difficult to keep
track of the scientific development and literature. 
Therefore, it is high time to write a review on traffic, which tries
to bring together the different approaches and to show up their 
interrelations. Doing so, I will take into account 
many significant contributions by traffic engineers as well.
Hopefully, this will
stimulate more intense discussions and cooperations among the
different disciplines involved.
\par
In the following sections, I will try to give an overview of the field of traffic dynamics
from the perspective of many-particle physics.

\subsection{Driven many-particle systems} \label{IC}

A big challenge for physicists are the dynamics and pattern formation
in systems far from equilibrium, especially living systems
(Nicolis and Prigogine, 1977; Haken, 1977, 1983, 1988; 
Pasteels and Deneubourg, 1987; Weid\-lich, 1991;
Feistel and Ebeling, 1989; Vicsek, 1992; Kai, 1992;
DeAngelis and Gross, 1992; Vallacher and Nowak, 1994;  Cladis and Palffy-Muhoray, 1995;
Helbing, 1995a). 
Non-equilibrium systems are characterized by not being closed, i.e.,  by having
an exchange of energy, particles, and/or information with their environment. 
Consequently, they often show complex behavior and, normally, there are no
general results such as the laws of thermodynamics and statistical mechanics
for closed systems of
gases, fluids, or solids in equilibrium. 
Nevertheless,  they can be cast in a general mathematical 
framework (Helbing, 2001), as I will show here.

\subsubsection{Classical mechanics, fluids, and granular media}\label{IC1}

Let us start with Newton's equation of motion from classical mechanics,
which describes the acceleration $\ddot{\vec{x}}_\alpha(t)$ of a body $\alpha$ 
of mass $m_\alpha$ subject to pair interactions with other bodies $\beta$:
\begin{equation}
 m_\alpha 
\ddot{\vec{x}}_\alpha(t) = \sum_{\beta(\ne \alpha)} \vec{F}_{\alpha\beta}(t) \, .
\end{equation}
The interaction forces $\vec{F}_{\alpha\beta}(t)$ are mostly dependent on the locations
$\vec{x}_\alpha(t)$ and $\vec{x}_\beta(t)$ of the interacting bodies $\alpha$ and $\beta$ at
time $t$. Often, they depend only on the distance vector
$\vec{d}_{\alpha\beta} = (\vec{x}_\beta - \vec{x}_\alpha)$, but in special cases, they are also
functions of the velocities $\vec{v}_\alpha(t) = \dot{\vec{x}}_\alpha(t)$ and
$\vec{v}_\beta(t) = \dot{\vec{x}}_\beta(t)$. For potential forces, the 
above many-body system can be
characterized by a Hamilton function. A typical example is the description of
the motion of celestial bodies.
\par
In {\em driven} many-body systems such as fluids under the influence of pressure
gradients and boundary forces or vibrated granular media like sand, 
we have to consider additional interactions
with the environment. Therefore, we need to
consider additional terms. This concerns (external) driving forces 
$\vec{F}_0(\vec{x},t)$ due to boundary interactions and
gravitational or electrical fields, 
(sliding) friction forces $\vec{F}_{\rm fr}(t) =
-\gamma_\alpha \vec{v}_\alpha(t)$ with friction coefficient $\gamma_\alpha$, and individual
fluctuations $\vec{\zeta}_\alpha(t)$ reflecting thermal interactions with the
environment (boundaries, air, etc.) or a variation of the surface structure of
the particles:
\begin{eqnarray}
 m_\alpha \ddot{\vec{x}}_\alpha(t) &=&
\vec{F}_0\bbox(\vec{x}_\alpha(t),t\bbox) 
 - \gamma_\alpha \vec{v}_\alpha(t)
 + \sum_{\beta(\ne \alpha)} \vec{F}_{\alpha\beta}(t) \nonumber\\
 &+& \vec{\zeta}_\alpha(t) \, .
\label{two}
\end{eqnarray}
From these so-called ``microscopic'', molecular-dynamics type of
equations, one can systematically derive ``macroscopic'', fluid-dynamic
equations for the spatio-temporal evolution of the particle density,
the  momentum density or average velocity, and the energy density or
velocity variance (related to the temperature). The 
methods for the construction of this micro-macro link will be sketched in
Sec.~\ref{IIB5}. 
\par 
In driven systems, the ongoing competition between the driving forces 
and the dissipative friction forces leads to a spatio-temporal redistribution
of energy, which produces a great variety of self-organization
phenomena. This results from 
non-linearities in the equations of motion, which allow small
initial perturbations to be enhanced and non-equilibrium patterns to be
dynamically stabilized. In fluids one can find the
formation of waves or vortices, bifurcation scenarios like period doubling behavior,
the Ruelle-Takens-Newhouse route to chaos, intermittency, or  
turbulence, depending on the particular boundary conditions
(Joseph, 1976; Landau and Lifshits, 1987; Drazin and Reid, 1981; 
Swinney and Gollub, 1985; Schuster, 1988;  Gro{\ss}mann, 2000).
In this review, it is important to know that 
turbulence normally requires three- or higher-dimensional systems, so it 
is not expected to appear in one- or two-dimensional vehicle or
pedestrian traffic. However, the instability mechanism 
for the explanation of the {\em subcritical
transition} to turbulence in the Hagen-Poiseuille experiment 
(Gebhardt and Gro{\ss}mann, 1994; Gro{\ss}mann, 2000) may
also be relevant to traffic systems, as pointed out
by Krug (cf. the discussion of metastability in
Sec.~\ref{IIC4}).
\par
In vibrated granular media, one can find emergent convection
patterns (Ehrichs \etal, 1995; Bourzutschky and Miller, 1995; P\"oschel and Herrmann, 1995), 
collective oscillating states (so-called oscillons, see 
Umbanhowar \etal, 1996), spontaneous {\em segregation} of different granular materials
(P\"oschel and Herrmann, 1995; Santra \etal, 1996; Makse \etal, 1997),
or {\em self-organized criticality} with power-law distributed 
avalance sizes in growing sand heaps (Bak \etal, 1987, 1988; Bak, 1996) 
or in the outflow from 
hoppers (Schick and Verveen, 1974; Peng and Herrmann, 1995). 
\par
In spite of many differences between flows of fluids,
granular media, and vehicles or pedestrians due to different conservation laws
and driving terms,  
one can apply similar methodological approaches, e.g.\\
(i) {\em microscopic, molecular dynamic models} (see, e.g., Hoover, 1986; 
Goldhirsch \etal, 1993; Buchholtz and P\"oschel, 1993; 
Hirshfeld \etal, 1997; Sec.~\ref{IIB1}),\\
(ii) {\em lattice gas automata} (Frisch \etal, 1986; Chen \etal, 1991;
Peng and Herrmann, 1994; Tan \etal, 1995) 
or {\em cellular automata} (see Sec.~\ref{IIB2}),\\
(iii) {\em gas-kinetic (Boltzmann- and Enskog-like) models} (Boltzmann, 1964; Enskog, 1917;
Chapman and Cowling, 1939; Cohen, 1968, 1969; 
Lun \etal, 1984; Jenkins and Richman, 1985; Kobryn \etal, 1996; Dufty \etal, 1996;
Lutsko, 1997;
Cercignani and Lampis, 1988; McNamara and Young, 1993; Sela \etal, 1996;
see Sec.~\ref{IIB5}), and\\
(iv) {\em fluid-dynamic models} (Haff, 1983; Goldhirsch, 1995; Sela and Goldhirsch, 1995;
Hayakawa \etal, 1995; Du \etal, 1995; see Sec.~\ref{IIB4}). 

\subsection{Self-driven many-particle systems}\label{ID}

In self-driven many-particle systems, the driving force is not of
external origin (exerted from outside), but associated with each 
single particle and self-produced. This
requires each particle to have some kind of internal energy reservoir 
(Schweitzer \etal, 1998b; Ebeling \etal, 1999).  
\par
Self-driven ``particles'' are
a paradigm for many active or living systems, where they are
a simplified and abstract representation of the most important
dynamic behavior of cells, animals, or even humans.
In order to reflect this, I will generalize Eq.
(\ref{two}) a little, replacing the external driving force  $\vec{F}_0(\vec{x}_\alpha,t)$
by an individual driving force $\vec{F}_\alpha^0(t)$. 
Moreover, Newton's third law $\vec{F}_{\beta\alpha}(t) = - \vec{F}_{\alpha\beta}(t)$ ({\em
  actio = reactio}) does not necessarily apply anymore to the ``self-driven'',
``self-propelled'', ``motorized'', or ``active'' ``particles'' we have in mind. I will show that these
minor changes will imply various interesting phenomena observed in 
nature, for example in biological, traffic, or socio-economic systems. In this
context, the masses $m_\alpha$ are sometimes not well-defined, and
it is better to rewrite the resulting equation by means of
the scaled quantities $\vec{F}_\alpha^0(t) = \gamma_\alpha v_\alpha^0(t) \vec{e}_\alpha^0(t)$, 
$\gamma_\alpha = m_\alpha/\tau_\alpha$, $\vec{F}_{\alpha\beta}(t) = m_\alpha \vec{f}_{\alpha\beta}(t)$, and 
$\vec{\zeta}_\alpha(t) = \gamma_\alpha \vec{\xi}_\alpha(t)$, where the accelerations
$\vec{f}_{\alpha\beta}(t)$ are often loosely called forces as well:
\begin{equation}
 \frac{d\vec{v}_\alpha(t)}{dt} = \frac{v_\alpha^0(t)\vec{e}_\alpha^0(t) + \vec{\xi}_\alpha(t) -
   \vec{v}_\alpha(t)}{\tau_\alpha} + \sum_{\beta(\ne \alpha)} \vec{f}_{\alpha\beta}(t) \, .
\label{three}
\end{equation}
From this equation we can see that, with a
relaxation time of $\tau_\alpha$,  the driving term $v_\alpha^0(t)\vec{e}_\alpha^0(t)/\tau_\alpha$
and friction term $-\vec{v}_\alpha(t)/\tau_\alpha$ together
lead to an exponential-in-time adaptation of the velocity
$\vec{v}_\alpha(t)$ to the desired speed $v_\alpha^0(t)$ 
and the desired direction $\vec{e}_\alpha^0(t)$ of motion. 
This is, however, disturbed by fluctuations
$\vec{\xi}_\alpha(t)$ and interactions $\vec{f}_{\alpha\beta}(t)$ with other particles $\beta$.
It is clear that attractive forces  $\vec{f}_{\alpha\beta}(t)$ will lead to
agglomeration effects. Therefore, we will usually investigate systems with
vanishing or repulsive interactions, for which one can find various
surprising effects. 
\par
A further simplification of the model equations can be reached in the
overdamped limit $\tau_\alpha \approx 0$ of fast (adiabatic) relaxation. 
In this case and with the abbreviations $\vec{v}_{\alpha\beta}(t) = \tau_\alpha
\vec{f}_{\alpha\beta}(t)$, $\vec{\xi}_\alpha(t) = v_\alpha^0(t) \vec{\chi}_\alpha(t)$, we obtain
\begin{eqnarray}
  \vec{v}_\alpha(t) &=& v_\alpha(t) \vec{e}_\alpha(t)  = v_\alpha^0(t) \vec{e}_\alpha^0(t) 
 + \!\!\sum_{\beta (\ne \alpha)} \!\!\vec{v}_{\alpha\beta} (t) + \vec{\xi}_\alpha(t) 
 \nonumber \\
 &=& v_\alpha^0(t) \bbox[ \vec{e}_\alpha^0(t) +
  \vec{\chi}_\alpha(t) \bbox] + \!\!\sum_{\beta (\ne \alpha)} \!\!\vec{v}_{\alpha\beta} (t) \, .
\label{four}
\end{eqnarray}

\subsubsection{The concept of social (behavioral) forces}\label{ID1}

Human behavior often seems to be ``chaotic'', irregular, and unpredictable. So, 
why and under which conditions can we apply the above force equations?
First of all, we need to be confronted with a phenomenon of motion in some
(quasi-)continuous space, which may be also an abstract behavioral space or opinion
scale (Helbing, 1992a, 1993b, 1994, 1995a). 
It is favourable to have a system where the fluctuations due to
unknown influences are not large compared to the systematic,
deterministic part of motion. This is usually the case in pedestrian and vehicle 
traffic, where people are confronted with standard situations and react
automatically rather than making complicated decisions between various possible
alternatives. For example, an experienced driver would not have to think about the
detailled actions to be taken when turning, accelerating, or changing lanes.
\par
This automatic behavior can be interpreted as the result of a
{\em learning process} based on trial and error
(Helbing \etal, 2001c), which can be simulated with {\em evolutionary algorithms} 
(Klockgether and Schwefel, 1970; Rechenberg, 1973; Schwefel, 1977;
Baeck, 1996). For example, pedestrians have a preferred side of walking (Oeding, 1963;
Older, 1968; Weidmann, 1993), since an asymmetrical avoidance behavior 
turns out to be profitable (Bolay, 1998). The related
{\em formation of a behavioral convention} can be described by means 
of {\em evolutionary game theory} (Helbing, 1990, 1991, 1992a, c, 1993a, 1995a, 1996c).
\par
Another requirement is the vectorial additivity of the separate force terms
reflecting different environmental influences and psychological factors. 
This is probably an approximation, but
there is some experimental evidence for it. Based on quantitative measurements for animals and
test persons subject to separately or simultaneously applied
stimuli of different nature and strength, one could show that the behavior in
conflict situations can be described by a superposition of forces (Miller, 1944, 1959; Herkner, 1975).
This fits well into a concept by Lewin (1951), according to which behavioral changes
are guided by so-called {\em social fields} or {\em social forces}, which has
been put into mathematical terms by Helbing (1991, 1992a, 1993b, 1994, 
1995a; see also Helbing and Moln\'{a}r, 1995). The third Newtonian law, however,
is usually not valid.

\subsection{What this review is about} \label{IB}

In the following sections,\\
(i) I will focus on the {\em phenomena actually observed} in traffic and their
characteristic properties, while I will 
discuss models only to the extend, to which they are helpful and necessary for an
interpretation and a better understanding of the observations,\\
(ii) I will discuss the main methods from statistical physics 
  relevant for modelling and analyzing traffic dynamics (see the table of contents),\\
(iii) 
I will discuss how far one can get with a
physical, many-particle description of traffic, neglecting socio-psychological factors
and human behavior. 
The limitations of this approach will be shortly discussed in Sec.~\ref{IIH}.
\par
This review is intended to serve both, experts and newcomers in the field,
so some matters will be simplified or explained in more detail for didactical reasons.
Moreover, I will try to identify
open problems and to shed new light on some controversial topics presently
under discussion. The main focus will be on the various kinds of
phenomena occuring in self-driven many-particle systems and the conditions, under which
they appear. I will start with the subject of one-dimensional vehicle traffic,
continue with two-dimensional pedestrian traffic and three-dimensional
air traffic by birds. 
\par
Those physicists who don't feel comfortable with
non-physical systems may instead imagine special kinds of granular, colloidal, or spin
systems driven by gravitational or electrical forces, or imagine particular systems
with {\em Brownian motors} (H\"anggi and Bartussek, 1996;
Astumian, 1997; J\"ulicher \etal, Prost, 1997; Reimann, 2000). Moreover, I would like to
encourage everyone to perform analogous experiments in these related physical systems.
\par
Despite the complexity of traffic and complications by unknown, 
latent, or hardly measurable human factors, physical traffic theory is meanwhile a prime
example of a highly advanced quantitative description of a living system.
An agreement with empirical data has been reached not only 
on a qualitative, but even on a semi-quantitative level (see Sec.~\ref{IID1}).
Moreover, there are even ``natural constants'' of traffic,
emerging from the non-linear vehicle interactions, see Sec.~\ref{IIC5}. 
But, before getting into all this, I would like to mention some other
interesting (self-)driven many-particle systems loosely
related to traffic.

\subsection{Particle hopping models, power-law scaling, and SOC} \label{ID2}

Many-particle systems in equilibrium can be very well understood with
methods from thermodynamics (Keizer, 1987)  and statistical physics
(Uhlenbeck and Ford, 1963; Landau and Lifshits, 1980; Ma, 1985;
Klimontovich, 1986; Huang, 1987).
Examples are phase transitions between different
aggregate states of matter like vapor, water, and ice, or the
magnetization of spin systems composed of many 
``elementary magnets''. It is known that the phase transition of
disordered many-particle systems with short-range interactions
into states with long-range order is easier in
higher dimensions, because of the greater number of neighbors to
interact with. Often, there is a certain upper dimension, above which
the system can be described by a mean field approach so that the
fluctuations can be neglected. For lower-dimensional spaces, one
can usually develop approximate theories for the influence of noise
by means of suitable expansions and renormalization-group treatments
based on scaling arguments. These give universal power-law scaling exponents for
the (``critical'') behaviour of the system close to the phase transition point
(critical point) (Stanley, 1971; Domb and Green, 1972--1976; 
Ma, 1976; Hohenberg and Halperin, 1977; Stinchcombe, 1983; 
Domb and Lebowitz, 1983--2000; Voss, 1985; Schuster, 1988).
In non-equilibrium systems, one can frequently find {\em self-organized
criticality} (Bak \etal, 1987, 1988; Bak, 1996)
and {\em fractal properties} (Mandelbrot, 1983; Family and Vicsek, 1991; Vicsek, 1992;  
Barab\'{a}si and Stanley, 1995). Self-organized criticality (SOC)
means that the respective system drives itself to the critical point, which is normally characterized
by long-range correlations and scale-free power laws in analogy to thermodynamics.
\par
In the considered many-particle systems,
there is often also a certain lower dimension, below
which the system is always disordered, because of the small number 
of neighbors. For example, it is known that
neither ferromagnetic nor antiferromagnetic order is possible in
one-dimensional equilibrium spin systems (Mermin and Wagner, 1966). The discovery
that the situation can be very different in
{\em driven}, non-equilibrium spin systems, has initiated an
intense research activity regarding the phase transitions in 
one-dimensional driven diffusive systems far from equilibrium.
It turned out that many properties of
equilibrium systems can be generalized to non-equilibrium systems, but
others not. 
\par
By means of particle hopping models, it was possible to gain a
better understanding of {\em directed percolation} (Domany and Kinzel,
1984), of {\em spontaneous structure formation} (Vilfan \etal, 
1994), of {\em spontaneous symmetry breaking} (Evans \etal, 1995), of the
{\em roughening transition} in certain growth processes (Alon \etal, 1996),
of the non-equilibrium {\em wetting transition} (Hinrichsen \etal, 1997),
and of {\em phase separation} (Evans \etal, 1998; Helbing \etal, 1999b). Nevertheless, 
for these non-equilibrium transitions there is still no general
theory available which would be of comparable generality as 
equilibrium thermodynamics or statistical physics. 
For further reading, I recommend the books by
Spohn (1991), Schmittmann and Zia (1995, 1998), Derrida and Evans
(1997), Liggett (1999),
and Sch\"utz (2000).
\par
Note that the above mentioned models usually assume a {\em random sequential 
(asynchronous) update}, i.e. the state
of each particle is randomly updated after an exponentially distributed 
waiting time $\Delta t$. However, it is known that this is not realistic for traffic systems,
which rather require a {\em parallel (synchronous) update} of the vehicle locations 
(Schreckenberg \etal, 1995), as in other flow problems. To reflect the driver 
reaction to the corresponding change of the traffic situation, it is common
to update the vehicle speeds synchronously as well. However, this is debatable, since
some models of spatio-temporal interactions in social systems show
artefacts, if updated in parallel (Huberman and Glance, 1993). For example,
a parallel update excludes the dynamical reachability of certain states 
(Schadschneider and Schreckenberg, 1998), which are called
paradisical or Garden-of-Eden states (Moore, 1962).

\subsubsection{The asymmetric simple exclusion process (ASEP)} \label{IE1}

The only particle hopping model I will discuss here is
the asymmetric simple exclusion process. Particularly relevant
contributions to this model go back to Spitzer (1970),
Liggett (1975), 
Domany and Kinzel (1984), Katz {\em et al.} (1984),
Liggett (1985),
Krug (1991), Derrida {\em et al.} (1992), Janowski and Lebowitz (1992), 
Derrida \etal\ (1993),
Ferrari (1993), Sch\"utz and Domany (1993), Stinchcombe and Sch\"utz (1995a, b),
Kolomeisky \etal (1998), and Sch\"utz (1998). For overviews see
Spohn (1991), Schmittmann and Zia (1995, 1998), 
Liggett (1999), and Sch\"utz (2000). 
In the following, I will restrict to the
{\em totally} asymmetric simple
exclusion process (TASEP). This model is defined by $L$ sites 
$j$ of a one-dimensional lattice, which can be either empty
(corresponding to the occupation number $n_j = 0$) or
occupied by one particle ($n_j=1$). The particle locations
and occupation numbers are updated every time step $\Delta t$.
When updated, a particle in cell $j$
hops to the right neighboring cell $(j+1)$ with probability $q$, if
this is empty, otherwise it stays in cell $j$. The total rate
of moving to the right is, therefore, given by $qn_j (1-n_{j+1})/\Delta 
t$. The boundaries are 
characterized as follows: A particle enters the system at the
left-most cell $j=1$ with probability $q_0$, if this is empty.
Moreover, a particle in the right-most cell $j=L$ leaves the system
with probability $q_L$. This corresponds to particle reservoirs
at the boundaries which can be described by constant occupation
probabilities $n_0 = q_0/q$ and $n_{L+1} = (1 - q_L/q)$.  
Although it takes enourmous efforts, the stationary states 
and even the dynamics of the TASEP can be analytically determined. For this, one has to solve the
corresponding master equation (see Secs.~\ref{TASEP} and \ref{IIB4a}).    

\subsection{Active Brownian particles}\label{ID4}

Like usual Brownian particles, {\em active} Brownian particles perform
a random walk as well. However, they are not only reactive
to an external potential $U(\vec{x},t)$, but they are driven by an internal
energy reservoir (``pumped particles'') or can actively change
the potential $U(\vec{x},t)$ while moving (``active walkers''). 
Therefore, I have recently suggested 
to call them {\em Brownian agents}. For an overview see Schweitzer (2001).

\subsubsection{Pumped Brownian particles}\label{IF1}

Schweitzer {\em et al.} (1998b) suggest a model 
describing how self-driven particles may take up and consume the 
internal energy behind their driving force. Their model corresponds 
to Eq.~(\ref{two}) with the specifications $\vec{F}_{\alpha\beta} = 0$
(i.e. no direct interactions), but $\vec{F}_0\bbox(\vec{x}_\alpha(t)\bbox)$
is replaced by the expression $ - \vec{\nabla} U(\vec{x}_\alpha) 
+ \gamma E_\alpha(t) \vec{v}_\alpha(t)$,
where the first term 
is an {\em external potential force} and the last term an {\em internal driving
force} ($\gamma$ being the friction coefficient). The
dimensionless (scaled) energy reservoir $E_\alpha(t)$ is assumed to
follow the equation
\begin{equation}
 \frac{dE_\alpha(t)}{dt} = Q_0\bbox(\vec{x}_\alpha(t)\bbox) 
 - \{c + d \, [\vec{v}_\alpha(t)]^2 \} E_\alpha(t) \, .
\end{equation}
Herein, $Q_0(\vec{x})$ reflects the exploitation rate of energy resources, 
while the last term describes the consumption rate of energy,
which grows quadratically with the speed ($c$ and $d$ being suitable
parameter values). If the relaxation of this
energy equation is fast, we can approximate the driving term by
\[
 \frac{v_\alpha^0(t)\vec{e}_\alpha^0(t)}{\tau_\alpha}
 = \frac{\gamma}{m_\alpha} E_\alpha(t) \vec{v}_\alpha(t)
 = \frac{\gamma}{m_\alpha} 
 \frac{Q_0\bbox(\vec{x}_\alpha(t)\bbox)\vec{v}_\alpha(t)}{c+d\,[\vec{v}_\alpha(t)]^2}\, ,
\]
so that the internal driving and the dissipative friction together
can be represented by an {\em active} friction coefficient
\(
\gamma'(v) = \gamma [1 - Q_0/(c+d v^2)].
\)
Moreover, the driving direction $\vec{e}_\alpha^0(t)$
is given by the normalized actual velocity 
$\vec{e}_\alpha(t) = \vec{v}_\alpha(t)/\|\vec{v}_\alpha(t)\|$, and
the desired velocity $v_\alpha^0(t)$ depends on the speed as
well:
\(
 v_\alpha^0(t) = 
 Q_0\|\vec{v}_\alpha(t)\|/\{c+d[\vec{v}_\alpha(t)]^2\} \ge 0 \, .
\) 
Notice that particle $\alpha$ takes up
energy only when it moves with some finite speed $v_\alpha(t) =
\|\vec{v}_\alpha(t) \| \ne 0$ 
(i.e. exploits its environment). Therefore, in the absence of a
potential $U(\vec{x},t)$, we find the stationary solution
$v_\alpha = v_\alpha^0 = 0$. However, this is only stable under the
condition $\gamma Q_0 < c m_\alpha$. Otherwise, particle $\alpha$ will
spontaneously start to move with average speed
\begin{equation}
 v_\alpha = v_\alpha^0 = \left( \frac{\tau_\alpha \gamma Q_0}{m_\alpha d}
 - \frac{c}{d} \right)^{1/2} \, .
\end{equation}
Depending on $Q_0(\vec{x})$ and $U(\vec{x})$, 
this spontaneous motion displays interesting dynamics such as
limit cycles, deterministic chaos, or intermittency 
(Schweitzer \etal, 1998b; see also Chen, 1997).
It also allows the particles to climb potential hills, e.g. in a
periodic ratchet potential $U(\vec{x})$ (Schweitzer \etal, 2000). 

\subsubsection{Dissipative Toda and Morse chains}\label{IF2}
 
Toda and Morse chains are particles coupled to a heat bath and
moving on a ring with particular {\em asymmetrical} 
springs among neighbors, which are described by non-linear Toda or Morse potentials
$U'(x_\beta-x_\alpha)$  (Bolterauer and Opper, 1981; Toda, 1983;
Jenssen, 1991; Ebeling and Jenssen, 1992; Dunkel \etal, 2001).
Assuming pumped particles with an active friction coefficient $\gamma'(v)$
similar to the previous paragraph,
Ebeling \etal \ (2000) have found interesting dynamical patterns for
overcritical pumping. This includes uniform rotations, one- and
multiple soliton-like excitations, and relative oscillations. For Morse potentials,
one also observes clustering effects which partly remind of jamming
(Dunkel \etal, 2001). 

\subsubsection{Active walker models}\label{IF3}

By modifying their environment {\em locally}, active walkers
have indirect interactions with each other, which may lead to the
formation of {\em global} structures. The versatility of this
concept for the description of physical, chemical, biological, and socio-economic
systems is discussed by Schweitzer (2001). For example, a simple model is given by
\(
 m_\alpha d\vec{v}_\alpha/dt = [ - \vec{\nabla} 
 U(\vec{x}_\alpha,t)
 - \gamma \vec{v}_\alpha(t) + \vec{\zeta}_\alpha(t) ]
\)
and changes of the environmental potential $U(\vec{x},t)$ according to
\[
 \frac{\partial U(\vec{x},t)}{\partial t} \!=\!
 - \!\! \sum_\alpha\! B\, \delta \bbox(\vec{x} - \vec{x}_\alpha(t)\bbox)
 - C U(\vec{x},t) + D\,\Delta U(\vec{x},t)  .
\]
Herein, $C$ and $D$ are constants, and $\delta (\vec{x} - \vec{x}_\alpha)$ denotes Dirac's delta
function, giving contributions only at $\vec{x} = \vec{x}_\alpha$.
Therefore, the first term on the right hand side 
reflects particles $\alpha$ leaving attractive markings at their respective
locations $\vec{x}_\alpha$.
The last term describes a diffusion  of the
field $U(\vec{x},t)$ (e.g., a chemical one), and the previous term its decay.
\par
The result of the dynamics is a spatial {\em agglomeration process}. 
While in the first stage, one finds the
emergence of localized clusters at random locations, the clusters
eventually merge with each other, resulting in one big cluster.
The growth and competition of these clusters can be described 
by an {\em Eigen-Fisher-like selection equation}
(Schweitzer and Schimansky-Geier, 1994; Fisher, 1930; Eigen, 1971; Eigen and Schuster, 1979).

\subsubsection{Pattern formation of bacterial colonies}\label{IF4}

Already in 1994, Eshel Ben-Jacob \etal \ 
have proposed a so-called {\em communicating walker model} for
the beautiful pattern formation in certain bacterial colonies
(for reviews see Ben-Jacob, 1997; Ben-Jacob \etal, 2000). This model
takes into account the consumption, diffusion, and depletion of
food, the multiplication under good growth conditions, and the
transition to an immobile spore state under starvation, as well as
the effect of {\em chemotaxis}, i.e. the attractive or repulsive reaction 
to self-produced chemical fields. The model allows one to reproduce 
the observed growth patterns of bacterial colonies as a function of
the nutrient concentration and the agar concentration
determinining how hard the bacteria can move.
The same bacterial patterns have recently been reproduced by means of
a macroscopic reaction-diffusion model (Golding \etal, 1998). The 
formation of the observed dendritic structures is due to a so-called
{\em diffusion instability} (Ben-Jacob, 1993, 1997).

\subsubsection{Trail formation by animals and pedestrians}\label{IF5}

Another example
is the observed formation of trunk trail systems by
certain ant species, which is also based on chemotaxis.
It can be modelled by two kinds of
chemical markings, one of which is dropped during the search for food,
while the other is dropped on the way back to the nest of the ant
colony (Schweitzer \etal, 1997). Other models for ant-like swarm formation
have been developed by Deneubourg \etal \ (1989) and Bonabeau (1996). 
\par
Further kinds of active walker models have been proposed for
the self-organization of trail systems by pedestrians or animals, where the
markings correspond to ``footprints'' or other kinds of
modifications of the ground which make it more comfortable to 
move (Helbing \etal, 1997a, b; Helbing, 1998c).
Interestingly enough, the model yields {\em fair solutions} and 
{\em optimal compromises} between
short and comfortable ways, see Fig.~\ref{tr} (Helbing, 1998c; Helbing and Vicsek, 1999).
The corresponding computer simulations are, therefore, a
valuable tool for developing optimized pedestrian facilities and
way systems.
\unitlength1cm
\begin{figure}[htbp]
\caption[]{Schematic representation of a human trail system (black solid lines)
evolving between four entry points and destinations (full circles)
on an initially homogeneous ground (after Helbing \etal, 1997a,
b, 2001c; Helbing, 1998c; Helbing and Vicsek, 1999). 
When the frequency of trail usage
is small, the direct way system (consisting of the four
ways along the edges and the two diagonal connections) is too long to
be maintained in competition with the regeneration
of the vegetation. Here, by bundling of trails, the frequency of usage
becomes large enough to support the depicted trail system. It corresponds
to the optimal compromise between the diagonal ways and the
ways along the edges, supplying maximum walking comfort at a minimal
detour of 22\% for everyone, which is a fair solution. 
\label{tr}}
\end{figure}

\subsection{Vehicle and pedestrian traffic}\label{II}

The next sections will mainly focus on unidirectional freeway traffic, but
bidirectional traffic and two-dimensional
city traffic are shortly sketched as well.
I will start with an overview over the most
important empirical findings.
Afterwards, I will concentrate on the different modelling
approaches and discuss the properties of these models,
starting with one-lane traffic of identical vehicles and then adding
more and more details, including heterogeneous traffic on multi-lane
roads. Finally, I will discuss two-dimensional pedestrian traffic
and three-dimensional flocks of birds.

\section{Empirical findings for freeway traffic} \label{IIA}

As physics deals with phenomena in our physical world, a physical
theory must face the comparison with empirical data. A good theory
should not only reproduce all the empirically known properties of
a system, but it should also make predictions allowing us to verify or 
falsify the theory. To gain on overview of data analyses, I recommend to read
the review articles and books by Gerlough and Huber (1975),
Koshi \etal\ (1983), the Transportation Research Board (1985),
May (1990), Daganzo (1997a, 1999a, b), Bovy (1998), and
Kerner (1998b, c, 1999a, b, 2000a, b). 

\subsection{Measurement techniques} \label{IIA1}

Probably the most refined technique to gain empirical data is based on 
{\em aerial photography}, allowing
us to track the trajectories of many interacting vehicles and even their
lane changing maneuvers (Treiterer and Taylor, 1966; 
Treiterer and Myers, 1974). 
A method suitable for
experimental investigations are {\em car-following data}. Depending on the equipment of the cars,
it is possible to determine the location and speed, possibly
the acceleration and {clearance} (netto distance), and sometimes even lane changing 
maneuvers of the equipped car or of the respective following vehicle 
(see, for example, Koshi \etal, 1983; Bleile, 1997a, b, 1999).
\par
However, most data are obtained by
detectors located at certain cross sections $x$ of the freeway. 
For example, {\em single induction-loop detectors} measure the numbers $\Delta N$ of
crossing vehicles $\alpha$
during a certain {\em sampling interval} $\Delta T$  as well as the 
times $t_\alpha^0$ and $t_\alpha^1$ when a vehicle $\alpha$ reaches and leaves the
detector. This facilitates to determine 
the {\em time headways} (brutto time separations)
\[
 \Delta t_\alpha = (t_{\alpha}^0 - t_{\alpha-1}^0) 
\] 
and the {\em time clearances} (netto time separations or time gaps) 
\(
t_\alpha^0 - t_{\alpha-1}^1 
\)
including their respective distributions, as well as 
the {\em vehicle flow} 
\begin{equation}
 Q(x,t) = \frac{\Delta N}{\Delta T} 
\label{FLOW}
\end{equation}
and the {\em time occupancy}
\(
 O(x,t) = \sum_\alpha (t_\alpha^1 - t_\alpha^0)/\Delta T 
\)
(where $\alpha_0 < \alpha \le \alpha_0+\Delta N$ and  $\alpha_0$ denotes the last vehicle before
the sampling interval begins). 
The newer {\em double induction-loop detectors} additionally 
measure the vehicle velocities $v_\alpha$ and the vehicle lengths
$l_\alpha$, allowing us to estimate the {\em headways} (brutto distances)
\(
d_\alpha = v_\alpha \, \Delta t_\alpha
\) 
(assuming constant vehicle speeds) and the {\em clearances} (netto distances)
\(
s_\alpha = (d_\alpha - l_{\alpha -1}).
\) 
The velocity  measurements are normally used to
obtain the (arithmetic) {\em average velocity}
\begin{equation}
 V(x,t) = \langle v_\alpha \rangle 
 = \frac{1}{\Delta N} \sum_{\alpha = \alpha_0 +1}^{\alpha_0+\Delta N} v_\alpha 
\label{arithm}
\end{equation}
but the {\em velocity variance}
\begin{equation}
 \theta(x,t) = \Big\langle 
 [ v_\alpha - \langle v_\alpha \rangle ]^2 \Big\rangle =
 \langle (v_\alpha)^2 \rangle - \langle v_\alpha \rangle^2 
\end{equation}
and the local velocity distribution may be determined as well.
The vehicle density $\rho(x,t)$ is often calculated via the fluid-dynamic formula
(\ref{densdef}), i.e. 
\( 
\rho(x,t) = Q(x,t)/V(x,t) .
\)
Another method is to relate the density with the time occupancy, as in 
the formula
\(
 \rho(x,t) = O(x,t)/[L(x,t)+L_D] ,
\)
where $L(x,t)$ is the average vehicle length during the measurement
interval and $L_D$ the detector length (i.e. the loop extension) in driving direction (May, 1990). 
\par
The problem of the above measurement methods is that the velocity
distribution measured over a time interval $\Delta T$ differs from 
the one measured on a freeway section of length $\Delta X$. In other
words, temporal and spatial averaging yield different results, since
fast vehicles cross a section of the freeway more frequently than slow 
ones, which is intuitive for a circular multi-lane freeway with fast
and slow lanes. The problem of determining the empirical density via
the formula $\rho = Q/V$ is
that it mixes a temporal average (the flow) with a spatial one (the
density). This can be compensated for by the harmonic
average velocity $V(x,t)$ defined by
\begin{equation}
 \frac{1}{V(x,t)} = \left\langle \frac{1}{v_\alpha} \right\rangle
 \, ,
\label{harmav}
\end{equation}
giving a greater weigth to small velocities
(Gerlough and Huber, 1975; Leutzbach, 1988). Using formula (\ref{harmav}) instead
of the commonly applied formula (\ref{arithm}) 
together with the relation $\rho = Q/V$  results in similarly shaped
velocity-density relations, but the density range is
much higher (see Fig.~\ref{VDENS}). Moreover, the velocity and flow
values at high densities are somewhat lower.
The disadvantage of Eq.~(\ref{harmav}) is its great sensitivity
to errors in the measurement of small velocities $v_\alpha$.
\begin{figure}[htbp]
\caption[]{Empirical velocity-density relations for different 
definitions of the average velocity. Symbols represent averages of
1-minute data determined via the harmonic velocity formula $V =
1/\langle 1/v_\alpha \rangle$, while the solid line is a fit function to the
velocity averages determined via the conventionally applied arithmetic
formula $V =  \langle v_\alpha \rangle$.
(After Helbing, 1997a.)\label{VDENS}}
\label{empvel}
\end{figure}

\subsection{Fundamental diagram and hysteresis} \label{IIA2}

Functional relations between the vehicle flow $Q(x,t)$, the
average velocity $V(x,t)$, and the vehicle density $\rho(x,t)$ or 
occupancy $O(x,t)$ have been measured for decades, beginning
with Greenshields (1935), who found a linear relationship.
The name ``fundamental diagram'' is mostly used for some fit function 
\begin{equation}
 Q_{e}(\rho) = \rho V_{e}(\rho)
\end{equation}
of the  empirical flow-density relation, where $V_{e}(\rho)$ stands for the fitted 
empirical velocity-density relation (see Fig.~\ref{VDENS}), which is monotonically
falling:
\( 
 dV_{e}(\rho)/d\rho \le 0  .
\)
Most measurements confirm the following features:%
\\
(i) At low densities $\rho$, there is a clear and quasi
one-dimensional relationship between
the vehicle flow and the density. It starts almost linearly and is bent
downwards. The slope at low densities
corresponds to the average free velocity $V_0$, which can 
be sustained at finite densities as long as there are sufficient
possibilities for overtaking. That is, the velocity-density relation
of a multi-lane road starts horizontally, while it tends to have a negative
slope on a one-lane road.%
\\
(ii) With growing density, the velocity decreases monotonically,
and it vanishes together with the flow
at some {\em jam density} $\rho_{\rm jam}$, which is 
hard to determine because of the above mentioned measurement problems. Estimates by different
researchers from various countries reach from 120 to 200 vehicles per
kilometer and lane, but values between 140 and 160 vehicles per
kilometer are probably most realistic.%
\\
(iii) The vehicle flow has one maximum $Q_{\rm max}$ at medium
densities.
\\
(iv) The empirical flow-density relation is discontinous and looks
comparable to a mirror image of the Greek letter lambda. The two
branches of this reverse lambda are 
used to define free low-density and congested
high-density traffic (see, e.g., Koshi \etal, 1983;
Hall \etal, 1986;
Neubert \etal, 1999b; Kerner, 2000a; cf. also Edie and Foote, 1958). In congested
traffic, the average vehicle velocity
is significantly lower than in free traffic. Therefore,
the free and congested part of the flow-density relation can be
approximately separated by a linear flow-density relation
$\rho V_{\rm sep}$, where $V_{\rm sep}$ is the average velocity
below which traffic is characterized as congested. Around this line,
the density of data points is reduced (see Fig.~\ref{FUNDAM}a).
\begin{figure}[htbp]
\caption[]{Empirical time series of the hysteretic breakdown of traffic flow plotted
(a) as a function of the density and 
(b) as a function of the time. High-flow states are only observed shortly before the
breakdown of traffic. Note that, immediately after the breakdown, the flow drops
temporarily below the typical bottleneck flow (see Persaud \etal, 1998; Cassidy and Bertini, 1999).
While free traffic is characterized by a quasi one-dimensional
flow-density relationship, in the congested regime the flow-density data are widely 
scattered. The free and congested traffic regimes can be separated by a
line $\rho V_{\rm sep}$, where we have chosen
$V_{\rm sep} = 70$~km/h, here.\label{FUNDAM}}
\end{figure}
\par 
The reverse-lambda-like structure of the flow-density data implies
several things: First, we have a discontinuity at some
{\em critical density} $\rho_{\rm cr}$ (Edie, 1961; Treiterer and Myers, 1974; Ceder and May, 1976;
Payne, 1984), which has animated researchers  (Hall, 1987; Dillon and Hall, 1987;
Persaud and Hall, 1989; Gilchrist and Hall, 1989)
to relate traffic dynamics with {\em catastrophe theory} (Thom, 1975; Zeeman, 1977).
Second, there is a certain density 
region $\rho_{\rm c1}\le \rho \le \rho_{\rm c2}$ 
with $\rho_{\rm c2} = \rho_{\rm 
cr}$, in which we can have either free or congested traffic, which
points to {\em hysteresis} (Treiterer and Myers, 1974). The tip of
the lambda corresponds to {\em high-flow states} of free traffic (in the passing lanes), which
can last for many minutes (see, e.g., Cassidy and Bertini, 1999). However, these are
not stable, since it is only a matter of time until
a transition to the lower, congested branch of the lambda takes place
(Persaud \etal, 1998; see also Elefteriadou \etal, 1995). The 
transition probability from a free to a congested state 
within a long time interval is 0 below the flow $Q_{\rm c1}= Q_{e}(\rho_{\rm c1})$
and 1 above the flow $Q_{\rm c2} = Q_{e}(\rho_{\rm c2})$. Between $Q_{\rm c1}$ and
$Q_{\rm c2}$, it is monotonically increasing, see Fig.~\ref{BREAK}
(Persaud \etal, 1998).
The transition from congested to free traffic does not go back to
high-flow states, but to flows $Q_{e}(\rho)$ belonging to densities $\rho < \rho_{\rm 
c2}$, typically around $\rho_{\rm c1}$.  The high-flow states in the density range 
$\rho_{\rm c1}\le \rho \le \rho_{\rm c2}$ are only {\em metastable} 
(Kerner and Rehborn, 1996b, 1997; Kerner, 1998b, 1999a, b, c, 2000a, b).\\
\begin{figure}[htbp]
\caption[]{Probability of breakdown of free traffic as a function of the flow
for different waiting times (after Persaud \etal, 1998).\label{BREAK}}
\end{figure}
{\noindent (v)} The flow-density data in the congested part are widely scattered 
within a two-dimensional area (see, e.g., Koshi \etal, 1983;
Hall \etal, 1986; K\"uhne, 1991b).
This complexity in traffic flow is usually interpreted as
effect of fluctuations, of jam formation, or of an instability of vehicle
dynamics. The scattering is reduced by increasing
the sampling interval $\Delta T$ of the data averaging (Leutzbach, 1988).\\
(vi) By removing the data belonging to wide moving jams (see Fig.~\ref{WIDE}), Kerner and Rehborn
(1996b) could demonstrate that the remaining data of congested traffic
data still display a wide and two-dimensional scattering (see Fig.~\ref{FUNDAM}a), thereby
questioning the applicability of a fundamental diagram and 
defining the state of {\em ``synchronized flow''}  (``synchronized'' because of the typical
synchronization between lanes in congested traffic, see Fig.~\ref{suppl}a,
and ``flow'' because of flowing in contrast to standing traffic in fully developed jams).
Therefore, Kerner and Rehborn classify three traffic phases:\\
-- free flow,\\
-- synchronized flow (see Secs.~\ref{IIA3}, \ref{IIA4}, and \ref{IIA5b} for details), and\\
-- wide moving jams (i.e. moving jams whose width in longitudinal
direction is considerably higher than the width of the jam fronts).\\
In contrast to synchronized flow, a wide moving jam propagates either
through free or any kind of sychronized flow and through
bottlenecks (e.g. on- and off-ramps), keeping the propagation
velocity of the jam's downstream front (Kerner, 2000a, b).
In contrast to wide moving jams, after synchronized flow has occurred
upstream of an on-ramp, the downstream front of synchronized flow is fixed
at the on-ramp. In an initially free flow, two types of transitions
are observed: either to synchronized flow or to (a) wide moving jam(s).
Both of them appear to be first-order phase transitions accompanied by 
different hysteresis and nucleation effects (Kerner and Rehborn,
1997; Kerner, 1998a, 1999a, 2000c).\\
(vii) The slopes 
\(
 S_x^{\Delta T}(t) = [Q(x,t+\Delta T) - Q(x,t)]/[\rho(x,t+\Delta T) - \rho(x,t)] 
\)
of the lines connecting successive data points are always positive
in free traffic. In synchronized flow, however, they 
erratically take on positive and negative values, characterizing
a complex spatio-temporal dynamics (Kerner and Rehborn, 1996b). This
erratic behavior 
is quantitatively characterized by a weak cross-correlation between
flow and density (Neubert \etal, 1999b). Banks (1999) showed that
it could be interpreted as a result of random variations in the 
time clearances (partly due to acceleration and deceleration maneuvers
in unstable traffic flow). These variations are, in fact, very large
(cf. Fig.~\ref{TGAPS}). Banks points out that, if drivers would keep a save 
clearance $s^*(v) = s' + Tv$ in congested
traffic, the flow would grow with decreasing (effective) 
{\em safe time clearance} $T$ according to
\begin{equation}
 Q_{e}(\rho) = \frac{1}{T} \left( 1 - \frac{\rho}{\rho_{\rm jam}} \right)
\, ,
\label{GREEN}
\end{equation}
where $s'$ denotes the minimum {\em front-bumper-to-rear-bumper distance} 
and 
\(
 \rho_{\rm jam} = 1/l' = 1/(l+s')
\)
the jam distance with the (average) {\em effective vehicle length} $l' = (l + s')$.
Simplifying his argument, in the congested regime positive slopes can result from a
reduction in the safe time clearance $T$ with growing density, 
while negative slopes normally correspond to a
reduction in the speed. Hence, the slopes $S_x^{\Delta T}(t)$ do not
necessarily have the meaning of a speed of wave propagation.%
\\ 
(viii) The flow-density data depend on the measurement cross section.
While the congested branch is very pronounced upstream of a bottleneck,
it is virtually not present far downstream of it. However,
immediately downstream of a bottleneck, one finds a positively
sloped congested area slightly below
the free branch {\em (``homogeneous-in-speed states'')}. It
looks similar to the upper part of the free branch, but with a 
somewhat lower desired velocity. I suggest that this may indicate
a transition to free traffic in the course of the road
(see Figs.~\ref{SPEED} and \ref{SCATTER}b) and that this  {\em
``recovering traffic''} bears already some signatures of free traffic  
(cf. also Persaud and Hurdle, 1988; Hall \etal, 1992).
For example, on-ramp flows just
add to the flows on the freeway, which may produce states with high flows
(Kerner, 2000b). Some observations, however, question the interpretation
of homogeneous-in-speed states as recovering traffic, as they can
extend over stretches of more than 3 kilometers (Kerner, 1999a).
\begin{figure}[htbp]
\caption[]{Homogeneous-in-speed states scatter around a line through
the origin, indicating constant speeds. The data are typical for ``recovering traffic'' at
cross sections located downstream of a bottleneck, where congested
traffic relaxes to free traffic.\label{SPEED}}
\end{figure}

\subsection{Time headways, headways, and velocities} \label{IIA3}

Time headways show an astonishing individual variation, supporting
Banks' theory discussed above. When distinguished for different density ranges,
time-headway distributions have an interesting property. It turns out that the
distribution becomes sharply peaked around approximately 1~s
for densities close to congestion (Smulders, 1990; Hoogendoorn and Bovy, 1998c;
Neubert \etal, 1999b). For lower and higher densities,
the distribution is considerably broader, see Fig.~\ref{TGAPS} 
(Tilch and Helbing, 2000). This indicates that 
congestion is an overcritical phenomenon occuring when the maximum
efficiency in terms of time headways (related to high-flow states) cannot be supported any longer.
One may conclude that the effective time clearance has, then, reached the safe time clearance
$T_\alpha$. A further increase in the density forces to reduce the speed, which automatically
increases the time headway $\Delta t_\alpha = (T_\alpha + l'_\alpha / v_\alpha)$, even if the
effective clearance remains $T_\alpha$ (cf. Banks, 1991b). 
\par
The distributions of vehicle headways $d_\alpha =v_\alpha \, \Delta t_\alpha$ 
can be determined as well, but it is {\em always}
surprisingly broad (see Fig.~\ref{HEADWAYS}). Therefore, Bleile (1999) suggests 
to consider this by an individual distance scaling of the velocity-clearance relation. 
\par
\begin{figure}[htbp]
    \caption{Empirical time-headway distributions for
different vehicle densities and traffic regimes. (a) Due to the larger proportion of
long vehicles (see Sec.~\ref{CarTr}), the time-headway distribution is broader on the right
lane compared to the left one.
(b) Before the breakdown of free traffic flow, the time-headway distribution is particularly
narrow, afterwards it becomes wide. (After Tilch and Helbing, 2000.)}
\label{fig:tg}\label{TGAPS}
\end{figure}
\begin{figure}[htbp]
\caption[]{The distribution of vehicle distances $d$ is rather broad at all densities, 
although the average vehicle distance decreases with the inverse density $1/\rho$
(after Tilch and Helbing, 2000).\label{HEADWAYS}}
\end{figure}
What are the reasons and consequences of the wide 
scattering of time headways and clearances, apart from driver-dependent preferences?
By detailed investigations of clearance-over-speed relations,
Koshi \etal\ (1983) have observed that vehicles keep larger
distances in real congested traffic than in free traffic or
undisturbed congested traffic under experimental conditions (cf. Fig.~\ref{BOVY}).
\par\begin{figure}[htbp]
\caption[]{\protect Average vehicle distances $d$
as a function of the {\em individual} vehicle velocity $v$, separately 
for time intervals with free and congested traffic (according to the classification
introduced in Fig.~\ref{FUNDAM}a).
Vehicles keep increased distances in congested traffic flow (--~--)
upstream of a bottleneck (a), but not in an undisturbed section
far enough away from it (b). The broken and solid lines are fit curves in order
to guide the eyes. (After Tilch and Helbing, 2000; note 
the related studies by Koshi \etal, 1983;
Dijker \etal, 1998.)\label{BOVY}}
\end{figure}
Plotting the speed over the vehicle distance, one can find a density-dependent
reduction of the speed (see Fig.~\ref{SCHRECK}), which has been called
``frustration effect''. 
\par\begin{figure}[htbp]
    \caption[]{\protect Average vehicle velocities $v$ as a function 
of the headways $d = v \, \Delta t$ for various density regimes, 
separately for free traffic (f) and congested traffic (c) on 
(a) the left and (b) the right lane for a cross section upstream of a bottleneck. 
(From Tilch and Helbing, 2000; note 
the related study by Neubert \etal, 1999b.)
Headways of less than 5~m are due to measurement errors.
The dependence of the $v(d)$ relation on the vehicle density and traffic regime 
is called {\em frustration effect} (see also Brilon and Ponzlet, 1996).
\label{SCHRECK}}
\end{figure}
In my opinion, this is partly an effect of a 
reduction in the vehicle speed with decreasing distance due to safety requirements,
combined with the wide distance scattering illustrated in Fig.~\ref{HEADWAYS}
(Tilch and Helbing, 2000).  Another relevant factor is the 
influence of the relative velocity $\Delta v$ on driver behavior (Bleile, 1997a, 1999). 
The average clearances of vehicles which are driving faster or slower than the
respective leading vehicles are naturally increased compared to 
vehicles driving at the same speed, see Fig.~\ref{DELTA} (Neubert \etal, 1999b; Tilch, 2001).
In summary, driver behavior is not only influenced by the clearance 
to the next car, but also by the relative velocity and the driver velocity 
(Koshi \etal, 1983; Bleile, 1997, 1999; Dijker \etal, 1998;
Neubert \etal, 1999b).
\par\begin{figure}[htbp]
    \caption[]{Empirical clearance $s=(d-l)$ as a function of the relative velocity $\Delta v$,
for free traffic (f) and congested traffic (c) in different density regimes. The clearance
is minimal for identical vehicle velocities. (From Tilch 2001, after Neubert \etal, 1999b.)}
    \label{DELTA}
\end{figure}
\par
Note that the relative velocity in congested traffic is oscillating because of
the instability of traffic flow, see Fig.~\ref{HOEFS} (Hoefs, 1972; Helbing and Tilch, 1998).
As a consequence, the relative velocity variance $\theta$, 
as compared to the average velocity $V$, is considerably higher in congested 
than in free traffic (see Figs.~\ref{VARIA} and \ref{ARHO}). I suggest that
this, together with the factors mentioned in the previous paragraph, may
explain the empirical relations depicted in Fig.~\ref{SCHRECK}. If this is confirmed,
one can drop the assumption of a frustration of drivers in congested traffic.
\par
\begin{figure}[htbp]
    \caption[]{Measured oscillations of the clearance $s = (d - l)$ and of the relative velocity 
$\Delta v$ around $\Delta v = 0$ (cf. Hoefs, 1972) indicate an instability in
car-following behavior (after Helbing and Tilch, 1998).}
 \label{HOEFS}
\end{figure}
\begin{figure}[htbp]
    \caption[]{Empirical standard deviation $\sqrt{\theta(t)}$ of vehicle velocities
divided by the average velocity $V(t)$ as a function of time
(from Tilch, 2001). The {\em relative} 
variation $\sqrt{\theta(t)}/V(t)$ is particularly large during the rush hour, when
traffic is congested. This reflects the oscillations in the following
behavior due to unstable traffic flow (cf. Fig.~\ref{HOEFS})
and implies that measurements of average velocities $V$ have a
small reliability in the congested traffic regime.}
    \label{VARIA}
\end{figure}
\par
Measurements of velocity distributions for traffic with small truck fractions
are in good agreement with
a {\em Gaussian distribution}, see Fig.~\ref{VELDIST} (Pampel, 1955; Leong, 1968; May, 1990). However, 
if the sampling intervals $\Delta T$ are taken too large,
one may also observe bimodal distributions (Phillips, 1977; K\"uhne, 1984a, b),
reflecting a transition from free to congested traffic. As expected for
a Gaussian distribution, 1-minute averages
of single-vehicle data for the skewness $\langle (v_\alpha - V)^3\rangle/
\theta^{3/2}$ and the kurtosis
$[\langle (v_\alpha - V)^4\rangle/ \theta^2 - 3]$ are
scattering around the zero line 
(Helbing, 1997e, a, c, 1998a). 
\begin{figure}[htbp]
\caption[]{Comparison of empirical velocity distributions at different
vehicle densities $\rho$ (---) with frequency polygons of
grouped normal distributions having
the same mean value and variance (--~--). A significant deviation is only observed
for $\rho = 40$~veh./km/lane, probably because of the instability of traffic flow.
(After Helbing, 1997a, c, e, 1998a.)\label{VELDIST}}
\end{figure}
\par
Like the velocity distribution itself, the higher velocity moments are 
sensitive to the choice of the sampling interval $\Delta T$. Large
values of $\Delta T$ may lead to peaks in the variance $\theta$ when the
average velocity $V$ changes abruptly. The additional contribution can 
be estimated as
\(
\overline{[\partial V(x,t)/\partial t ]^2}  (\Delta T)^2/4  ,
\)
where the overbar indicates a time average over the sampling interval
(Helbing, 1997a, b, e). 1-minute data of the variance $\theta$ are more or less
monotonically decreasing with the vehicle density (see Fig.~\ref{emp_theta}), while 5-minute data
have maxima in the density range between 30 and 40 vehicles per
kilometer and lane, indicating particulary unstable traffic 
(Helbing, 1997a). 
\par
\begin{figure}[htbp]
\caption[]{Density-dependent standard deviation $\sqrt{\theta}$ of vehicle velocities in a single
lane of the freeway ($\cdot$: 1-minute data, $\Diamond$: average values for a
given density $\rho$). The velocity variance $\theta$ is more or less monotonically falling with
increasing density. The greater scattering above 30 vehicles per kilometer 
indicates unstable traffic flows. (After Helbing, 1997a, b, e.)}
\label{emp_theta}
\end{figure}
Note that the average $\theta = \sum_i \rho_i \theta_i / (\sum_i \rho_i)$ 
of the velocity variances $\theta_i$ in the 
different lanes $i$ is by an amount of $\langle (V_i - V)^2\rangle$ lower than the velocity variance
evaluated over all lanes, in particular at low densities $\rho_i$.
This is due to the different average velocities $V_i$ in the neighboring lanes.
For a two-lane freeway, the difference in average speeds decreases
almost linearly up to a density of 35 to 40 vehicles per kilometer,
while it fluctuates around zero at higher densities, see Fig.~\ref{suppl}a (Helbing, 1997a, b;
Helbing \etal, 2001b).
This reflects a {\em synchronization} of the velocities $V_i(x,t)$
in neighboring lanes in congested traffic, both in wide
moving jams and synchronized flow 
(Koshi \etal, 1983; Kerner and Rehborn, 1996b; for reduced speed differences
in congested traffic see also Edie and Foote, 1958; 
Forbes \etal, 1967; Mika \etal, 1969).
However, according to Figs.~\ref{suppl}b, c, there is a difference in the 
speeds of cars and long vehicles (``trucks''), which indicates that vehicles 
can still sometimes overtake, 
apart from a small density range around 25 vehicles per
kilometer, where cars move as slow as trucks (Helbing and Huberman,
1998). However, direct measurements of the number of lane changes or
of overtaking maneuvers as a function of the macroscopic variables are 
rare (Sparman, 1978; Hall and Lam, 1988;
Chang and Kao, 1991). It would be particularly interesting to look
at the dependence of lane-changing rates on the velocity-difference among lanes.
\begin{figure}[htbp]
\caption[]{(a) Difference between the average velocity $V_2(\rho)$ in the
left lane and $V_1(\rho)$ in the right lane as a function of the
lane-averaged vehicle density $\rho$ (from Helbing \etal, 2001b;
see also Helbing, 1997a, b; Helbing \etal, 2001a).  
(b), (c) Average velocities of cars and trucks (long vehicles)
in the left and the right lane as a function of the density. The
difference of these empirical curves shows a minimum around 25
vehicles per kilometer, where cars are almost as slow as trucks
(after Helbing and Huberman, 1998; Helbing \etal, 2001b; Helbing, 2001). However,
at higher densities, cars are faster again, which shows that there
must be overtaking maneuvers in the congested density regime.\label{suppl}}
\end{figure}

\subsection{Correlations} \label{IIA4}

In congested traffic, the average velocities in neighboring lanes 
are synchronized, see Fig.~\ref{spurvergl} (Mika \etal, 1969; Koshi \etal, 
1983; cf. also Kerner and Rehborn, 1996b, for the non-linear features
of synchronized flow). The synchronization is probably a result of
lane changes to the faster lane, until the speed difference is
equilibrated (for recent results see, e.g., Lee \etal, 1998;
Shvetsov and Helbing, 1999; Nelson, 2000).  
This equilibration process leads to a higher vehicle density in
the so-called ``fast'' lane, which is used by less trucks 
(see, for example, Hall and Lam, 1988;
Helbing, 1997a, b). Nevertheless, density changes in congested neighboring lanes are
correlated as well. Less understood and, therefore, even more surprising
is the approximate synchronization of the lane-specific velocity variances
$\theta_i$ over the whole density range (Helbing, 1997a, b, e). It may be a sign of adaptive
driver behavior.
\par
\begin{figure}[htbp]
\caption[]{Comparison of the temporal evolution of
different aggregate (macroscopic) traffic variables on neighboring lanes $i$
upstream of a bottleneck (from Helbing, 1997a, b, e). 
(a) The average velocities $V_i(t)$ synchronize 
in the congested traffic regime during the rush hours, and the densities
$\rho_i(t)$ are varying in a correlated manner. While in the left lane, the speed is higher
in free traffic, in congested traffic it is the density (both of which is a consequence
of the smaller truck fraction). (b) The
standard deviations $\sqrt{\theta_i}(t)$ of individual vehicle velocities
show {\em always} a tendency to be synchronized. If (and only if) the
time interval of data averaging is large (here it is 5 min), the variance
shows peaks due to sudden changes in the average velocity $V_i(t)$.}
\label{spurvergl}
\end{figure}
Correlations are also found between the density and flow or average
velocity. At small densities, there is a strong positive correlation
with the flow, which is reflected by the almost linear free branch 
of the fundamental diagram. In contrast, at large densities, the velocity and density
are strongly anticorrelated (see Fig.~\ref{spurvergl}) because of the monotonically
decreasing velocity-density relation (see Fig.~\ref{VDENS}). The average velocity 
and variance are positively correlated, since both quantities are
related via a positive, but density-dependent prefactor, 
see Fig.~\ref{ARHO} (Helbing, 1996b, 1997a, c, 
Treiber \etal, 1999;
Shvetsov and Helbing, 1999; Helbing \etal, 2001a, b). 
\par\begin{figure}[htbp]
    \caption[]{The density-dependent variance prefactors $A_i(\rho_i) = \theta_i(\rho_i) / 
[V_i(\rho_i)]^2$  (reflecting something like squared {\em relative} individual velocity variations)
have significantly increased values in the congested traffic regime, which
may be a result of unstable traffic flow (see Figs.~\ref{HOEFS} and \ref{VARIA}).
The empirical data can be approximated by fit functions of the form
$A_i(\rho_i) = A_{0i} + \Delta A_i \{ 1 +
    \exp [ -(\rho_i-\rho_{{\rm c}i})/ \Delta\rho_i ] 
\}^{-1}$, where $A_{0i}$ and $A_{0i}+\Delta A_i$ are the variance prefactors
for free and congested traffic, respectively,  
$\rho_{{\rm c}i}$ is of the order of the critical density for the transition from 
free to congested traffic, and $\Delta\rho_i$ denotes the width of the transition.
(From Shvetsov and Helbing, 1999.)
\label{ARHO}}
\end{figure}
Finally, it is interesting to look at the velocity correlations among
successive cars. Neubert \etal \ (1999b) have found a long-range velocity correlation
in synchronized flow, while in free traffic vehicle velocities are almost statistically 
independent. This finding is complemented by results from 
Helbing \etal \ (2001b), see Fig.~\ref{COR}. The observed velocity correlations
may be interpreted in terms of {\em vehicle platoons} 
(Wagner and Peinke, 1997). 
Such vehicle platoons have been tracked by Edie and Foote (1958) and
Treiterer and Myers (1974).
\begin{figure}[htbp]
\caption[]{\label{COR}
(a) Velocity correlation in synchronized flow as a function of the
number of consecutive vehicles. The data indicate
long-range correlations, which are not observed in free traffic or for
traffic variables such as the time headway or headway. (After Neubert \etal, 1999b.)
(b) Correlation between 
the velocities of successive vehicles 
in the right lane as a function of the density (thin lines), 
determined from single-vehicle data (from Helbing \etal, 2001b).
As expected, the correlation coefficient is about
zero at small densities. It reaches a maximum at around 20
to 30 vehicles per kilometer, where the transition from free to
congested traffic occurs. Afterwards, it stays on a high level
(around 0.65). In the left lane (not displayed), the velocity correlation is a
little bit higher, probably because of the smaller fraction of trucks (long vehicles).
The correlation coefficient is not sensitive to the type 
of the leading vehicle (car or truck),
but to that of the following vehicle, which is reasonable. The thick line
(-~$\cdot$~-) indicates that, even at small densities of 10 veh./km/lane, the velocity
correlation between cars depends strongly on the headway to the next vehicle ahead and
becomes almost one for very small headways (after Helbing \etal, 2001b). 
This is, because different velocities would imply a high danger of accidents, then.}
\end{figure}

\subsection{Congested traffic} \label{IIA5}

\subsubsection{Jams, stop-and-go waves, and power laws}\label{IIA5a}

The phenomenon of stop-and-go waves (start-stop waves) has been
empirically studied by a lot of authors, including Edie and Foote
(1958), Mika {\em et al.} (1969), and 
Koshi {\em et al.} (1983). The latter have found that the parts of the velocity-profile,
which belong to the fluent stages of stop-and-go waves,
do not significantly depend on the flow (regarding their height and length), 
while the frequency does. 
Correspondingly, there is no {\em characteristic} frequency
of stop-and-go traffic, 
which indicates that we are confronted with {\em non-linear waves}. 
The average duration of one wave period is normally between 4 and 20
minutes for wide traffic jams
(see, e.g., Mika \etal, 1969; K\"uhne, 1987; Helbing, 1997a, c, e), 
and the average wave length between
2.5 and 5~km (see, e.g., Kerner, 1998a). Analyzing the {\em power spectrum} 
(Mika \etal, 1969; Koshi \etal, 1983) points to
{\em white noise} at high wave frequencies $\omega$ (Helbing, 1997a, c, e), 
while a {\em power law} $\omega^{-\phi}$ with exponent
$\phi =1.4$ is found at low frequencies (Musha and Higuchi, 1976, 1978). The latter has
been interpreted as sign of {\em self-organized criticality} (SOC) in the
formation of traffic jams (Nagel and Herrmann, 1993; Nagel and Paczuski, 1995).
That is, congested traffic would drive itself towards the critical density $\rho_{\rm cr}$,
reflecting that it tries to re-establish the largest vehicle density associated with free flow
(see below regarding the segregation between free and congested
traffic and Sec.~\ref{IIC5} regarding the constants of traffic flow).
\par
Based on evaluations of aerial photographs, Treiterer (1966, 1974) 
has shown the existence of {\em ``phantom traffic jams''}, i.e. the
spontaneous formation of traffic jams with no obvious
reason such as an accident or a bottleneck (see Fig.~\ref{TREITERER}).
According to Daganzo (1999a), the breakdown of free traffic 
``can be traced back to a lane change in front of a
highly compressed set of cars'', which shows that there is actually a
reason for jam formation, but its origin can be a rather small
disturbance. However, disturbances do not always lead to traffic jams,
as del Castillo (1996a) points out. Even under comparable conditions, some
perturbations grow and others fade away, which is in accordance with
the {\em metastability} of traffic mentioned above (Kerner and Konh\"auser, 1994;
Kerner and Reh\-born, 1996b, 1997; Kerner, 1999c). 
\par
\begin{figure}[htbp]
\caption[]{Vehicle trajectories obtained by Treiterer and Myers (1974) by aerial photography
techniques (reproduction from Leutzbach, 1988). The location (in units of 100 ft) is displayed
over the time (in seconds). While the slopes of the trajectories 
reflect the individual vehicle velocities, their density represents the spatial
vehicle density. Correspondingly, the figure shows the formation of a ``phantom
traffic jam'', which stops vehicles temporarily. Note that the 
downstream jam front propagates upstream with constant velocity.}
\label{TREITERER}
\end{figure}
While small perturbations in free traffic travel downstream 
with a density-dependent velocity $C(\rho) < V$ (Hillegas \etal, 1974),
large perturbations 
propagate upstream, i.e., against the direction of the vehicle flow
(Edie and Foote, 1958; Mika \etal, 1969). The disturbances have been found to propagate
without spreading (Cassidy and Windover, 1995; Windower, 1998; Mu\~noz and Daganzo, 1999;
see also Kerner and Rehborn, 1996a, and the flow and speed data reported by Foster, 1962;
Cassidy and Bertini, 1999).
The propagation velocity $C$ in congested traffic seems to be roughly 
comparable with a ``natural constant''. In each country, it has a typical 
value in the range $C_0 = 15\pm 5$~km/h, depending
on the accepted safe time clearance and average vehicle length 
(see, e.g.,  Mika \etal, 1969; Kerner and Rehborn, 1996a; Cassidy and Mauch, 2000).
Therefore, fully developed traffic
jams can move in parallel over a long time periods and road
sections. Their propagation speed is not even influenced by
ramps, intersections, or synchronized flow upstream of bottlenecks,
see Fig.~\ref{WIDE} (Kerner and Rehborn, 1996a; Kerner, 2000a, b).
\par 
\begin{figure}[htbp]
\caption{Example of two wide moving traffic jams propagating in parallel with constant speed
through free and congested traffic and across three intersections I1, I2, and I3
(from Kerner, 2000a, b; see also Kerner and Rehborn, 1996a; Kerner, 1998b).
\label{WIDE} }
\end{figure}
Wide moving jams are characterized by stable wave
profiles and by kind of ``universal'' parameters (Kerner and Rehborn, 1996a). 
Apart from\\
(i) the propagation velocity $C_0$, these are\\
(ii) the density $\rho_{\rm jam}$ inside of jams,\\
(iii) the average velocity and flow inside of traffic jams,
both of which are approximately zero,\\
(iv) the outflow $Q_{\rm out}$ from jams (amounting
to about 2/3 of the maximum flow reached in free traffic,
which probably depends on the sampling interval $\Delta T$),\\
(v) the density $\rho_{\rm out}$ downstream of jams, if these are propagating
through free traffic ({\em ``segregation effect''} 
between free and congested traffic). When propagating
through synchronized flow, the outflow $Q_{\rm out}^{\rm sync}
= Q_{e}(\rho_{\rm out}^{\rm sync})$ of wide moving jams is
given by the density $\rho_{\rm out}^{\rm sync}$ of the surrounding traffic
(Kerner, 1998b).
The concrete values of the characteristic parameters slightly depend
on the accepted safe time clearances, the average vehicle length,
truck fraction, and weather conditions (Kerner and Rehborn, 1998a).

\subsubsection{Extended congested traffic}\label{IIA5b}

The most common form of congestion is not localized like a
wide moving jam, but spatially extended and often persisting over several hours.
It is related with a {\em capacity drop} to a flow, which is (at least in the United States) typically
10\% or less below the {\em ``breakdown flow''} of the previous high-flow
states defining the {\em theoretically possible capacity} 
(Banks, 1991a; Kerner and Rehborn, 1996b, 1998b;
Persaud \etal, 1998; Westland, 1998; Cassidy and Bertini, 1999;
see also May, 1964, for an idea how to exploit this phenomenon with ramp metering;
Persaud, 1986; Banks, 1989, 1990; Agyemang-Duah and Hall, 1991; Daganzo, 1996).
For statistical reasons, it is not fully satisfactory to determine capacity drops 
from maximum flow values, as these depend on the sampling interval $\Delta T$. Anyway,
more interesting are probably the {\em bottleneck flows} $Q_{\rm bot}$ after the 
breakdown of traffic. Because of relation $(\ref{botflow})$, these may be more than
30\% below the maximum free flow $Q_{\rm max}$
(see Fig.~\ref{FUNDAM}b). Note that bottleneck flows
depend, for example, on ramp flows $Q_{\rm rmp}$ and may, therefore, vary with 
the location. The congested 
flow immediately downstream of a bottleneck defines the {\em discharge flow}
$\tilde{Q}_{\rm out} \ge Q_{\rm bot}$, which appears to be
larger than (or equal to) the characteristic outflow $Q_{\rm out}$ from wide jams: 
$\tilde{Q}_{\rm out} \ge Q_{\rm out}$.
\par
In extended congested traffic, the velocity drops much more than the flow, but it
stays also finite. The velocity profiles $V(x,t)$ can differ considerably from one
cross section to another (see Fig.~\ref{PINCH}). In contrast, 
the temporal profiles of congested traffic flow $Q(x,t)$, when measured at subsequent
cross sections $x$ of the road, are often just shifted by some time interval which is varying 
(see Fig.~\ref{SHIFT}). This may be explained with
a linear flow-density relation of the form (\ref{GREEN})
with $dQ_{e}(\rho)/d\rho = C_0 = - 1/(T\rho_{\rm jam})$ (Hall \etal, 1986;
Ozaki, 1993; Hall \etal, 1993; 
Dijker \etal, 1998; Westland, 1998), 
together with the continuum equation for the
conservation of the vehicle number
(see the kinematic wave theory in Sec.~\ref{IIB4a}; Cassidy and Mauch, 2000;
Smilowitz and Daganzo, 1999;  Daganzo, 1999a). However, a linear flow-density
relation in the congested regime is questioned by the ``pinch effect'' (see Sec.~\ref{IIA5c} and 
Fig.~\ref{PINCH}).
\par\begin{figure}[htbp]
\caption[]{Example of time-dependent traffic flows at three subsequent cross sections
of a 3 kilometer long freeway section without ramps (including a hardly noticeable breakdown
from free to ``synchronized'' congested flow during the rush hours between 7:30 and 9:30 am, 
cf. the corresponding Fig.~\ref{spurvergl}.) The flow profiles at subsequent cross sections
are mainly shifted by some time interval, although fluctuations appear in
the congested regime (after Helbing, 1997a).\label{SHIFT}}
\end{figure}
The extended form 
of congestion is classical and mainly found upstream of bottlenecks, so that it
normally has a spatially fixed (``pinned'') downstream front. In contrast, the
upstream front is moving against the flow direction, 
if the {\em (dynamic) capacity} $Q_{\rm bot}$
of the bottleneck is exceeded, but downstream, if the traffic volume,
i.e. the inflow to the freeway section is lower than the capacity. Hence,
this spatially extended form of
congestion occurs regularly and reproducible during rush hours. 
Note, however, that bottlenecks may have many different
origins: on-ramps, reductions in the number of lanes, accidents
(even in opposite lanes because of rubbernecks), speed limits, road works,
gradients, curves, bad road conditions (possibly due to rain, fog, or ice),
bad visibility (e.g., because of blinding sun), diverges 
(due to ``negative'' perturbations, see Sec.~\ref{IID},
weaving flows by vehicles trying to switch to the slow
exit lane, or congestion on the off-ramp, see  Daganzo \etal, 1999;
Lawson \etal, 1999; Mu\~noz and Daganzo, 1999; Daganzo, 1999a).
Moving bottlenecks due to 
slow vehicles are possible as well (Gazis and Herman, 1992), 
leading to a forward movement of the respective downstream 
congestion front. Finally, in cases of {\em two} subsequent inhomogeneities 
of the road, there are forms of congested traffic in which both,
the upstream and downstream fronts are locally 
fixed (Treiber \etal, 2000; Lee \etal, 2000; Kerner, 2000a). 
\par
While Kerner calls the above extended forms of congested traffic
{\em ``synchronized flow''} because of the synchronization among lanes
(see Secs.~\ref{IIA3}, \ref{IIA4}), Daganzo (1999b) speaks of 1-pipe flow.
Kerner and Rehborn (1997) have pointed out that the transition from free to extended
congested traffic is of hysteretic nature
and looks similar to the first-order
phase transition of supersaturated vapor to water {\em (``nucleation effect'')}.
It is often triggered by
a small, but overcritical 
peak in the traffic flow (playing the role of a nucleation germ).
This perturbation travels downstream in the
beginning, but it grows eventually and changes its propagation
direction and speed, until it travels upstream with velocity $C_0$.
When the perturbation reaches the bottleneck, it triggers a breakdown
of traffic flow to the bottleneck flow $Q_{\rm bot}$. In summary,
{\em synchronized flow typically starts to form downstream of the bottleneck}
(cf. Fig.~\ref{TR}b).
\par 
Kerner (1997, 1998a) has attributed the capacity drop to increased time headways due
to delays in acceleration. This has recently been
supported by Neubert \etal \ (1999b). At a measurement section
located directly at the downstream front of a congested region,
they found the appearance of a peak at about 1.8 seconds
in the time-headway distribution, when congestion set in, see Fig.~\ref{Neubert2}. 
The reachable {\em saturation flows} of accelerating traffic are, therefore,
$Q_{\rm out} \approx 2000$ vehicles per hour. 
This is compatible with measurements of stopped vehicles accelerating
at a traffic light turning green (Androsch, 1978) or after an
incident (Raub and Pfefer, 1998). 
\par
\begin{figure}[htbp]
\caption[]{Time-headway distributions for different density 
regimes at a cross section where vehicles accelerate out of congested
traffic. While freely flowing vehicles use to have short time headways around 1~s,
the time headway of accelarating vehicles is typically around 1.8~s.
(After Neubert \etal, 1999b.) \label{Neubert2} }
\end{figure}
Kerner (1998b) points out that the above transition to ``synchronized'' congested flow is,
in principle, also found on one-lane roads, 
but then it is not anymore of hysteretic nature and not connected with synchronization. 
Moreover, he and Rehborn (1996b)
distinguish three kinds of synchronized flow:\\
(i) Stationary and homogeneous states where both the average
speed and the flow rate are stationary during a relatively long time
interval (see, e.g., also Hall and Agyemang-Duah, 1991; Persaud \etal,
1998; Westland, 1998). 
I will later call these {\em ``homogeneous congested traffic''} (HCT).\\
(ii) States where only the average vehicle speed is stationary, 
named {\em ``homogeneous-in-speed states''} (see also Kerner, 1998b; Lee \etal, 2000).
I interpret this state as {\em ``recovering traffic''}, 
since it bears several signatures of free traffic
(cf. Secs.~\ref{IIA2} and \ref{IIE3}).\\
(iii) Non-stationary and non-homogeneous states
(see also Kerner, 1998b; Cassidy and Bertini, 1999; Treiber \etal, 2000). 
For these, I will also use the term {\em ``oscillating congested traffic''} (OCT).
\par
Not all of these states are long-lived, since there are
often spatio-temporal sequences of these different types of
synchronized flow, which indicates {\em continuous
(second-order) transitions}. However, at least states (i) and (iii)
are characterized by
a wide, two-dimensional scattering of the flow-density data, i.e.
an increase in the flow can be either related with an increase or with 
a decrease in the density, in contrast to free flow (see
Sec.~\ref{IIA2}).

\subsubsection{Pinch effect}\label{IIA5c}

Recently, the spontaneous appearance of stop-and-go traffic has been
questioned by Kerner (1998a) and Daganzo {\em et al.} (1999).
In his empirical investigations, Kerner (1998a) finds that jams can be born from extended
congested traffic, which presupposes the previous transition from 
free to synchronized flow. The alternative mechanism
for jam formation is as follows (see Fig.~\ref{PINCH}): Upstream of a section with
homogeneous congested traffic close to a bottleneck, there is a
so-called {\em ``pinch region''} characterized by the spontaneous 
birth of small narrow 
density clusters, which are growing while they travel further
upstream. Wide moving jams are eventually formed by the merging or disappearance of
narrow jams, which are said to move faster upstream than
wide jams. Once formed, the wide jams seem to suppress the occurence of new
narrow jams in between. Similar findings were reported by Koshi {\em et al.} (1983), 
who observed that ``ripples of speed grow larger
in terms of both height and length of the waves 
as they propagate upstream''.  Daganzo (1999b) notes as well that
``oscillations exist in the 1-pipe regime and that these oscillations may grow in amplitude as one
moves upstream from an active bottleneck'', which he interprets as a {\em pumping effect}
based on ramp flows.
The original interpretation by Koshi \etal \ suggests 
a non-linear self-organization phenomenon, assuming a concave, non-linear 
shape of the congested flow-density branch
(cf. Sec.~\ref{IID2}). Instead of forming wide jams,  narrow jams may coexist 
when their distance is larger than about 2.5~km 
(Kerner, 1998a; Treiber \etal, 2000).
\begin{figure}[htbp]
\caption{Example for the time-dependent average velocities in the three lanes 
of a freeway at subsequent cross sections
(after Kerner, 1998a). The figures show a transition from free to synchronized
flow at cross section D5, 
the emergence of narrow jams in the ``pinch region'' at the upstream cross section D4,
and the formation of a few stable wide jams (see further upstream cross section D1).
The wide moving jams arise by the merging or disappearance of 
narrow jams between cross sections D4 and D1.
\label{FS}\label{PINCH}}
\end{figure}

\subsection{Cars and trucks}\label{CarTr}

Distinguishing vehicles of different lengths (``cars'' and ``trucks''), one 
finds surprisingly strong variations of the truck fraction, see Fig.~\ref{TRUCKFRAC}
(Treiber and Helbing, 1999a). This point may be quite relevant for the
explanation of some observed phenomena, as quantities characterizing the
behavior of cars and trucks are considerably different. 
For example, this concerns the distribution of desired velocities 
as well as the distribution of time headways (see Fig.~\ref{PARAMETERS2}).
\begin{figure}[htbp]
    \caption{The truck fraction is varying considerably in the course of time
and should, therefore, be taken into account in empirical analyses of traffic flow.
(From Treiber and Helbing, 1999a.)}
    \label{TRUCKFRAC}
\end{figure}
\begin{figure}[htbp]
\caption{Separate time-headway distributions for cars and trucks on the right lane
(after Tilch and Helbing, 2000). The time headways in front of trucks (long vehicles)
are considerably longer than those of cars, in particular in congested traffic.
}
\label{PARAMETERS2}
\end{figure}

\subsection{Some critical remarks} \label{IIA7}

The collection and evaluation of empirical data is a subject with often 
underestimated problems. To make reliable conclusions, in original investigations
one should specify\\
(i) the measurement site and conditions (including applied control
measures),\\
(ii) the sampling interval,\\
(iii) the aggregation method,\\
(iv) the statistical properties
(variances, frequency distributions, correlations, survival times of traffic states, etc.),\\
(v) data transformations,\\
(vi) smoothing procedures,\\
and the respective dependencies on them.
\par
The measurement conditions include ramps and road sections with their respective
in- and outflows,
speed limits, gradients, and curves with the respectively related capacities, 
furthermore weather conditions, 
presence of incidents, and other irregularities. Moreover, one
should study the dynamics of the long vehicles separately, which
may also have a significant effect (see Sec.~\ref{CarTr}).
\par 
A particular attention has also to be 
paid to the fluctuations of the data, which requires a {\em statistical} investigation.
For example, it is not obvious whether the
scattering of flow-density data in synchronized flow 
reflects a complex dynamics due to non-linear interactions
or whether it is just because of random fluctuations in the system.
In this connection, I  remind of the power laws found in the
high-frequency variations of macroscopic quantities (see Sec.~\ref{IIA5a}). 
\par
The statistical variations of traffic flows imply that all
measurements of macroscopic quantities should be complemented by 
error bars (see, e.g., Hall \etal, 1986).  Due to the relatively small ``particle'' numbers 
behind the determination of macroscopic quantities, the error bars are 
actually quite large. Hence, many temporal variations are within one
error bar and, therefore, not significant. As a consequence,
the empirical determination
of the dynamical properties of traffic flows is not a simple task.
Fortunately, many hard-to-see effects are in agreement with
predictions of plausible traffic models, in particular with deterministic 
ones (see Secs.~\ref{IID} and \ref{IIE3}). 
\par
Nevertheless, I would like to call for more refined measurement
techniques, which are required for more reliable data. 
These must take into account correlations between 
different quantities, as is pointed out
by Banks (1995). Tilch and Helbing (2000) have, therefore, used the
following measurement procedures: In order to have
comparable sampling sizes, they have averaged over a {\em fixed} number
$\Delta N$ of cars, as suggested by Helbing (1997e). 
Otherwise the {\em statistical} error at small traffic flows (i.e. at small and
large densities) would be quite large. This is compensated for by 
a flexible measurement interval $\Delta T$. 
It is favourable 
that $\Delta T$ becomes particularly small in the (medium) density range of unstable
traffic, so that the method yields a good representation of traffic
dynamics. However, choosing small values of $\Delta N$ does not make sense,
since the temporal variation of the aggregate values will mainly reflect
statistical variations, then.
In order to have a time resolution of about 2 minutes on each lane,
one should select $\Delta N=50$, while $\Delta N=100$ can be chosen when averaging
over both lanes. Aggregate values over both lanes for $\Delta N=50$ are
comparable with 1-minute averages, but show a smaller statistical
scattering at low densities (compare the results in Helbing 1997a, c with those in Helbing, 1997e).
\par
Based on the passing times $t_\alpha^0$ of successive vehicles $\alpha$ in the same 
lane, we are able to calculate the time headways $\Delta t_\alpha$.
The (measurement) time interval
\begin{equation}
  \label{eq:T_N}
  \Delta T = \sum_{\alpha=\alpha_0+1}^{\alpha_0+\Delta N} 
\Delta t_\alpha
\end{equation}
for the passing of $\Delta N$ vehicles defines the (inverse of the)
traffic flow $Q$ via
\(
 1/Q = \Delta T/\Delta N = \langle \Delta t_\alpha \rangle,
\)
which is attributed to the time
\(
  t =  \langle t_\alpha^0 \rangle  = \sum_\alpha t_\alpha^0/\Delta N. 
\)
\par\begin{figure}[htbp]
\caption[]{\label{COVAR}The covariance between headways $d_\alpha$ and inverse
  velocities $1/v_\alpha$
    shows significant deviations from zero in congested traffic, while it approximately
    vanishes in free flow. Even after traffic has recovered, there seem to remain 
    weak correlations between headways and vehicle speeds for a considerable time.
    These are probably a reminiscence of congestion due to platoons which have
    not fully dissolved. (After Tilch, 2001.)}
\end{figure}
Approximating the vehicle headways by $d_\alpha = v_\alpha \Delta t_\alpha$,
one obtains
\[
  \frac{1}{Q} = \langle \Delta t_\alpha \rangle 
  = \left\langle \! \frac{d_\alpha}{v_\alpha} \! \right\rangle
  = \langle  d_\alpha \rangle \left\langle \! \frac{1}{v_\alpha} 
  \! \right\rangle + \mbox{cov}\left(\!d_\alpha,\frac{1}{v_\alpha} \!\right) ,
\]
where $\mbox{cov}(d_\alpha,1/v_\alpha)$
is the covariance between the headways $d_\alpha$
and the inverse velocities $1/v_\alpha$. We expect that this covariance 
is negative and particularly relevant at large vehicle densities, which
is confirmed by the empirical data (see Fig.~\ref{COVAR}). Defining the
density $\rho$ by
\begin{equation}
  1/\rho = \left\langle d_\alpha \right\rangle
\label{proposed}
\end{equation}
and
the average velocity $V$ via Eq.~(\ref{harmav}), 
we obtain the fluid-dynamic flow relation (\ref{densdef}) by the
conventional assumption $\mbox{cov}(d_\alpha,1/v_\alpha) = 0$. 
This, however,
overestimates the density systematically, since the covariance tends to be negative
due to the speed-dependent safety distance of vehicles.
In contrast, the common method of determining the density via $Q/\langle v_\alpha\rangle$
systematically underestimates the density (see Fig.~\ref{fig:C_rho}).
Consequently, errors in the measurement of the flow and the density due to
a neglection of correlations partly account for the observed scattering
of flow-density data in the congested regime. However, a considerable amount of
scattering is also observed in flow-occupancy data, which avoid the above measurement
problem.
\par
\begin{figure}[htbp]
     \caption[]{\protect 50-vehicle averages of (a)
the density $Q\langle 1/v_\alpha\rangle$ and (b) the conventionally determined density
$Q/\langle v_\alpha\rangle$, both times as a function of the 
density $\rho$ according to the proposed definition (\ref{proposed}). The
usually assumed relations are indicated by solid lines. 
(From Tilch and Helbing, 2000.)\label{fig:C_rho}}
\end{figure}

\section{Modelling approaches for vehicle traffic}\label{IIB}

In zeroth order approximation, I would tend to say that each decade was
dominated by a certain modelling approach:\\ 
(i) In the 50ies, the propagation of shock waves, i.e. of density jumps in
traffic, was described by a {\em fluid-dynamic model} for kinematic waves.\\
(ii) The activities in the 60ies concentrated on {\em ``microscopic''
car-following models}, which are often called follow-the-leader models.\\ 
(iii) During~the 70ies,~{\em gas-kinetic}, so-called {\em Boltz\-mann-like models} 
for the spatio-temporal change of the velocity distribution were florishing.\\
(iv) The simulation of {\em ``macroscopic'',  fluid-dynamic
models} was common in the 80ies.\\
(v) The 90ies were dominated by discrete {\em cellular automata models} of vehicle traffic
and by systematic investigations of the dynamical solutions of the 
models developed in the previous 50 years.\\
(vi) Recently,
the availability of better data brings up more and more experimental studies 
and their comparison with traffic models. 
\par
Altogether, researchers from engineering, mathematics, operations 
research, and physics have probably suggested more than 100 different
traffic models, which unfortunately cannot all be covered by this review. 
For further reading, I therefore recommend the books and proceedings by
Buckley (1974),
Whitham (1974), 
Gerlough and Huber (1975),
Gibson (1981), May (1981),
Vumbaco (1981),
Hurdle \etal (1983),
Volmuller and Hamerslag (1984),
Gartner and Wilson (1987),
Leutzbach (1988),
Brannolte (1991),
Daganzo (1993),
Pave (1993),
Snorek \etal\ (1995),
Wolf \etal\ (1996),
Lesort (1996),
the Transportation Research Board (1996),
Gartner \etal\ (1997),
Helbing (1997a),
Daganzo (1997a),
Rysgaard (1998),
Schreckenberg and Wolf (1998),
Bovy (1998),
Brilon {\em et al.} (1999),
Ceder (1999), 
Hall (1999), and
Helbing {\em et al.} (2000c, 2001a, b).
Readers interested in an exhaustive discussion of cellular
automata should consult the detailed review by
Chowdhury \etal\ (2000b). 
\par
In the following subsections, 
I can only introduce a few representatives for each modelling approach
(selected according to didactical reasons) and try to show up the relations among them 
regarding their instability and other properties (see Secs.~\ref{IIC} to \ref{IIF}). 
\par
Before, I would like to formulate some
criteria for good traffic models:
For reasons of robustness and calibration, such models should only
contain a few variables and parameters which have an intuitive
meaning. Moreover, these should be easy to measure, and the
corresponding values should be realistic.
In addition, it is not satisfactory to selectively reproduce subsets
of phenomena by different models. Instead,
a good traffic model should at least qualitatively
reproduce all known features of traffic flows, including the 
localized and extended forms of congestion. Furthermore, the observed
hysteresis effects, complex dynamics,
and the existence of the various self-organized 
constants like the propagation velocity of stop-and-go waves
or the outflow from traffic jams should all be reproduced.
A good model should also make new predictions allowing us to 
verify or falsify it. Apart from that,
its dynamics should not lead to vehicle collisions
or exceed the maximum vehicle density. Finally, 
the model should allow for a fast numerical simulation.

\subsection{Microscopic follow-the-leader models} \label{IIB1}

The first microscopic traffic models were proposed by Reuschel (1950a, b) and
the physicist Pipes (1953).
Microscopic traffic models assume that the acceleration of a driver-vehicle
unit $\alpha$ is given by
the neighboring vehicles. The dominant influence on driving behavior comes 
from the next vehicle $(\alpha-1)$ ahead, called the {\em leading vehicle}.
Therefore, we obtain the following 
model of driver behavior from Eq. (\ref{three}):
\begin{equation}
 \frac{dv_\alpha(t)}{dt} = \frac{v_\alpha^0 +
 \xi_\alpha(t)-v_\alpha(t)}{\tau_\alpha}
 + f_{\alpha,(\alpha-1)}(t) \, . 
\label{genforce}
\end{equation}
Herein, $f_{\alpha,(\alpha-1)}(t)\le 0$ describes the repulsive effect of
vehicle $(\alpha-1)$, which is generally a function of\\
(i) the relative velocity
$\Delta v_\alpha(t) = [v_\alpha(t) - v_{\alpha-1}(t)]$,\\
(ii) the own velocity $v_\alpha(t)$ due to the velocity-dependent safe
distance kept to the vehicle in front,\\
(iii) of the headway (brutto distance) 
$d_\alpha(t) = [x_{\alpha-1}(t)-x_\alpha(t)]$ or the clearance (netto
distance) 
$s_\alpha(t) = [d_\alpha(t) - l_{\alpha-1}]$, with $l_\alpha$
meaning the length of vehicle $\alpha$.\\
Consequently, for identically behaving vehicles with
$v_\alpha^0 = v_0$, $\tau_\alpha = \tau$, and $f_{\alpha,(\alpha-1)} = 
f$, we would have
\begin{equation}
 f_{\alpha,(\alpha-1)}(t) = f\bbox(s_\alpha(t),v_\alpha(t),\Delta
 v_\alpha(t)\bbox) \, .  
\label{agreement}
\end{equation}
If we neglect fluctuations for the time being and
introduce the traffic-dependent velocity
\begin{equation}
 v^{e}(s_\alpha,v_\alpha,\Delta v_\alpha) = v_0 + \tau
f(s_\alpha,v_\alpha, \Delta v_\alpha) \, ,
\end{equation}
to which driver $\alpha$ tries to adapt,
we can considerably simplify the {\em ``generalized force model''}
(\ref{genforce}):
\begin{equation}
 \frac{dv_\alpha}{dt} = \frac{v^{e}(s_\alpha,v_\alpha,\Delta
 v_\alpha) - v_\alpha}{\tau} \, .
\label{type}
\end{equation}

\subsubsection{Non-integer car-following model}\label{IIB1a}

Models of the type (\ref{type}) are called {\em follow-the-leader 
models} (Reuschel 1950a, b; Pipes, 1953;
Chandler \etal, 1958; Chow, 1958; Kometani and Sasaki, 1958, 1959, 1961;
Herman \etal, 1959;
Gazis \etal, 1959; Gazis \etal, 1961; Newell, 1961;
Herman and Gardels, 1963; Herman and Rothery, 1963;
May and Keller, 1967; Fox and Lehmann, 1967; Hoefs, 1972). One of the
simplest representatives results from the assumption that the 
netto distance is given by the velocity-dependent safe distance
\(
 s^*(v_\alpha) = s' + T v_\alpha ,
\)
where $T$ has the meaning of the (effective) safe time clearance. This implies
$s_\alpha(t) = s^*\bbox(v_\alpha(t)\bbox)$
or, after differentiation with respect to time, 
\(
 dv_\alpha(t)/dt = [ds_\alpha(t)/dt]/T =  [d\,d_\alpha(t)/dt]/T
 = [v_{\alpha -1}(t) - v_{\alpha}(t)]/T .
\)
Unfortunately, this model does not explain the empirically observed
density waves (see Sec.~\ref{IIA5a}). Therefore, one has to introduce
an additional time delay $\Delta t \approx 1.3$~s 
in adaptation, reflecting the finite
reaction time of drivers. This yields the following 
{\em stimulus-response model:}
\begin{equation}
 \underbrace{\frac{dv_\alpha(t+\Delta t)}{dt}}_{\rm Response} = 
 \frac{1}{T} \underbrace{[ v_{\alpha -1}(t) - v_{\alpha }(t) ]}_{\rm Stimulus} \, .
\label{einfach}
\end{equation} 
Herein, $1/T$ is the sensitivity to the stimulus. 
This equation belongs to the class of {\em delay
differential equations}, which normally have an unstable solution for
sufficiently large delay times $\Delta t$. 
For the above case, Chandler \etal\ (1958) 
could show that, under the instability condition
\(
 \Delta t/T > 1/2 ,
\)
a variation of individual vehicle velocities will be amplified.
The experimental value is $\Delta t/T \approx 0.55$.
As a consequence, the non-linear vehicle dynamics
finally gives rise to stop-and-go waves, but also to accidents. In order to
cure this, to explain the empirically observed fundamental diagrams,
and to unify many other model variants, Gazis \etal\ 
(1961) have introduced a generalized sensitivity factor with two parameters
$m_1$ and $m_2$:
\begin{equation}
 \frac{1}{T} = \frac{1}{T_0} \frac{ [v_{\alpha}(t+\Delta t)]^{m_1}}
 {[x_{\alpha -1}(t) - x_{\alpha }(t)]^{m_2}} \, .
\end{equation}
The corresponding non-integer car-following model can be rewritten in the form
\(
 [d v_{\alpha}(t+\Delta t)/dt]/[v_{\alpha}(t+\Delta
t)]^{m_1} = (1/T_0) [d\,d_\alpha(t)/dt]/[d_\alpha(t)]^{m_2} ,
\)
which is solved by
\(
 f_{m_1}\bbox(v_{\alpha}(t+\Delta t)\bbox) = c_0 + c_1 f_{m_2}\bbox(d_\alpha(t)\bbox)
\) 
with $f_k(z) = z^{1-k}$ if $k\ne 1$ and $\ln z$ otherwise 
($c_0$, $c_1$ being integration constants). For $m_1\ne 1$ and $m_2\ne 1$ 
we will discuss the stationary  case related to identical
velocities and distances. The vehicle density $\rho$ is then given by the
inverse brutto distance $1/d_\alpha$, and the corresponding equilibrium
velocity $V_{e}$ is identical with $v_{\alpha}$. Therefore, we
obtain the velocity-density relation
\(
 V_{e}(\rho) = V_0 [ 1 - ( \rho/\rho_{\rm max})^{m_2-1}]^{1/(1-m_1)}
\) 
with the free velocity $V_0$ and the maximum density $\rho_{\rm max}$.
Most of the velocity-density relations that were under discussion at
this time are special cases of the latter formula for different values of
the model parameters $m_1$ and $m_2$. Realistic
fundamental diagrams result, for example, 
for the non-integer values $m_1\approx 0.8$ and  $m_2\approx 2.8$ (May and Keller, 1967;
Hoefs, 1972) or $m_1=0.953$ and $m_2 = 3.05$
(K\"uhne and R\"odiger, 1991; K\"uhne and Kroen, 1992).
\par
Finally, note that the above non-integer car-following model is used in the traffic 
simulation package MITSIM and has recently been
made more realistic with acceleration-dependent parameters (Yang and Koutsopoulos, 1996; 
Yang, 1997). However, other car-following models deserve to be mentioned
as well (Gipps, 1981; Benekohal and Treiterer, 1988;
del Castillo, 1996b; Mason and Woods, 1997). For a historical review 
see Brackstone and McDonald (2000).

\subsubsection{Newell and optimal velocity model}\label{IIB1b}

One of the deficiencies of the non-integer car-following models is that
it cannot describe the driving behavior of a {\em single} vehicle. Without
a leading vehicle, i.e. for $d_{\alpha}\rightarrow \infty$, vehicle
$\alpha$ would not accelerate at all. Instead, it should approach its
desired velocity $v_\alpha^0$ in free traffic.
Therefore, other car-following models do not assume an adaptation to
the velocity of the leading vehicle, but an adaptation to 
a distance-dependent velocity $v'_{e}(d_\alpha)$ which should
reflect the safety requirements and is
sometimes called the {\em ``optimal velocity''}. While Newell (1961)
assumes a delayed adaptation of the form
\begin{equation}
 v_\alpha(t+\Delta t) = v'_{e}\bbox(d_\alpha(t)\bbox) = v_{e}\bbox(s_\alpha(t)\bbox) \,,
\label{NEWELL}
\end{equation}
Bando {\em et al.} (1994, 1995a, b) suggest to use the relation
\begin{equation}
 v'_{e}(d) = (v_0/2) [ \mbox{tanh} (d - d_{c}) + \mbox{tanh}\, d_{c} ] 
\label{Band}
\end{equation}
with constants $v_0$ and $d_{c}$, together with the optimal velocity model
\begin{equation}
 \frac{dv_\alpha(t)}{dt} = \frac{v'_{e}\bbox(d_\alpha(t)\bbox)
- v_\alpha(t)}{\tau} 
\, ,
\label{BANDO}
\end{equation}
which may be considered as a first-order Taylor approximation 
$v_\alpha(t+\Delta t) \approx [ v_\alpha(t) + \Delta t\; dv_\alpha(t)/dt ]$ of
Eq.~(\ref{NEWELL}) with $\tau = \Delta t$. 
For the optimal velocity model
one can show that small perturbations are eventually amplified 
to traffic jams, if the {\em instability condition}
\begin{equation}
 \frac{dv'_{e}(d_\alpha)}{d\, d_\alpha} = \frac{dv_{e}(s_\alpha)}{ds_\alpha} > \frac{1}{2 \tau}
\label{bandinst}
\end{equation}
is satisfied (Bando \etal, 1995a), i.e., if we have large relaxation times
$\tau$ or big changes of the velocity $v_{e}(s_\alpha)$ with the 
clearance $s_\alpha$ (see Fig.~\ref{BAN}). Similar investigations have been
carried out for analogous models with an additional 
explicit delay $\Delta t$ (Bando \etal, 1998; Wang \etal, 1998b).
\begin{figure}[htbp]
\caption[]{Trajectories according to the optimal velocity model by Bando \etal \ (1995a) for
each fifth vehicle, if the overall density $\varrho$ on the circular road 
falls into the regime of linearly unstable traffic (from Helbing, 1997a). As in
Fig.~\ref{TREITERER}, small perturbations are amplified and lead to the formation
of ``phantom traffic jams''.\label{BAN}}
\end{figure}

\subsubsection{Intelligent driver model (IDM)}\label{IIB1c}

As the optimal velocity model does not contain a driver response to the
relative velocity $\Delta v_\alpha$ with respect to the leading vehicle,
it is very sensitive to the concrete choice of the
function $v'_{e}(d_\alpha)$ and produces accidents, when fast cars
approach standing ones (Helbing and Tilch, 1998). To avoid this, one has to assume
particular velocity-distance relations and choose a very small
value of $\tau$, which gives unrealistically large accelerations (Bleile, 1997b, 1999).
In reality, however, the acceleration times are
about five to ten times larger than the braking times. Moreover,
drivers keep a larger safe distance and decelerate earlier, when the
relative velocity $\Delta v_\alpha(t)$ is high. These aspects have,
for example, been taken into account in models by Gipps (1981),
Krau{\ss} \etal\ (1996, 1997), Helbing (1997a), Bleile (1997b, 1999), Helbing and Tilch
(1998), Wolf (1999), or Tomer \etal\ (2000).
\par
For illustrative reasons, I will introduce the so-called intelligent driver
model (Treiber and Helbing, 1999b; Treiber \etal, 2000). 
It is easy to calibrate, robust, accident-free, and numerically efficient, yields
realistic acceleration and braking behavior and reproduces the
empirically observed phenomena. Moreover, for a certain specification of 
the model parameters, its fundamental diagram is related to that of the
gas-kinetic-based, non-local traffic (GKT) model,
which is relevant for the micro-macro link we have 
in mind (see Sec.~\ref{IIB5}). For other specifications, the IDM is related with the
Newell model or the Nagel-Schreckenberg model (see Sec.~\ref{IIB2a}).
\par
The acceleration assumed in the IDM is a continuous function 
of the velocity $v_{\alpha}$, the clearance 
$s_{\alpha}=(d_\alpha - l_{\alpha-1})$,
and the velocity difference (approaching rate)
$\Delta v_{\alpha}$ of vehicle $\alpha$ to the leading vehicle: 
\begin{equation}
\label{IDMv}
\frac{dv_\alpha}{dt} = a_{\alpha}
         \left[ 1 -\left( \frac{v_{\alpha}}{v_{\alpha}^0} 
                  \right)^{\delta} 
                  -\left( \frac{s_{\alpha}^*(v_{\alpha},\Delta v_{\alpha})}
                                {s_{\alpha}} \right)^2
         \right].
\end{equation}
This expression is a superposition of the acceleration tendency
$a_{\alpha}[1-(v_{\alpha}/v_{\alpha}^0)^{\delta}]$ 
on a free road, and the deceleration tendency
$f_{\alpha,(\alpha-1)}
= -a_{\alpha}[s_{\alpha}^*(v_{\alpha},\Delta v_{\alpha})/s_{\alpha}]^2$
describing the interactions with other vehicles. The parameter
$\delta$ allows us to fit the acceleration behavior. While $\delta = 1$
corresponds to an exponential-in-time acceleration on a free road, as
assumed by most other models, in the limit
$\delta \rightarrow \infty$, we can describe a constant acceleration
with $a_\alpha$, until the desired velocity $v_\alpha^0$ is reached.
The deceleration term
depends on the ratio between the ``desired clearance'' $s_{\alpha}^*$ 
and the actual clearance $s_\alpha$, where the desired clearance 
\begin{equation}
\label{sstar}
s^*_{\alpha}(v_\alpha, \Delta v_\alpha) 
    = s'_{\alpha} + s_{\alpha}^{\prime\prime} \sqrt{\frac{v_\alpha}{v_{\alpha}^0}}
    + T_{\alpha} v_\alpha
    + \frac{v_\alpha \Delta v_\alpha }  {2\sqrt{a_{\alpha} b_{\alpha}}}
\end{equation}
is dynamically varying with the velocity $v_\alpha$
and the approaching rate $\Delta v_\alpha$, reflecting
an intelligent driver behavior.
The IDM parameters can be chosen individually for each vehicle
$\alpha$, but for the moment, we will assume identical vehicle
parameters and drop the index $\alpha$ for readability. These are 
the desired velocity $v_0$,
the safe time clearance $T$,
the maximum acceleration $a$,
the comfortable deceleration $b$,
the acceleration exponent $\delta$,
the jam distances $s'$ and $s^{\prime\prime}$,
and the vehicle length $l$, which has no
dynamical influence. To reduce the number of parameters, 
one can assume $\delta = 1$, $s^{\prime\prime}=0$, and $l=0$, which still 
yields good results.
\par
In equilibrium traffic with
$dv_{\alpha}/dt=0$ and $\Delta v_{\alpha}=0$, drivers tend to keep 
a velocity-dependent
equilibrium clearance $s_{e}(v_{\alpha})$ to the front vehicle
given by
\(
s_{e}(v) = s^*(v,0) [ 1 -
        ( v/v_0 )^{\delta} ]^{-1/2}\, . 
\)
Solving this for the equilibrium velocity $v=v_{e}$
leads to simple expressions
only for $s^{\prime\prime} =0$ and $\delta=1$, $\delta=2$, or $\delta\to\infty$.
In particular, the equilibium velocity for the special case
$\delta=1$ and $s'=s^{\prime\prime}=0$ is
\(
v_{e}(s)  = \{- 1 + \sqrt{1+[4 T^2 (v_0)^2]/s^2} \} s^2/(2 v_0 T^2) .
\)
From this equation and the micro-macro relation 
\(
s = (d - l) = (1/\rho - l)  = (1/\rho - 1/\rho_{\rm max})
\)
between clearance and density follows 
the corresponding equilibrium traffic flow $Q_{e}(\rho)=\rho V_{e}(\rho)$ 
as a function of the traffic density $\rho$.
The acceleration coefficient $\delta$ 
influences the transition region between the free and
congested regimes. For $\delta\to\infty$ and $s^{\prime\prime}=0$, 
the fundamental diagram becomes triangular-shaped:
\(
Q_{e}(\rho) = \mbox{min}\bbox(\rho v_0, [1-\rho(l+s')]/T\bbox) .
\)
For decreasing $\delta$, it becomes smoother and smoother.

\subsection{Cellular automata (CA)} \label{IIB2}

Cellular automata are interesting for their speed and their complex
dynamic behavior (Wolfram, 1984, 1986, 1994; Stauffer, 1991), including such fascinating 
phenomena as {\em self-organized criticality} (Bak \etal, 1987, 1988;
Bantay and Janosi, 1992; Olami \etal, 1992), 
formation of spiral patterns (Markus and Hess, 1990), or 
oscillatory and chaotic sequences of
states (Wolfram, 1984; Markus and Hess, 1990; Nowak and May, 1992). 
Their enormous computation speed
and efficiency is a consequence of the following properties, which
are ideal preconditions for parallel computing: 
(i) discretization of 
space into identical cells (sites) $j$ of size $\Delta x$, 
(ii) a finite number of possible states
$g(x)$ (iii) the (parallel) update at times $t = i\,\Delta t$
with an elementary time step $\Delta t$, and (iv) globally applied update rules,
based on (v) short-range interactions
with a finite (small) number of neighbouring sites. 
Despite these simplifications, cellular automata 
and related {\em lattice gas automata} have a broad range
of applications, reaching from realistic
simulations of granular media (Peng and Herrmann,  1994)
or fluids (Frisch \etal, 1986; Chen \etal, 1991)
(including interfacial phenomena and magnetohydrodynamics),
over the computation of chemical reactions (Markus and Hess, 1990; Dab \etal, 1991),
up to the modeling of avalanches (Bantay and Janosi, 1992).
\par
Their application to traffic dynamics 
has stimulated an enormous research activity, 
aiming at an understanding and control of traffic instabilities, which
are responsible for stop-and-go traffic and congestion, both on freeways 
and in cities. The first cellular automata for freeway traffic go back to
Cremer and coworkers (Cremer and Ludwig, 1986; Sch\"utt, 1990), 
and Nagel and Schreckenberg (1992). Since then,
there have been overwhelmingly many proposals and publications in
this field. Nevertheless, I will keep my discussion short, 
since it has recently been reviewed by Chowdhury \etal\ (2000b).
Complementary, it is worth checking out the Java applets supplied at the
web pages {\tt http:/$\!$/rcswww.urz.tu-dresden.de/$\tilde{\hphantom{n}}$helbing/Road Applet/} 
and {\tt http://www.traffic.uni-duisburg. de/model/index.html}, which allow 
the interested reader to compare the 
dynamics of various cellular automata. 

\subsubsection{The Nagel-Schreckenberg model and its slow-to-start variant}\label{IIB2a}
 
Cellular automata describe the vehicle dynamics less detailed than
follow-the-leader models, but their simplifications are in favour of an 
extremely fast simulation of huge numbers of interacting vehicles.
Nagel and Schreckenberg (1992) suggest to divide the street into 
cells $j$ of length $\Delta x$ and the time $t$ into
intervals $i$ of duration $\Delta t = 1$~s. Each cell is either empty or
occupied by one vehicle with speed
\begin{equation}
 v_i = \hat{v}_i \frac{\Delta x}{\Delta t} \, ,
\end{equation}
where $\hat{v}_i \in \{ 0,1,\dots,\hat{v}_{\rm max}\}$. For freeways,
one frequently chooses the cell size $\Delta x = 7.5$~m and the scaled desired velocity
$\hat{v}_{\rm max} = v_0 \Delta t /\Delta x = 5$. 
The car positions are updated in parallel according to the following
rules:\\
(i) {\em Motion:} Move the vehicle forward by $\hat{v}_i$
cells.\\
(ii) {\em Acceleration:} If a vehicle has not yet reached its
maximum velocity, it would like to drive with the increased velocity
$\hat{v}'_i = (\hat{v}_i +1)$, corresponding to a constant acceleration
$\Delta x/(\Delta t)^2$.\\
(iii) {\em Deceleration:} If the distance (i.e., the number of cells) to the
next vehicle ahead is $\hat{d}_i \le
\hat{v}'_i$, the velocity is reduced to $\hat{v}_i^{\prime\prime}
= (\hat{d}_i-1)$, otherwise $\hat{v}_i^{\prime\prime} = \hat{v}'_i$.\\
(iv) {\em Randomization:} With probability $p$, the
velocity is reduced to $\hat{v}_{i+1} = (\hat{v}_i^{\prime\prime} - 1)$,
if this yields a non-negative velocity, otherwise $\hat{v}_{i+1}=0$.
\par
According to steps (ii) to (iv), the updated velocity as a function
of the previous velocity $\hat{v}_i$ can be summarized by the formula
\begin{equation}
 \hat{v}_{i+1} =
 \max\bbox(0,\min(\hat{v}_{\rm max},\hat{d}_i-1,\hat{v}_i+1) 
 -\xi_i^{(p)}\bbox)
\, ,
\end{equation}
where the Boolean random variable $\xi_i^{(p)} = 1$ 
with probability $p$ and 0 otherwise. 
\par
Note that, according
to rule (iii), drivers move always below the speed which advances them 
by the clearance to the car ahead within $\Delta t$. Hence, the parameter
$\Delta t$ plays, at the same time, the role of an updating step, of
the adaptation time $\tau$, and of the safe time clearance $T$. 
Moreover, the discretization length $\Delta x$ agrees with 
the effective vehicle length $l'$, i.e.,  with the inverse jam 
density. This makes the Nagel-Schreckenberg model so extremely compact
and elegant.
\par
\begin{figure}[htbp]
\caption[]{Trajectories of each tenth vehicle for
unstable traffic in the Nagel-Schreckenberg model, if the slowdown probability is 
(a) $p=0.5$ and (b) $p=0.001$. In the limit $p \rightarrow 0 $,
the jam amplitude goes to zero.\label{NS}}
\end{figure}
The model parameter (``slowdown probability'')
$p$ describes individual velocity fluctuations 
due to delayed acceleration (imperfect driving).
For freeway traffic, Nagel and Schreckenberg have 
often set $p=0.5$, which leads to a relatively noisy dynamics (see Fig.~\ref{NS}),
while $p=0.2$ and $\hat{v}_{\rm max} = 2$ are suitable values for city traffic
(Esser and Schreckenberg, 1997). 
\par
Another model variant with a {\em slow-to-start rule}, the
velocity-dependent randomization model (Barlovic \etal, 1998), assumes
$p= 0.01$ for finite velocities, while $p_0=0.5$ for $\hat{v}_i = 0$.
A similar model was proposed by Benjamin \etal\ (1996).
The fundamental diagram of the resulting model combines the properties
of the Nagel-Schreckenberg model with $p=0.01$ at low densities
with the properties of the one with $p=0.5$ at high densities. 
In between, one finds some ``crossover region'', in which the solutions
of both models can exist, but the high-flow states are {\em metastable}.
We will come back to this and other topics later (see Sec.~\ref{IIC}). 

\subsubsection{Some other cellular automaton models}\label{IIB2b} 

An interesting variant of the Nagel-Schreckenberg model results in the 
{\em cruise control limit} (Nagel and Paczuski, 1995), where the fluctuations are turned off for $\hat{v}_i
= \hat{v}_{\rm max}$. Then, the life-time distribution of traffic jams changes  
from an exponential (Nagel, 1994) to a {\em power-law distribution}, which implies fractal
self-similarity and points to {\em self-organized criticality} (Nagel, 1994; Nagel and Rasmussen, 1994).
\par
The cellular automaton by Brilon and Wu (1999) modifies the Nagel-Schreckenberg
model by additionally introducing an acceleration probability and requiring a minimal
time clearance for acceleration.
\par
Takayasu and Takayasu (1993) were the first to introduce a {\em slow-to-start rule}. 
According to a generalization of their TT or T2 model
(Schadschneider and Schreckenberg, 1997a), a standing vehicle with velocity
$\hat{v}_i = 0$ will accelerate with probability $q = (1-p)$, if there is exactly one
empty cell in front, for $\hat{d}_i > 2$ it will deterministically accelerate to
$\hat{v}_{i+1}=1$.
\par
The model by Nagel and Herrmann (1993) can be viewed as a continuum version of the
Nagel-Schreckenberg model. Its slightly generalized version by
Sauermann and Herrmann (1998) reads
\[
 \hat{v}_{i+1} = \left\{
\begin{array}{ll}
\max(\hat{d}_i - \hat{\delta},0) & \mbox{for } \hat{v}_i > \hat{d}_i - \hat{\alpha}, \\
\min(\hat{v}_i + \hat{a}, \hat{v}_{\rm max}) & \mbox{for } \hat{v}_i < \hat{d}_i - \hat{\beta}, \\
\hat{v}_i & \mbox{otherwise.} 
\end{array}\right.
\]
Herein, the acceleration coefficient is determined by $\hat{a}=\hat{a}_{\rm max} \min
(1,\hat{d}_i/\hat{\gamma})$, where  
$\hat{\alpha}$, $\hat{\beta}$, $\hat{\gamma}$, and $\hat{\delta}$ are the 
parameters of the model. 
\par
Another collision-free continuum-in-space version of the Nagel-Schreckenberg model 
is related to the model by Gipps (1981) and given
by the equation
\[
  v_\alpha(t+1) = \max \Big( 0, \mbox{rnd}\bbox(v_\alpha^{\rm des}(t) - a\,\Delta t,
 v_\alpha^{\rm des}(t)\bbox) \Big) \, ,
\]
where $\mbox{rnd}(z_1,z_2)$ represents a random number uniformly distributed in the interval
$[z_1,z_2)$. The desired velocity is determined by
\(
 v_\alpha^{\rm des}(t) = \min\bbox(v_\alpha^{\rm max}, v_\alpha^{\rm safe}(t), v+a
 \,\Delta t\bbox) ,
\)
while the safe velocity is calculated via
\(
  v_\alpha^{\rm safe}(t) = v_{\alpha-1}(t) + b [ s_\alpha 
- v_{\alpha -1}(t)\, \Delta t]/[v_\alpha(t) + b\,\Delta t ] .
\)
This model was developed and investigated in detail by Krau{\ss} and coworkers
(Krau{\ss} \etal, 1996, 1997; Krau{\ss}, 1998a, b).
\par
The model developed by  Ianigro (1994) is based on {\em Petri nets}.
Yukawa and Kikuchi (1995, 1996) have studied traffic models based on 
{\em coupled maps} (see also Tadaki \etal, 1998). Although there are many more
cellular automata models of traffic flow (e.g., Wolf, 1999), 
I close with pointing to the model 
\(
 \hat{v}_{i+1} =
 \max\bbox(0,\min(\hat{v}_{\rm max},\hat{d}_i-1) 
 -\xi_i^{(p)}\bbox) 
\)
with unlimited acceleration capabilities
by Fukui and Ishibashi (1996a; see also Braude, 1996;
Wang \etal, 1998a). 
\par
Most cellular  automata models have a principal difference to other
models of traffic flow. In contrast to deterministic approaches
such as the above car-following models, they require
fluctuations for the explanation of traffic jams. That is, in the
limit $p \rightarrow 0$, traffic jams will normally disappear
(see Fig.~\ref{NS}). This does not disqualify cellular automata
models, since driver behavior is certainly imperfect and traffic flow
subject to random influences. However, this raises the following
questions: (i) Are fluctuations a dominant or a subordinate effect?
(ii) Which kinds of observations require the consideration of noise?
(iii) Can cellular automata models of traffic flow be mathematically
connected with other traffic models? 

\subsubsection{The discrete optimal velocity model}\label{IIB2c}

To construct a mathematical link between the
Nagel-Schreckenberg model and the optimal velocity model,
let us discretize the latter:
\(
v_\alpha(t+\Delta t) =
 v_\alpha(t) + 
[v'_{e}\bbox(d_\alpha(t)\bbox) - v_\alpha(t)]  \Delta t/\tau  .
\)
Dropping the vehicle index $\alpha$ and scaling the time by $\Delta t = 1$~s,
distances by the cell length $\Delta x$ (to be specified later on), 
and velocities by $\Delta x/\Delta t$ results in 
\(
\hat{v}_{i+1} = \hat{v}_i 
+ \hat{\lambda} [ \hat{v}'_{e} (\hat{d}_i ) - \hat{v}_i ] ,
\)
where $\hat{\lambda} = \Delta t/\tau$. In order to have integer-valued
velocities $\hat{v}$ and locations, we need to introduce a
tabular function $h(z)$. Additionally, we will add some noise
for comparison with the Nagel-Schreckenberg model. The finally
resulting discrete and noisy optimal velocity model reads (Helbing and Schreckenberg, 1999):
\begin{equation}
 \hat{v}_{i+1} = \max \Big( 0, \hat{v}_i 
+ h\bbox( \hat{\lambda} [ \hat{v}'_{e} (\hat{d}_i ) - \hat{v}_i ] \bbox) - \xi_i^{(p)} \Big)
\, .
\end{equation}
For the floor function $h(z) = \lfloor z \rfloor$, the argument $z$ is
rounded down to the largest natural number $n\le z$.
Then, the above equation implies
\(
 \hat{v}_{i+1} \le \hat{\lambda} \hat{v}'_{e}\bbox(\hat{d}_i\bbox) 
 + (1-\hat{\lambda}) \hat{v}_i .
\)
\par\bigskip\begin{figure}[htbp]
\caption[]{Trajectories of each tenth vehicle in the congested regime
according to the discrete version of the optimal velocity model for
slowdown probability (a) $p=0.5$ and (b) $p=0.001$. Jam formation does
not disappear for small values of $p$. These just reduce the average
distance between traffic jams, while their amplitude remains 
constant (see Helbing and Schreckenberg, 1999).\label{HS}}
\end{figure}

\subsubsection{Comparison}\label{IIB2d}
 
The above model is practically as efficient and fast as the
Nagel-Schreckenberg model, but it 
allows a fine-grained description
of vehicle velocities and locations by selecting small values of
$\Delta x$. Moreover, it can be specified in a way that vehicles will
normally not jump into the same cell or over each other. However, 
like in the continuous optimal velocity model itself, the discrete version 
is sensitive to the choice of the parameter $\hat{\lambda}=\Delta t/\tau$ and the
integer-valued velocity-distance function $\hat{v}'_{e}(\hat{d})$. 
A value $\hat{\lambda} \approx 0.77$ seems to be optimal.
For city traffic, the additional specifications $\hat{v}'_{e}(\hat{d}) = \min(\hat{d}-1,3)$ and
$\Delta x = 6.25$~m yield realistic results. The favourite
properties of this model are that (i) jam formation is not affected by
reducing the fluctuation strength (see Fig.~\ref{HS}),
(ii) the characteristic constants of traffic are reproduced (see
Secs.~\ref{IIA5a} and \ref{IIC5}), (iii) the critical densities and the characteristic
constants can be calculated, (iv)
a relation with other cellular automata can be constructed.
\par
For example, the Nagel-Schreckenberg model corresponds to the selection
$\hat{\lambda} = 1$ and $\hat{v}'_{e}(\hat{d}) = \min(\hat{d}-1,\hat{v}_{\rm max})$.
Moreover, $h(z) = \lfloor z \rfloor$, if $z < 1$, otherwise $h(z)=1$,
which limits acceleration, while deceleration capabilities are
unlimited. Nevertheless, these two mathematically
related models lead to a different dynamics
(compare Figs.~\ref{NS} and \ref{HS}), probably because the assumed velocity
changes $d\hat{v}'_{e}(\hat{d})/d\hat{d}$ with the distance 
assumed in the Nagel-Schreckenberg model are not sufficiently large for a 
linear instability, and the ratio between
acceleration and braking capabilities is too small (see Sec.~\ref{IIC4}, Fig.~\ref{KRAUSS1}).

\subsection{Master equation}
\label{IIB3}

Some particle hopping models (see Sec.~\ref{ID2})
simplify the description of transport
processes even further than the above cellular automata do. This is for 
two reasons: (i) to identify the minimal requirements for certain 
phenomena in self-driven many-particle systems and (ii) 
to facilitate analytical investigations, although their treatment is not necessarily
easier. However, the main difference is that cellular automata are {\em discrete} in time rather
than {\em continuous}, allowing us to specify deterministic processes with transition probability 1.
\par\label{DERI}
For a stochastic description, let us consider a 
system with $L$ states $j$ occupied by $n_j$ particles.
The distribution $P(\vec{n},t)$ shall denote the probability
of finding the system configuration, i.e. the occupation vector
\begin{equation}
 \vec{n} = (n_1,\dots,n_j,n_{j+1},\dots,n_L)
\end{equation}
at time $t$. This probability is reduced by transitions to other
configurations $\vec{n}'$, the frequency of which is proportional to $P(\vec{n},t)$.
The proportionality factor is the {\em conditional transition probability}
$P(\vec{n}',t+\Delta t|\vec{n},t)$ of finding the configuration
$\vec{n}'$ at time $(t+\Delta t)$, given that we have the configuration
$\vec{n}$ at time $t$.
Conversely, the probability $P(\vec{n},t)$ increases by transitions
from configurations $\vec{n}'$ to $\vec{n}$, which are proportional
to the occurence probability $P(\vec{n}',t)$ of $\vec{n}'$ and
to the transition probability $P(\vec{n},t+\Delta t|\vec{n}',t)$.
Considering the normalization
\(
\sum_{\vec{n}'} P(\vec{n}',t+\Delta t|\vec{n},t) = 1 ,
\)
the resulting balance equation govering the
dynamics of so-called {\em Markov chains} reads
\begin{equation}
P(\vec{n},t+\Delta t) = \sum_{\vec{n}'}
 P(\vec{n},t+\Delta t|\vec{n}',t) P(\vec{n}',t) 
\label{markov}
\end{equation}
or, in the continuous limit $\Delta t \rightarrow 0$,
\begin{eqnarray}
\frac{dP(\vec{n},t)}{dt} &=& \sum_{\vec{n}'(\ne \vec{n})}
 W(\vec{n}|\vec{n}'; t) P(\vec{n}',t) \nonumber \\
 &-& \sum_{\vec{n}'(\ne \vec{n})}
 W(\vec{n}'|\vec{n};t) P(\vec{n},t)  \, ,
\label{master}
\end{eqnarray}
where we have introduced the {\em transition rates}
\[
 W(\vec{n}|\vec{n}';t) = \lim_{\Delta t \rightarrow 0}
 \frac{P(\vec{n},t+\Delta t|\vec{n}',t)}{\Delta t} \mbox{\ \ } (\vec{n}'\ne \vec{n}) 
\, .
\]
Note that the {\em master equation} (\ref{master}) assumes
the conditional probabilities $P(\vec{n},t+\Delta t|\vec{n}',t)$
to depend on $t$  and $\Delta t$ only, but not on previous time steps. 
\par
For the TASEP (see Sec.~\ref{IE1}) we have
\begin{equation}
P(\vec{n}',t+\Delta t|\vec{n},t) = \left\{
\begin{array}{ll}
q n_j(1-n_{j+1}) & \mbox{if } \vec{n}'=\vec{n}_{j,(j+1)} \\
0 & \mbox{otherwise,}
\end{array}\right.
\label{trate}
\end{equation}
where
\[
 \vec{n}_{jj'} = (n_1,\dots,n_j-1,\dots,n_{j'}+1,\dots,n_L) \, .
\]
In the continuous limit, we assume $\lim_{\Delta t\rightarrow 0}
q/\Delta t = \nu$.

\subsubsection{Solution methods, mapping to spin chains, and matrix product ansatz}\label{MPA}
 
Since the master equation is linear, many solution methods are
available for it (cf. the books and reviews by Feller, 1967; 
van Kampen, 1981; Haken, 1977; Gardiner, 1985; 
Haus and Kehr, 1987; Weiss, 1994; Helbing, 1995a; Sch\"utz, 2000; see also Helbing and Molini, 1995). 
For example, one can map the TASEP to a spin system by identifying
empty sites with down spins and occupied ones with up spins.
Following Doi (1976; see also Sandow and Trimper 1993b; Kaulke and
Trimper, 1995), the
probability distribution $P(\vec{n},t)$ can be related to a state
vector $|P(t)\rangle$ in Fock space, and with respect to its 
orthonormal basis $\{|\vec{n}_k\rangle\}$ we have the relation
\(
 |P(t)\rangle = \sum_k P(\vec{n}_k,t) |\vec{n}_k \rangle .
\)
The linear master equation becomes
\(
 \partial  |P(t)\rangle / \partial t = {\cal L} |P(t) \rangle ,
\)
where the linear operator ${\cal L}$ can, similar to a {\em second
quantization}, be expressed in terms of
creation and annihilation operators usually satisfying Bose commutation rules
(Doi, 1976; Grassberger and Scheunert, 1980; Peliti, 1985; Sandow and
Trimper, 1993a). When the occupation numbers per lattice site
are limited to a maximum value, one has to take into account the {\em exclusion principle},
leading to commutation rules of Pauli operators or similar ones
(Gwa and Spohn, 1992; 
Sandow and Trimper, 1993b; Rudavets, 1993; Sch\"utz and Sandow, 1994;
Alcaraz \etal, 1994).
\par
Alternatively, one may rewrite the master equation
as {\em Schr\"odinger equation} in imaginary time by setting
${\cal H} = - {\cal L}$ (Felderhof and Suzuki, 1971; Siggia, 1977; Alexander and Holstein, 1978;
Sch\"utz, 2000).
The {\em stochastic Hamiltonian} ${\cal H}$ is defined by its matrix elements
\(
\langle \vec{n}_k | {\cal H} | \vec{n}_k \rangle = \sum_{l(\ne k)}
 W(\vec{n}_l|\vec{n}_k) ,
\)
and 
\( 
 \langle \vec{n}_k | {\cal H} | \vec{n}_l \rangle = -
 W(\vec{n}_k|\vec{n}_l) .
\)
The stationary state corresponds to the eigenvector of ${\cal H}$ or
${\cal L}$ with eigenvalue 0, i.e. it can be interpreted as ground state. 
For discrete time dynamics, the corresponding equation (\ref{markov}) can 
be analogously rewritten as
\(
 |P(t+\Delta t)\rangle = {\cal T} |P(t) \rangle .
\)
Here, the stationary state is the eigenvector of the so-called
{\em transfer matrix} ${\cal T}$ with eigenvalue 1. It can be determined by
a so-called {\em matrix product ansatz}
(Derrida \etal, 1993; Hinrichsen, 1996; Rajewsky and Schreckenberg, 1997; Derrida and 
Evans, 1997; Rajewsky \etal, 1998; Klauck and
Schadschneider, 1999; Karimipour, 1999c; see also Krebs and Sandow, 1997,  for an
existence theorem). The stationary solution is 
\begin{equation}
 P_{\rm st}(\vec{n}) = \frac{1}{Z_L} \bigg\langle \! W \bigg| \! \prod_{j=1}^L [ n_{j} {\cal D}
 + (1-n_{j}) {\cal E} ] \bigg| V \! \bigg\rangle ,
\end{equation}
where $|V\rangle$ and $|W\rangle$ are vectors characterizing the boundary conditions, 
${\cal D}$ and ${\cal E}$ are suitable matrices, 
and $Z_L = \langle W | ({\cal D} + {\cal E})^L | V \rangle$
is a normalization constant.
More general formulas are available for 
dynamic solutions, using Bethe ansatz equations
(Stinchcombe and Sch\"utz, 1995a, b; Sasamoto and
Wadati, 1997; Sch\"utz, 1998).

\subsubsection{Mean field approach and Boltzmann equation}\label{MEAN}
 
It is often useful to consider the mean value equations for the
expected values $\langle n_j \rangle = \sum_{\vec{n}} n_j
P(\vec{n},t)$, 
which we obtain by multiplying Eq. (\ref{master}) with $n_j$, summing up over
$\vec{n}$, and suitably interchanging $\vec{n}$ and $\vec{n}'$:
\begin{equation}
 \frac{d\langle n_j \rangle}{dt} = \sum_{\vec{n}}
 n_j \frac{dP(\vec{n},t)}{dt} 
 = \langle m_j(\vec{n},t) \rangle \, .
\end{equation}
Here, we have introduced the first jump moments
\begin{equation}
 m_j(\vec{n},t) = \sum_{\vec{n}'} (n'_j-n_j) W(\vec{n}'|\vec{n};t) \, .
\label{jump1}
\end{equation}
Let us assume spontaneous transitions with individual
transition rates $w(j'|j)$ from state $j$ to $j'$
and pair interactions changing the states from $j$ and $k$ to 
$j'$ and $k'$ with
transition rates $w_2(j',k'|j,k)$. Defining 
\[
 \vec{n}_{jj'}^{kk'}\! =\! (\dots n_j\!-\!1 \dots 
 n_{j'}\!+\!1 \dots  n_k\!-\!1 \dots n_{k'}\!+\!1\dots)  \, ,
\]
the corresponding configurational transition rates are given by
\begin{equation}
 W(\vec{n}'|\vec{n}) = \left\{ 
\begin{array}{ll}
 w(j'|j)n_j & \mbox{if } \vec{n}' = \vec{n}_{jj'}\, ,\\
 w_2(j',k'|j,k)n_jn_k & \mbox{if } \vec{n}' = \vec{n}_{jj'}^{kk'} \, 
, \\
 0 & \mbox{otherwise.}
\end{array}
\right.
\label{pair}
\end{equation}
That is, the total rate of spontaneous transitions is proportional to the number
$n_j$ of systems which may change their state $j$ independently of
each other, while the total rate of pair interactions is proportional to the
number $n_jn_k$ of possible interactions between systems in states $j$ 
and $k$. Inserting Eq.~(\ref{pair}) into (\ref{jump1}) eventually leads
to
\begin{eqnarray}
 m_j(\vec{n}) &=& \sum_{j'} \bigg[ w(j|j') + \sum_{k,k'}
 w_2(j,k|j',k')n_{k'} \bigg] n_{j'} \nonumber\\
 &-& \sum_{j'} \bigg[ w(j'|j) + \sum_{k,k'}
 w_2(j',k'|j,k)n_{k} \bigg] n_{j} 
\end{eqnarray}
(see, for example, Helbing, 1992a, 1995a).
The {\em mean field approach} assumes 
\begin{equation}
 \langle m_j(\vec{n}) \rangle \approx m_j(\langle \vec{n} \rangle) \, ,
\label{MF}
\end{equation}
i.e., that the system dynamics is determined by the mean value $\langle
\vec{n} \rangle$, which is true for a
sharply peaked, unimodal distribution $P(\vec{n},t)$. This leads to
the generalized {\em Boltzmann equation}
\begin{eqnarray}
\frac{d\rho(j,t)}{dt}\! &=&\!\!
\sum_{j'}\!\! \bigg[ w(j|j') \!+\! \sum_{k,k'}
 \hat{w}_2(j,k|j',k')\rho(k',t) \bigg] \rho(j',t)\nonumber\\
 &-& \!\sum_{j'}\!\! \bigg[ w(j'|j) \!+\! \sum_{k,k'}
 \hat{w}_2(j',k'|j,k)\rho(k,t) \bigg] \rho(j,t) \, , \nonumber \\
 & & \label{genboltz}
\end{eqnarray}
where we have introduced the densities $\rho(j,t) = \langle n_j
\rangle/\Delta x$ and $\hat{w}_2(j',k'|j,k) =
w_2(j',k'|j,k)\, \Delta x$. Note that this Boltzmann equation neglects the covariances
\(
 \sigma_{jk}(t) = \bbox\langle (n_j - \langle n_j\rangle)
  (n_k - \langle n_k\rangle) \bbox\rangle 
 = \bbox( \langle n_j n_k \rangle - \langle n_j \rangle \langle n_k \rangle \bbox) 
\)
and the corresponding correlations 
\(
 r_{jk}(t) = \sigma_{jk}/ \sqrt{\sigma_{jj}\sigma_{kk}} .
\)

\subsubsection{TASEP and Nagel-Schreckenberg model}\label{TASEP}
 
For the TASEP with the transition rates (\ref{trate}), the mean field
approach (\ref{genboltz}) yields 
\begin{equation}
 \frac{d\langle n_j \rangle}{dt} = Q_{j-1}(t) - Q_j(t) 
\label{merk}
\end{equation}
with the particle flow
\begin{equation}
 Q_j(t) = \nu \langle n_j (1-n_{j+1}) \rangle \approx \nu \langle n_j \rangle
 (1-\langle n_j \rangle) \, .
\end{equation}
While in open systems exact calculations are difficult due to correlations, this
relation becomes exact for the stationary state of periodic systems
(Spohn, 1991; Schadschneider and Schreckenberg, 1993; 
Schreckenberg \etal, 1995; Liggett, 1999). A
spatially continuous approximation of Eq.~(\ref{merk}) results in the Burgers equation
(\ref{burg1}) or, in lowest-order approximation, 
in the Lighthill-Whitham equation (\ref{LiWi}).
\par
However, replacing the random sequential
update of the TASEP by a parallel update, we end up  with the
traffic model by Nagel and Schreckenberg with
$\hat{v}_{\rm max} = 1$ and $p=(1-q)$ (see Sec.~\ref{IIB2a}).
For a parallel update, the state changes of many particles are
coupled, leading to significant correlations. Consequently,
we have to take into account higher-order interactions 
in the respective transition rates $W(\vec{n}'|\vec{n})$.
It turns out that the above result for the TASEP corresponds to the
{\em site-oriented mean-field theory} (SOMF) for the Nagel-Schreckenberg model with
$\hat{v}_{\rm max} = 1$, which underestimates the flows by neglecting the correlations 
(Nagel and Schreckenberg, 1992; Schreckenberg \etal, 1995). This can be
corrected for by eliminating so-called paradisical or garden-of-Eden
states (Moore, 1962) that
cannot be reached dynamically, when a parallel update is 
applied (Schadschneider and Schreckenberg, 1998).
Alternatively, one can develop a {\em car-oriented mean-field
theory} (COMF), which calculates the probabilities $P_n(t)$ of finding 
$n$ empty sites in front of a vehicle (Schadschneider and
Schreckenberg, 1997b). A superior method (Schadschneider, 1999)
is a {\em site-oriented cluster-theoretic approach}
(Schadschneider and Schreckenberg, 1993; Schreckenberg \etal, 1995), calculating
occurence probabilities of states composed of $n$ successive sites (Kikuchi,
1966; Gutowitz \etal, 1987; ben-Avraham and K\"ohler,
1992; Crisanti \etal, 1993). Instead of
Eq.~(\ref{MF}), a 2-cluster approximation gives 
\begin{eqnarray*}
\hspace*{-3mm} & & \langle n_{j-1}(t) n_j(t) n_{j+1}(t) n_{j+2}(t) \rangle
\nonumber \\
\hspace*{-3mm} &\approx & \frac{\langle n_{j-1}(t) n_j(t) \rangle
 \langle n_j(t) n_{j+1}(t)\rangle \langle
 n_{j+1}(t) n_{j+2}(t) \rangle}{\langle n_j(t) \rangle 
 \langle n_{j+1}(t) \rangle} \, . 
\end{eqnarray*}
This allows us to calculate the time-dependent average flow
\(
 Q(t) = q \langle\!\langle n_j(t) \bbox(1 - n_{j+1}(t)\bbox) 
\rangle\!\rangle / \Delta t = \hat{Q}(t)/\Delta t
\)
(Wang and Hui, 1997),
if we take into account that the occupation numbers change according to 
\begin{eqnarray*}
 n_j(t+1) &=& \bbox(1-\xi_{j-1}^{(1-p)}(t)\bbox) n_{j-1}(t)
\bbox(1-n_j(t)\bbox)  \\ 
&+& n_j(t) n_{j+1}(t)
 + \xi_j^{(p)}(t) n_j(t) \bbox(1-n_{j+1}(t)\bbox) \, . 
\end{eqnarray*}
In the above formulas,
$\langle\!\langle n_j \rangle\!\rangle = \varrho \, \Delta x$ indicates an average 
over all lattice sites $j$. 
For the stationary case, one can finally derive 
the correct stationary flow-density relation
\begin{equation}
 \hat{Q} = \frac{1}{2} \{ 1 - [1 - 4q\langle\!\langle n_j \rangle\!\rangle 
 (1-\langle\!\langle n_j \rangle\!\rangle) ]^{1/2} \} 
\end{equation}
(Yaguchi, 1986; Schadschneider and Schreckenberg, 1993, 1997b, 1998; 
Schreckenberg \etal, 1995). The above results can be gained 
using $\langle \xi_j^{(p)}(t) \rangle = p$,
$\xi_j^{(p)}(t)\xi_j^{(1-p)}(t) = 0$,
$[\xi_j^{(p)}(t)]^2 = \xi_j^{(p)}(t)$, 
as well as the statistical independence of
$\xi_j^{(p)}(t)$, $\xi_{j+1}^{(p)}(t)$, and $\xi_j^{(p)}(t+1)$
(Wang and Hui, 1997). The dynamics and
correlation functions are also different from the TASEP 
(Schadschneider, 1999).

\subsubsection{Nucleation and jamming transition}

A master equation model for the jamming transition on a circular road
of length $L$ has been developed by Mahnke and Pieret (1997). It assumes that
congested regions are eventually merging to form one big ``megajam''.
If the spatial dynamics of the $N$ cars is neglected, 
we need only one single occupation number $n$ 
to count the number of cars in the traffic jam. Moreover, the model assumes
that cars leave a standing traffic jam with the constant rate
\(
 w(n-1|n) = 1/T  .
\)
They join the queue with the rate
\(
 w(n+1|n) = v_{e}\bbox(s(n)\bbox)/s(n) 
\)
given by the vehicle speed 
$v_{e}(s) = v_0 s^2/[(s_0)^2+s^2]$, divided by the average clearance 
\(
 s(n) = (L - Nl)/(N-n)
\)
of freely moving vehicles ($s_0$ being some constant). Mahnke and
Kaupu\v{z}s (1999) derive the stationary distribution 
$P_{\rm st}(n)$ of the queue
length $n$ and show a transition from free to congested traffic at
a certain critical vehicle density. While the maximum of the
probability distribution $P_{\rm st}(n)$ is located at $n=1$ in free
traffic, in congested traffic it is located at 
some finite value $n>1$, which is growing with the vehicle density.
\par
Mahnke and Kaupu\v{z}s compare the nucleation, growth, and condensation
of car clusters with the formation of
liquid droplets in a supersaturated vapor (Schweitzer \etal, 1988;
Ebeling \etal, 1990). 
Moreover, they recognize that, for fixed scaled
parameters and fixed vehicle length $l$, the fundamental 
flow-density diagram of their model is dependent on the 
road length $L$. This, however, results from the fact that they relate
the flow to the overall vehicle density $\varrho =N/L$ rather 
than to the local density $\rho = 1/[l+s(n)]$. 

\subsubsection{Fokker-Planck equation}\label{FOKKER}

K\"uhne and Anstett (1999) have obtained similar results like
Mahnke and coworkers, and they have managed to derive
the distribution $P_{\rm st}(n)$ and
the location of its maximum as well. Their calculations are
based on the related Fokker-Planck equation
\begin{equation}
 \frac{\partial \hat{P}(y,t)}{\partial t}
 = - \frac{\partial}{\partial y} [\hat{m}(y,t)p(y,t)]
 \!+\! \frac{1}{2} \frac{\partial^2}{\partial y^2}
 [\hat{m}_2(y,t)\hat{P}(y,t)] 
\end{equation}
for the scaled, quasi-continuous variable $y=n/N$.
This Fokker-Planck equation 
can be viewed as a second-order Taylor approximation of the
master equation, after this has been rewritten 
in the form
\(
 d\hat{P}(y,t)/dt
 = \sum_{y'} [ \hat{W}(y',y\!-\!y')\hat{P}(y\!-\!y',t) 
- \hat{W}(y',y)\hat{P}(y,t) ]  
\)
with $\hat{W}(y',y) = W\bbox((y+y') N|yN\bbox) = W(n+y'N|n)$.
Correspondingly, the first jump moment, which has
the meaning of a {\em drift coefficient}, is 
\(
 \hat{m}(y,t) = \sum_{y'} y' \hat{W}(y',y) ,
\)
while the second jump moment, which can be interpreted as
diffusion coefficient, is given by
\(
 \hat{m}_2(y,t) = \sum_{y'} (y')^2 \hat{W}(y',y) .
\)
Details regarding the Fokker-Planck equation and its solution
can be found in the books by van Kampen (1981), Horsthemke and Lefever (1984),
Gardiner (1985), and Risken (1989).

\subsection{Macroscopic traffic models} \label{IIB4}

As we will see in Sec.~\ref{IIC4}, the above findings of Mahnke, K\"uhne and coworkers 
regarding the fundamental diagram and the nucleation effect are well compatible with
those for macroscopic traffic models.
In contrast to microscopic traffic models, macroscopic ones
restrict to the description of the collective vehicle dynamics in
terms of the spatial vehicle density $\rho(x,t)$ per lane and the average
velocity $V(x,t)$ as a function of the freeway location $x$ and time
$t$. Macroscopic traffic models have been often preferred to car-following
models for numerical efficiency, but in terms of computation speed
they cannot compete with cellular automata. However, other
advantages are 
(i) their better agreement with empirical data,
(ii) their suitability for analytical investigations,
(iii) the simple treatment of inflows from ramps (see Sec.~\ref{IID}),
(iv) the possibility to simulate the traffic dynamics in
several lanes by effective one-lane models considering a certain
probability of overtaking (see Sec.~\ref{IIB5}). 

\subsubsection{The Lighthill-Whitham model (LW model)}\label{IIB4a}

The oldest and still most popular macroscopic traffic model goes back
to Lighthill and Whitham (1955). It appears that Richards (1956) 
developed the same model independently of them. Their fluid-dynamic
model is based on the fact that, away from on- or off-ramps, no
vehicles are entering or leaving the freeway (at least if we neglect
accidents). This conservation of the vehicle number leads to the
{\em continuity equation}
\begin{equation}
 \frac{\partial \rho(x,t)}{dt} + \frac{\partial Q(x,t)}{\partial x}
 = 0 \, .
\label{cont}
\end{equation}
Herein,
\begin{equation}
 Q(x,t) = \rho(x,t)V(x,t)
\label{densdef}
\end{equation}
is the {\em traffic flow} per lane, which is the product of the density and
the average velocity (see Sec.~\ref{IIB5}). 
\par
We may apply the so-called total or substantial derivative 
\[
 \frac{d_{_V}}{dt} = \frac{\partial}{\partial t} + V \frac{\partial}{\partial
   x} \, ,
\]
describing temporal changes in a coordinate system moving with velocity $V(x,t)$,
i.e. with ``the substance'' (the vehicles). With this, 
we can rewrite (\ref{cont}) in the form
\(
 d_{_V}\rho(x,t)/dt 
 = - \rho(x,t) \, \partial V(x,t)/\partial x  ,
\)
from which we conclude that the vehicle density increases in time ($d_{_V}\rho/dt >
0$), where the velocity decreases in the course of the road ($\partial 
V/\partial x < 0$), and vice versa. Moreover, the density can
never become negative, since $\rho(x,t) = 0$ implies $d_{_V}\rho(x,t)/dt = 0$.
\par
Eq.~(\ref{cont}) is naturally part of any macroscopic traffic model.
The difficulty is to specify the traffic flow $Q(x,t)$. Lighthill and
Whitham assume that the flow is simply a function of the density:
\begin{equation}
 Q(x,t) = Q_{e}\bbox(\rho(x,t)\bbox) = \rho V_{e}\bbox(\rho(x,t)\bbox) \ge 0 \, .
\label{LWass}\label{fund2}
\end{equation}
Herein, the so-called fundamental (flow-density) diagram $Q_{e}(\rho)$
and the equilibrium velocity-density relation $V_{e}(\rho)$ are thought to be suitable
fit functions of empirical data, for which there are
many proposals (see, for example, Sec.~\ref{IIB1}). 
The first measurements by Greenshields (1935) suggest
a linear relation of the form
\begin{equation}
 V_{e}(\rho) = V_0 ( 1 - \rho/\rho_{\rm jam} ) \, ,
\label{linv}
\end{equation}
which is still sometimes used for analytical investigations. 
\par
Inserting Eq.~(\ref{LWass}) into the continuity equation (\ref{cont}),
we obtain
\begin{equation}
 \frac{\partial \rho}{\partial t} + C(\rho) \frac{\partial \rho}{\partial x} = 0 \, .
\label{LiWi}
\end{equation}
This is a {\em non-linear wave equation} (Whitham, 1974, 1979), which describes the
propagation of so-called {\em kinematic waves} with the velocity
\begin{equation}
 C(\rho) = \frac{dQ_{e}}{d\rho} = V_{e}(\rho) + \rho
 \frac{dV_{e}}{d\rho} \, .
\end{equation}
Because of $dV_{e}(\rho)/d\rho \le 0$, we have $C(\rho) \le V_{e}(\rho)$.
Hence, the kinematic waves always propagate backwards with respect to
the average velocity $V_{e}(\rho)$ of motion, namely with the speed
\(
 c(\rho) = [C(\rho) - V_{e}(\rho)] = \rho dV_{e}(\rho)/d\rho \le 
 0 .
\)
\par\begin{figure}[htbp]
\caption[]{(a) Trajectories of each tenth vehicle and (b) corresponding spatio-temporal
density plot illustrating the
formation of a shock wave on a circular road according to the
Lighthill-Whitham model (from Helbing, 1997a). Although the initial condition
is a smooth sinusoidal wave, the upstream and downstream fronts are getting
steeper and steeper, eventually producing discontinuous jumps in the density profile,
which propagate with constant speeds.
The wave amplitude remains approximately unchanged.}
\label{BurgerS}
\end{figure}
Note that $C(\rho)$ is the speed of the {\em characteristic lines} 
(i.e. of local information propagation), which is density dependent. In contrast to linear waves, the
characteristic lines intersect, because their speed in congested areas is lower.
This gives rise to changes of the wave profile, namely to the 
formation of shock fronts, while the {\em amplitude} of kinematic
waves does not change significantly (see Fig.~\ref{BurgerS}). The property of 
a shock front is characterized by the 
upstream value $\rho_-$ and the downstream value $\rho_+$ of the density (see Fig.~\ref{ILLU}). 
The difference in the flow $Q_-$ upstream
and the flow $Q_+$ downstream causes the movement of the density
jump $(\rho_+-\rho_-)$ with speed $S$. Because of vehicle conservation, we have
\(
 (Q_+ - Q_-) = (\rho_+ - \rho_-) S .
\)
Consequently, with Eq.~(\ref{LWass}), the speed of shock propagation is
\begin{equation}
 S(\rho_+,\rho_-) = [Q_{e}(\rho_+)-Q_{e}(\rho_-)]/(\rho_+ -
 \rho_-) 
\label{shockprop}
\end{equation}
(see Fig.~\ref{ILLU}). 
\par
\begin{figure}[htbp]
    \caption[]{Illustration of the geometrical relations between the fundamental
diagram $Q_{e}(\rho)$ (solid line) with other quantities characterizing traffic flow.
The vehicle velocity $V_{e}(\rho)$ corresponds to the slope of the line connecting the
origin (0,0) and the point $\bbox(\rho,Q_{e}(\rho)\bbox)$ ($\cdots$), the
propagation speed $C(\rho)$ of density waves to the slope of the tangent
$dQ_{e}(\rho)/d\rho$ (-~-~-), and the speed $S$ of shock waves to the slope of the
connecting line between $\bbox(\rho_-, Q_{e}(\rho_-)\bbox)$ and
$\bbox(\rho_+,Q_{e}(\rho_+)\bbox)$ (--~--).\label{ILLU}}
\end{figure}
Based on this information, we can understand the phase diagram 
for the TASEP (see Fig.~\ref{SCHUETZ} and Secs.~\ref{IE1}, \ref{DERI},
\ref{TASEP}). Stimulated by a study by Krug (1991), the
corresponding results were obtained by Sch\"utz and Domany (1993) and
Derrida \etal\ (1993). A detailled overview is given by Sch\"utz (2000).
For a system with open boundary conditions with average occupation numbers
$\langle n_0 \rangle$ at the upstream end and $\langle n_{L+1} \rangle$ 
at the downstream end, they found the lower density to dominate within the system, if 
$\langle n_0 \rangle = q_0/q < 1/2$ and 
$\langle n_0 \rangle = q_0/q < q_{L+1}/q = (1-\langle n_{L+1} \rangle)$, 
while the higher density dominated in the system, if
$\langle n_0 \rangle = q_0/q > q_{L+1}/q = (1-\langle n_{L+1} \rangle)$ 
and  $(1-\langle n_{L+1} \rangle) = q_{L+1}/q < 1/2$. 
This is a consequence of downstream shock propagation
in the first case, while shocks propagate upstream in the second case.
When the line $\langle n_0 \rangle = q_0/q = q_{L+1}/q = (1-\langle n_{L+1} \rangle)$
is crossed, a {\em phase transition} from 
a low-density state to a high-density state occurs. Finally, under the condition
$\langle n_0 \rangle = q_0/Q > 1/2$ and $q_{L+1}/q > 1/2$, i.e.,
$\langle n_{L+1}\rangle < 1/2$, the system takes on a {\em maximum flow state}.
For a more detailled analysis see the review by Sch\"utz (2000). Related studies have been
also carried out by Nagatani (1995c).
\unitlength10mm
\begin{figure}[htbp]
\caption[]{Phase diagram of the TASEP with open boundaries
and other models that can be macroscopically
described by the Lighthill-Whitham or Burgers equation, 
with representative density profiles (see insets).
The diagonal line separates the areas $A_{\rm I}$ and $A_{\rm II}$ of downstream 
shock propagation from the areas $B_{\rm I}$ and $B_{\rm II}$ of upstream shock propagation
with corresponding changes in the density profile. Consequently, crossing this line is related to
a transition between a low-density phase controlled by the upstream boundary
and a high-density phase controlled by the downstream boundary. Area $C$ is
characterized by the maximum possible flow.
(After Sch\"utz and Domany, 1993; Sch\"utz, 2000.) 
\label{SCHUETZ}}
\end{figure}

\subsubsection{The Burgers equation}\label{IIB4b}

The Lighthill-Whitham model is very instructive and the basis of a
very elaborate shock wave theory (Haight, 1963; Whitham, 1974; Gazis, 1974; Hida, 1979;
Newell, 1993; Daganzo, 1999b), but 
the development of shock waves results in serious 
difficulties to solve the Lighthill-Whitham model numerically.
A suitable integration method is the {\em Godunov scheme}
(Godunov, 1959; Ansorge, 1990; Bui \etal, 1992; LeVeque, 1992; Lebaque, 1995). 
As an alternative, Newell (1993), Daganzo (1994, 1995a, b), and Lebacque (1997) have recently
developed variants of the Lighthill-Whitham model. 
\par
To avoid the development of shock fronts, one often adds a small diffusion term to 
the LW model to smoothen the wave fronts. Whitham himself (1974) has
suggested to generalize Eq.~(\ref{fund2}) according to
$Q = [Q_{e}(\rho) - D \partial \rho/\partial x]$ or
\begin{equation}
 V(x,t) = V_{e}\bbox(\rho(x,t)\bbox) 
 - \frac{D}{\rho(x,t)} \frac{\partial \rho(x,t)}{\partial x} \, .
\label{suggest}
\end{equation}
The resulting equation reads
\begin{equation}
 \frac{\partial \rho}{\partial t} + \left[ V_{e}(\rho) + \rho
\frac{dV_{e}}{d\rho} \right] \frac{\partial \rho}{\partial x}
 = \frac{\partial}{\partial x} \left( D \frac{\partial
\rho}{\partial x} \right)  \, .
\label{Burg2}
\label{burg1}
\end{equation}
The last term is a diffusion term. For a diffusion constant $D > 0$, 
it yields a significant contribution only when the 
curvature $\partial^2\rho/\partial x^2$ of the density is great.
In case of a density maximum, the curvature is negative, which results 
in a reduction of the density. A density minimum is smoothed out 
for analogous reasons.
\par
For the following analytical considerations, we assume the linear
velocity-density relation (\ref{linv}). With the resulting propagation speed 
\(
 C(x,t) = V_0 [ 1 - 2\rho(x,t)/\rho_{\rm jam} ]
\)
of kinematic waves, we can transform (\ref{burg1}) into
\begin{equation}
  \frac{\partial C(x,t)}{\partial t} + C(x,t) \frac{\partial
C(x,t)}{\partial x} = D \frac{\partial^2 C(x,t)}{\partial x^2} \, .
\label{BURG}
\end{equation}
This so-called {\em Burgers equation} is the simplest equation containing non-linear propagation 
and diffusion. Surprisingly, this equation can be solved exactly,
because it is related to the linear {\em heat equation}
\begin{equation}
 \frac{\partial \Psi(x,t)}{\partial t} = D\frac{\partial^2
\Psi(x,t)}{\partial x^2}
\end{equation}
by means of the Cole-Hopf transformation
\(
 C(x,t) = - [2D/\Psi(x,t)] \partial \Psi(x,t)/\partial x
\)
(Whitham, 1974). 
\par
The Burgers equation overcomes the problem of shock waves, but it cannot
explain the self-organization of phantom traffic jams or stop-and-go
waves (cf. Fig.~\ref{SELFORG}). Recently, the main research activity focusses on the noisy Burgers
equation containing an additional fluctuation term
(Forster \etal, 1977; Musha and Higuchi, 1978; Nagel, 1995, 1996).

\subsubsection{Payne's model and its variants}\label{PAY}\label{IIB4c}

Payne (1971, 1979a) complemented the continuity equation (\ref{cont}) by a
dynamic velocity equation, which he derived from Newell's car-following
model (\ref{NEWELL}) by means of a 
Taylor approximation. Recently, related approaches have been pursued by
Helbing (1998a), Nelson (2000), as well as Berg \etal\ (2000). 
\par
Payne identified microscopic and macroscopic velocities according to
\(
 v_\alpha(t+\Delta t) = V(x+V\,\Delta t, t+\Delta t) 
 \approx  [ V(x,t) + V\,\Delta t\, \partial V(x,t)/\partial x +
 \Delta t \, \partial V(x,t)/\partial t ] . 
\)
Moreover, he replaced the inverse of the headway $d_\alpha$
to the car in front by the density $\rho$ at the place
$x + d_\alpha(t)/2$ in the middle between the leading and the
following vehicle:
\(
 1/d_\alpha(t) = \rho\bbox(x+ d_\alpha(t)/2,t\bbox)
 = \rho\bbox( x + 1/(2\rho),t \bbox) 
 \approx  [ \rho(x,t) + 1/(2\rho) \partial
 \rho(x,t)/\partial x ]  .
\)
This yields
\(
 v'_{e}\bbox(d_\alpha(t)\bbox) = V_{e}\bbox(1/d_\alpha(t)\bbox) 
 \approx \{ V_{e}\bbox(\rho(x,t)\bbox) + 1/[2\rho(x,t)] [dV_{e}(\rho)/d\rho]
 \partial \rho(x,t)/\partial x \} .
\)
From the previous equations
we can obtain Payne's velocity equation
\begin{equation}
 \frac{\partial V}{\partial t} + V \frac{\partial V}{\partial x}
 =  \frac{1}{\Delta t} \left[ V_{e}(\rho) - \frac{D(\rho)}{\rho}
 \frac{\partial \rho}{\partial x}- V \right] \, ,
\label{payne}
\end{equation}
where we have introduced the density-dependent diffusion
\(
 D(\rho) = - 0.5 dV_{e}(\rho)/\partial \rho
 = 0.5 | d V_{e}(\rho)/d \rho | \ge  0 .
\)
The single terms of Eq.~(\ref{payne}) have the following
interpretation:%
\\
(i) Like in hydrodynamics,
the term $V\partial V/\partial x$ is called the {\em transport 
or convection term}. It describes a motion of the velocity profile 
with the vehicles. 
\\
(ii) The term $-[D(\rho)/(\rho \, \Delta t)]\partial \rho /\partial x$
is called {\em anticipation term}, since it reflects the reaction of identical
drivers to the traffic situation in their surrounding, in particular
in front of them.%
\\
(iii) The {\em relaxation term} $[V_{e}(\rho)-V]/\Delta t$ describes the
adaptation of the average velocity $V(x,t)$ to the density-dependent
{\em equilibrium velocity} $V_{e}(\rho)$. This adaptation is exponential in
time with relaxation time $\Delta t$. 
\par
The instability condition of Payne's model agrees exactly with the one of the
optimal velocity model, if we set $\tau = \Delta t$  (see Secs.~\ref{IIB1b} and \ref{IIC1}). 
In the limit $\Delta t \rightarrow 0$, one can derive the stable model (\ref{suggest}) as 
{\em adiabatic approximation} of Eq.~(\ref{payne}). 
Inserting this into the continuity equation, we obtain Eq.~(\ref{burg1})
with Payne's density-dependent diffusion function.
\par
For reasons of numerical robustness, Payne (1979a, b) used a
special discretization modifying his original model 
by numerical viscosity terms. His model was taken up by 
many people, for example, Papageorgiou (1983). Cremer and coworkers
multiplied the anticipation term by a factor
$\rho/(\rho + \rho_0)$ with some constant $\rho_0 > 0$
to get a more realistic behavior at small
vehicle densities $\rho \approx 0$ (Cremer, 1979; Cremer and Papageorgiou, 1981;
Cremer and May, 1986; Cremer and Mei{\ss}ner, 1993). 
A discrete version of this model is used in the simulation package
SIMONE (Mei{\ss}ner and B\"oker, 1996).
Smulders (1986, 1987, 1989) introduced additional fluctuation
terms in the continuity and velocity equations to
reflect the observed variations of the macroscopic traffic quantities.
Further modifications of Payne's freeway simulation package
FREFLO (1979b) were suggested by Rathi \etal\ (1987).

\subsubsection{The models by Prigogine and Phillips}\label{IIB4d}

An alternative model was proposed by Phillips (1979a, b), who derived
it from a modified version of Prigogine's Boltzmann-like traffic model 
(see Sec.~\ref{IIB5a}). Similar to Prigogine and Herman (1971), he obtained
the continuity equation (\ref{cont}) and the velocity equation
\begin{equation}
 \frac{\partial V}{\partial t} + V\frac{\partial V}{\partial x}
 = - \frac{1}{\rho}\frac{\partial P}{\partial x} +
\frac{1}{\tau(\rho)} [V_{e}(\rho) - V ] \, ,
\label{phillips}
\end{equation}
but with another density-dependent relaxation time $\tau(\rho)$. The quantity
\(
 P(x,t) = \rho(x,t) \theta(x,t)
\)
denotes the so-called {\em ``traffic pressure''}, where $\theta(x,t)$ is the
velocity variance of differently driving vehicles. For this variance, one can either
derive an approximate dynamic equation or assume a variance-density 
relation such as 
\(
 \theta(x,t) = \theta_0 [ 1 - \rho(x,t)/\rho_{\rm jam}
 ] \ge 0 
\)
(Phillips, 1979a, b).
According to this, the variance decreases with increasing density and
vanishes together with the equilibrium velocity $V_{e}(\rho)$ at the jam
density $\rho_{\rm jam}$, as expected.
\par
Like the Payne model, the model by Phillips is unstable in a certain
density range. Hence, it could explain emergent stop-and-go waves, but 
it is not numerically robust. In particular, at high densities $\rho$
the traffic pressure decreases with $\rho$, so that vehicles would accelerate
into congested regions, which is unrealistic.
\begin{figure}[htbp]
\caption[]{\label{SELFORG} Illustration of the typical traffic dynamics 
according to macroscopic models with a dynamic velocity equation. The simulations
were carried out with the non-local GKT model
introduced in Sec.~\protect\ref{IIB5d} and show the spatio-temporal evolution of the
traffic density $\rho(x,t)$ on a circular road of circumference 10 km,
starting with homogeneous traffic, to which
a localized initial perturbation (\ref{pertform}) of amplitude $\Delta \rho =10$ vehicles/km/lane
was added. 
(a) Free and stable traffic at low vehicle density.
(b) Formation of stop-and-go waves in the range of linearly unstable traffic
at medium vehicle density. (From Treiber \etal, 1999;
Helbing \etal, 2001a, b.)}
\end{figure}

\subsubsection{The models by Whitham, K\"uhne, Kerner, Konh\"auser,
and Lee}\label{IIB4e}

With relation (\ref{LWass}), the Lighthill-Whitham model
assumes that the traffic flow is always
in equilibrium $Q_{e}(\rho)$. This is at least very questionable for
medium densities (Kerner and Reh\-born, 1996b; Kerner \etal, 1997), 
where there is no unique empirical relation between
flow and density (see Secs.~\ref{IIA2} and \ref{IIC3}). 
Although more conservative researchers still believe that the
Lighthill-Whitham theory is correct, implying that there would be
no unstable traffic in which disturbances are amplified,
Daganzo (1999a) recently notes that ``large oscillations in flow, speed and cumulative count
increase in amplitude across the detectors spanning a long freeway queue 
and its intervening on-ramps'', cf. Figs.~\ref{PINCH} and \ref{COEXIST}. Moreover,
Cassidy and Bertini (1999) state that
''the discharge flows in active bottlenecks exhibit
near-stationary patterns that (slowly) alternate about a constant rate .... The onset of upstream
queueing was always accompanied by an especially low discharge flow followed by a recovery 
rate and  these are the effects of driver behavior we do not yet understand'',
cf. Figs.~\ref{FUNDAM}b and \ref{TR}d. 
\par
Whitham himself (1974) suggested a generalization of the
LW theory, which is related to Phillip's one, but without the mentioned problem of 
decreasing traffic pressure. He obtained 
it from the continuity equation together with the assumption that
the adaptation to the velocity (\ref{suggest}) can be delayed by
some relaxation time $\tau$:
\(
d_{_V} V/dt = [ V_{e}(\rho) -
(D/\rho) \partial \rho/\partial x - V ] / \tau .
\)
Therefore, Whithams velocity equation reads
\(
 \partial V/\partial t + V \, \partial V/\partial x
 = - (\theta_0/\rho) \partial \rho/\partial x
 + [V_{e}(\rho) - V]/\tau
\)
with suitable constants $\theta_0 = D > 0$ and $\tau > 0$.
\par
Unfortunately, this model shows the same problems with numerical
robustness as the models by Payne and Phillips, since it eventually
produces shock-like waves. Therefore, K\"uhne (1984a, b, 1987)
added a viscosity term
$\nu \partial^2 V/\partial x^2$, which has a similar effect 
as the diffusion term in the Burgers equation and resolves the problem (cf. Fig.~\ref{SELFORG}). 
In analogy with the Navier-Stokes equation for compressible fluids,
\underline{K}erner and \underline{K}onh\"auser (1993) preferred a viscosity term of the form
$(\eta/\rho)\partial^2 V/\partial x^2$, leading to the velocity
equation
\begin{equation}
 \frac{\partial V}{\partial t} + V \frac{\partial V}{\partial x}
 = - \frac{\theta_0}{\rho} \frac{\partial \rho}{\partial x}
 + \frac{\eta}{\rho} \frac{\partial^2 V}{\partial x^2} +
 \frac{V_{e}(\rho) - V}{\tau} \, .
\label{KK}
\end{equation}
A stochastic version of this {\em KK model}
has been developed by Kerner \etal\ (1995b) as well
to consider fluctuations in vehicle acceleration (K\"uhne, 1987).
\par
It turns out that the above KK model is rather sensitive to the choice of
the parameters and the velocity-density relation. 
This is probably for similar reasons as in the optimal velocity model,
which had also no dependence of the equilibrium velocity
on the relative velocity between cars.
The most suitable specification of parameters was probably made by
Lee \etal\ (1998). 

\subsubsection{The Weidlich-Hilliges model}\label{IIB4f}

The basic version of the model by Weidlich and Hilliges 
(Weidlich, 1992; Hilliges, Reiner and Weidlich, 1993; Hilliges and Weidlich, 1995) is discrete:
The freeway is divided into cells $j$ of equal length $\Delta x$,
which reminds of cellular automata models. Moreover, the density
$\hat{\rho}(j,t) = \rho(j\,\Delta x, t)$ is governed by the
spatially discretized continuity equation
\begin{equation}
 \frac{\partial \hat{\rho}(j,t)}{\partial t}
 + \frac{\hat{Q}(j,t) - \hat{Q}(j-1,t)}{\Delta x} = 0 \,.
\label{WH1}
\end{equation}
However, Weidlich and Hilliges assume for the flow the
phenomenologically motivated relation
\begin{equation}
 \hat{Q}(j,t) = \hat{\rho}(j,t) \hat{V}(j+1,t) \, ,
\label{WH2}
\end{equation}
according to which the drivers in cell $j$ adapt their speed to the
velocity in the next cell $(j+1)$. This looks like a generalization of the mean
value equations of the TASEP (see Sec.~\ref{TASEP}) and reflects an anticipatory
behavior of the drivers. The choice $\Delta x \approx 100$~m
of the cell length is, therefore, determined by the driver behavior.
Under the assumption
\(
 \hat{V}(j,t) = V_{e}\bbox(\hat{\rho}(j,t)\bbox) ,
\)
the model cannot describe emergent stop-and-go waves, 
since the continuous version of the Weidlich-Hilliges model corresponds to Eq.~(\ref{Burg2})
with the diffusion function
\(
 D(\rho) =  [ V_{e}(\rho) - \rho \,dV_{e}/d\rho ] \Delta x/2
\ge 0 ,
\)
as a Taylor expansion of $\rho\bbox((j-1)\Delta x,t\bbox)$ and 
$V\bbox((j+1)\Delta x,t\bbox)$ shows. For the linear
velocity-density relation (\ref{linv}), we have simply a diffusion
constant $D = V_0 \, \Delta x/2$, which implies an equivalence to 
Eq.~(\ref{burg1}) and the Burgers equation.
\par
To model self-organized stop-and-go waves, Hilliges and Weidlich
(1995) have supplemented Eqs.~(\ref{WH1}), (\ref{WH2}) by the dynamic velocity 
equation
\begin{eqnarray}
 \frac{\partial \hat{V}(j,t)}{\partial t} &+& \hat{V}(j,t)
 \frac{\hat{V}(j+1,t) - \hat{V}(j-1,t)}{2\,\Delta x} \nonumber \\
 &=& \frac{V_{e}\bbox(\hat{\rho}(j,t)\bbox) - \hat{V}(j,t)}{\tau} 
\end{eqnarray}
with $\tau \approx 5$~s. 
In the continuous limit $\Delta x \rightarrow 0$, 
the terms on the left-hand side correspond to the substantial time derivative
$d_{_V}V/dt = \partial V/\partial t + V\partial V/\partial x$.
\par
Moreover, Hilliges and Weidlich (1995) have shown that $\partial
\hat{\rho}(j,t)/\partial t\ge 0$ if $\hat{\rho}(j,t) = 0$,
and $\partial \hat{V}(j,t)/\partial t \ge 0$ if $\hat{V}(j,t)=0$.
This guarantees the required non-negativity of
the density and the average velocity. Additionally, they have developed
simple rules for the treatment of junctions,
which allow the simulation of large freeway networks.

\subsubsection{Common structure of macroscopic traffic models}\label{IIB4g}

The previous models are closely related with each other (Helbing, 1997a, e), as they can
all be viewed as special cases of the density equation
\begin{equation}
 \frac{\partial \rho}{\partial t} + V\frac{\partial \rho}{\partial x}
 = - \rho \frac{\partial V}{\partial x} + D(\rho) \frac{\partial^2 \rho}
 {\partial x^2} + \xi_1(x,t)
\label{common1}
\end{equation}
and the velocity equation
\begin{eqnarray}
 \frac{\partial V}{\partial t} + V\frac{\partial V}{\partial x}
 &=& - \frac{1}{\rho}\frac{d P}{d\rho} \frac{\partial \rho}{\partial x}
 + \nu \frac{\partial^2 V}{\partial x^2} \nonumber \\
 &+& \frac{1}{\tau} ( V_{e} - V ) + \xi_2(x,t) \, . 
\label{common2}
\end{eqnarray}
The differences are in the specification of the diffusion 
$D(\rho)$, the fluctuations $\xi_1(x,t)$, $\xi_2(x,t)$, the
traffic pressure $P(\rho)$, the viscosity-like
quantity $\nu(\rho)$, the relaxation time $\tau(\rho)$,
and the equilibrium velocity $V_{e}(\rho)$. The Lighthill-Whitham model,
for example, results in the limit $\tau \rightarrow 0$. Payne's model
is gained with $\tau = \Delta t$ and $P(\rho) = [V_0 - V_{e}(\rho)]/
(2\,\Delta t)$, while $P(\rho) = \rho \theta_0$ in the KK model (\ref{KK}). Some
parameters must be set to zero in an obvious way.
\par
Section~\ref{IIB5} will show how to derive macroscopic from microscopic traffic models.
We will see that  $D$ and $\xi_1$ should be non-vanishing only in {\em ``first-order models''}
restricting themselves to a density equation. In {\em ``second-order models''} 
with a dynamic velocity
equation, the pressure $P$ consists of a contribution  $\rho \theta$, if there is a variation
of individual velocities due to fluctuations or different driver-vehicle types, and
other contributions which are due to the non-locality of vehicle interactions
(Helbing 1996b, 1997a, 1998a, d; see also Secs.~\ref{NAVI} and \ref{PAY}). Daganzo
(1995c) has fundamentally criticized second-order macroscopic 
traffic models, but his arguments have been invalidated (Whitham, 1974;
Paveri-Fontana, 1975; Helbing, 1995b, 1996a, b, 1997a;
Helbing and Treiber, 1999; Treiber \etal, 1999;
Aw and Rascle, 2000; Zhang, 2000).

\subsection{Gas-kinetic traffic models and micro-macro link} \label{IIB5}

In the following paragraphs, I will show how the
car-following models of Sec.~\ref{IIB1} can be connected with
macroscopic traffic models via a mesoscopic level of description.

\subsubsection{Prigogine's Boltzmann-like model}\label{IIB5a}

Starting in 1960, Prigogine and coworkers
proposed, improved, and investigated a simple gas-kinetic model
(Prigogine and Andrews, 1960; Prigogine, 1961). The results are summarized in 
the book by Prigogine and Herman (1971), to which I refer 
in the following.
\par
Gas-kinetic theories are based on an equation for the {\em phase-space
density} 
\begin{equation}
  \tilde{\rho}(x,v,t) = \rho(x,t)\tilde{P}(v;x,t) \, , 
\end{equation}
which is the product of the vehicle density $\rho(x,t)$ and the distribution
$\tilde{P}(v;x,t)$ of vehicle speeds $v$ at location $x$ and time $t$.
Because of vehicle conservation, we find again kind of a 
continuity equation for the phase-space density:
\begin{equation}
 \frac{d_v\tilde{\rho}}{dt} = 
 \frac{\partial \tilde{\rho}}{\partial t}
 + v \frac{\partial \tilde{\rho}}{\partial x}
 = \left(\frac{d\tilde{\rho}}{d t}\right)_{\rm acc}
 \!\! + \left(\frac{d \tilde{\rho}}{d t}\right)_{\rm int} \, .
\label{gas1}
\end{equation}
However, compared to Eq.~(\ref{cont}),
the right-hand side of this conservation equation is not zero this time,
because of velocity changes. In a coordinate system moving with speed $v$,
the temporal change of the phase-space density is given by
the acceleration behavior and interactions of the vehicles.
\par
Prigogine suggested that the acceleration behavior can be described by
a relaxation of the velocity distribution $\tilde{P}(v;x,t)$ to some desired
distribution $\tilde{P}_0(v)$:
\begin{equation}
 \left(\frac{d\tilde{\rho}}{dt}\right)_{\rm acc} 
 = \frac{\rho(x,t)}{\tau\bbox(\rho(x,t)\bbox)} [ \tilde{P}_0(v) - \tilde{P}(v;x,t) ]
\, .
\label{gas2}\label{prigacc}
\end{equation}
The quantity $\tau(\rho)$ denotes again a density-dependent relaxation time.
By $\tilde{P}_0(v)$, the model intends to reflect the variation of the desired
velocities among drivers, causing the scattering of the
actual vehicle velocities $v$. The distribution $\tilde{P}_0(v)$ can be
obtained by measuring the velocity distribution of vehicles with large
clearances (Tilch, 2001). 
\par
Interactions among the vehicles are modelled by the Boltzmann-like term
\begin{eqnarray}
 \left(\frac{d\tilde{\rho}}{dt}\right)_{\rm int}
 &=& \!\!\int\limits_{w>v}\!\! dw \, [1-\hat{p}(\rho)] |w-v|\tilde{\rho}(x,w,t)
 \tilde{\rho}(x,v,t) \nonumber \\
 &-& \!\!\int\limits_{w<v}\!\! dw \, [1-\hat{p}(\rho)] |v-w|\tilde{\rho}(x,v,t)
 \tilde{\rho}(x,w,t) \, . \nonumber \\
 & & 
\label{prigint}
\end{eqnarray}
This assumes that fast vehicles with velocity $w$ interact with
slower ones with velocity $v<w$ at a rate $|w-v|\tilde{\rho}(x,w,t)
 \tilde{\rho}(x,v,t)$, which is proportional to the relative velocity
$|w-v|$ and to the product of the phase-space densities of the interacting vehicles,
describing how often vehicles with velocities $w$ and $v$ meet at
place $x$. Given that the faster vehicle can overtake with some
density-dependent probability $\hat{p}(\rho)$, it will have to decelerate to 
the velocity $v$ of the slower vehicle with probability $[1-\hat{p}(\rho)]$,
increasing the phase-space density $\tilde{\rho}(x,v,t)$
accordingly. However, the phase-space density is decreased when
vehicles of velocity $v$ meet slower vehicles with velocity $w < v$.
This is reflected by the last term of Eq.~(\ref{prigint}).
\par
The interaction term (\ref{prigint}) can be simplified to
\begin{equation}
 \left(\frac{d\tilde{\rho}}{dt}\right)_{\rm int} =
[1-\hat{p}\bbox(\rho(x,t)\bbox)] \rho(x,t) [V(x,t) - v] \tilde{\rho}(x,v,t)  
\label{gas3}
\end{equation}
by introducing the {\em average velocity} 
\begin{equation}
 V(x,t) = \int dv \; v \, \tilde{P}(v;x,t) = \int dv \; v \, \frac{\tilde{\rho}(x,v,t)}{\rho(x,t)} \,.
\end{equation}
Additionally, we define the {\em velocity variance}
\begin{equation}
 \theta(x,t) = \int dv \, [v-V(x,t)]^2 \tilde{P}(v;x,t) 
\end{equation}
and the average desired velocity
\(
 V_0 = \int dv \, \tilde{P}_0(v) .
\)
To derive macroscopic density and velocity equations, we multiply the
kinetic equation given by Eqs. (\ref{gas1}), (\ref{gas2}), and (\ref{gas3}) by
1 or $v$ and integrate over $v$, 
taking into account that the {\em vehicle density} is given by
\begin{equation}
 \rho(x,t) = \int dv \, \tilde{\rho}(x,v,t) \, .
\end{equation}
We obtain equations (\ref{common1}) and (\ref{common2}) 
with $D(\rho)= 0 = \nu(\rho)$, $\xi_1(t)= 0 = \xi_2(t)$, the {\em traffic pressure}
\begin{equation}
 P(\rho,\theta) = \rho \theta \, ,
\label{trafpress2}
\end{equation}
and the {\em equilibrium velocity}
\begin{equation}
 V_{e}(\rho,\theta) = V_0 - \tau(\rho) [1-\hat{p}(\rho)] \rho \theta \, .
\label{prigvau}
\end{equation}
Therefore, the gas-kinetic approach allowed Prigogine and his coworkers
to derive mathematical
relations for the model functions $P$ and $V_{e}$, which
other researchers had to guess by phenomenological considerations before.
Unfortunately, the above relations are only valid for small densities, but
theories for higher densities are meanwhile available (see Secs.~\ref{IIB5c} and \ref{IIB5d}).
\par
Analyzing the Boltzmann-like traffic model,
Prigogine (1961) found a transition from free to congested traffic
above a certain critical density, which he compared to the phase
transition from a gaseous to a fluid phase (see also Prigogine and Herman, 1971). This
congested state is characterized by the appearance of a second maximum
of the velocity distribution at $v=0$. That is, part of the vehicles move,
while others are completely at rest. This bimodal
equilibrium distribution is also found in modern variants of
Prigogine's model (Nelson, 1995; Nelson \etal, 1997), which is a consequence of the
special acceleration term (\ref{gas2}).
\par
To obtain a better agreement with empirical data,
Phillips (1977, 1979a, b) modified Prigogine's
relations for the overtaking probability $\hat{p}(\rho)$
and the relaxation time $\tau(\rho)$. Moreover, he assumed that the
distribution of desired velocities $\tilde{P}_0(v)$ depends on the density
$\rho$ and the average velocity $V$. Finally, he made several proposals
for the velocity variance $\theta$. Prigogine's model has been also
improved by Andrews (1970a, b, 1973a, b), Phillips (1977, 1979a), Paveri-Fontana (1975),
Nelson (1995) and others (Helbing, 1995c, d, 1996a, b, 1997a, d, 1998a;
Wagner \etal, 1996; Wagner, 1997a, b, c, 1998b; 
Nagatani, 1996a, b, c, 1997a, b; Wegener and Klar, 1996; Klar \etal, 1996;
Klar and Wegener, 1997, 1999a, b; Shvetsov and Helbing, 1999).
They modified the acceleration term,
introduced velocity correlations among successive vehicles, investigated interactions among
neighboring lanes, and/or considered space requirements.

\subsubsection{Paveri-Fontana's model}\label{IIB5b}

Paveri-Fontana (1975) carried out a very detailled investigation of 
Prigogine's gas-kinetic traffic model and recognized some strange
features. He critized that the acceleration term (\ref{prigacc}) would describe
discontinuous velocity jumps occuring with a rate related to
$\tau(\rho)$. Moreover, the desired
velocities of vehicles were a property of the road and not the
drivers, while there are actually different personalities of drivers
(Daganzo, 1995c):
aggressive ones, driving fast, and timid ones, driving
slowly.
\par
Paveri-Fontana resolved these problems by distinguishing different
driver-vehicle types. These were characterized by individual 
desired  velocities $v_0$. As a consequence, he introduced 
an extended phase-space density $\hat{\rho}(x,v,v_0,t)$ and the
corresponding gas-kinetic equation. By integration over $v_0$, one
can regain equation (\ref{gas1}). While Paveri-Fontata
specified the interaction term analogously to Prigogine's one, a
different approach was chosen for the acceleration term. It corresponds
to the microscopic acceleration law
\begin{equation}
 dv/dt = (v_0 - v)/\tau(\rho) \, ,
\label{micracc}
\end{equation}
finally resulting in the formula
\begin{equation}
 \left(\frac{d\tilde{\rho}}{dt}\right)_{\rm acc}
 = - \frac{\partial}{\partial v} \left[ \tilde{\rho}(x,v,t)
 \frac{\tilde{V}_0(v;x,t) - v}{\tau\bbox(\rho(x,t)\bbox)} \right] 
\label{pavacc}
\end{equation}
(Helbing, 1995d, 1996a; Wagner \etal, 1996). Herein, the quantity
\(
 \tilde{V}_0(v;x,t) = \int dv_0 \, v_0 \hat{\rho}(x,v,v_0,t)/
\tilde{\rho}(x,v,t) \approx  V_0(x,t) + [v- V(x,t)] C'(x,t)/\theta(x,t)
\)
denotes the average desired velocity of vehicles moving with velocity
$v$, which is greater for fast vehicles. While 
\(
 V_0(x,t) = \int dv_0 \int dv \, v_0 \hat{\rho}(x,v,v_0,t)/
 \rho(x,t)
\)
represents the average desired velocity at place $x$ and time $t$,
\(
  C'(x,t) = \int dv_0 \int dv \, (v-V)(v_0-V_0) 
 \hat{\rho}(x,v,v_0,t)/\rho(x,t)
\)
represents the covariance between the actual and desired velocities.
In spite of these differences, the macroscopic equations for the
density and the average velocity agree exactly with the ones by
Prigogine. However, the equations for the velocity variance,
the covariance, the average desired velocity and so on are different.
\par 
The existence, construction, and properties of solutions of the Paveri-Fontana equation
have been carefully studied (Barone, 1981; Semenzato, 1981a, b). In addition,
it should be noted that there are further alternatives to the
specification (\ref{pavacc}) of the acceleration term. For example, Alberti and
Belli (1978) assume a density-dependent driver behavior. In contrast,
Nelson (1995) and others (Wegener and Klar, 1996; Klar and Wegener, 1997) 
have proposed an {\em acceleration 
interaction} leading to mathematical expressions
similar to Prigogine's interaction term.

\subsubsection{Construction of a micro-macro link}\label{IIB5c}

For the derivation of macroscopic equations, one may start with the
master equation (\ref{master}) together with 
the transition rates (\ref{pair}), if the state $j$ is replaced by
the state vector $(x,v)$, and the sums are replaced by integrals
due to the continuity of the phase space. The spontaneous
state changes are described by the transition rate
\begin{eqnarray}
 w(x',v'|x,v) &=& \lim_{\Delta t \rightarrow 0} \frac{1}{\Delta t}
 \delta\bbox(x'- (x+v\,\Delta t)\bbox) \nonumber \\
&\times & 
\int d\xi  \, \frac{1}{\sqrt{2\pi\tilde{D}\Delta t}} e^{-(\xi\Delta t/\tau)^2/(2\tilde{D}\Delta t)}
\nonumber \\
 &\times & \delta\!\left(\!v'- \!\Big(v+\frac{v_0+\xi-v}{\tau} \Delta
 t\Big)\! \right)  ,
\end{eqnarray}
reflecting the motion and acceleration of identical vehicles with Gaussian-distributed
velocity fluctuations.
Inserting this into Eq.~(\ref{genboltz}) and reformulating the respective terms 
as a Fokker-Planck equation (see Sec.~\ref{FOKKER}) gives, together with
$\tilde{\rho}(x,v,t) = \langle n_{(x,v)} \rangle /\Delta x$, the following
kinetic equation:
\begin{equation}
 \frac{\partial \tilde{\rho}}{\partial t}
 + \frac{\partial (\tilde{\rho} v)}{\partial x}
 + \frac{\partial}{\partial v} \! \left( \tilde{\rho} \frac{v_0 -
 v}{\tau} \right) = \frac{1}{2} \frac{\partial^2
 (\tilde{\rho}\tilde{D})}{\partial v^2} + \! \left(\frac{\partial
 \tilde{\rho}}{\partial t} \right)_{\rm int} \!\!\! \, .
\label{kineq}
\end{equation}
Herein, the quantity $\tilde{D}$ represents 
a velocity diffusion coefficient, and
\begin{eqnarray}
& & \hspace*{-7mm}\left(\frac{\partial
 \tilde{\rho}}{\partial t} \right)_{\rm int} =
 \int dx' \int dv' \int dy \int dw \int dy' \int dw' \, \nonumber \\ 
&\times & [  \hat{w}_2(x,v,y,w|x',v',y',w')
\tilde{\rho}_2(x',v',y',w',t) 
\nonumber \\[1mm]
& & - \hat{w}_2(x',v',y',w'|x,v,y,w) 
\tilde{\rho}_2(x,v,y,w,t) ]
\label{complex}
\end{eqnarray}
is the interaction term, where we have introduced the so-called {pair-distribution function}
\(
 \tilde{\rho}_2(x,v,y,w,t) = \left\langle n_{(x,v)} n_{(y,w)}\right\rangle / (\Delta x)^2 .
\)
The transition rates $\hat{w}_2(x',v',y',w'|x,v,y,w)$ are proportional to the relative velocity
$|v-w|$ and proportional to the probability of transitions from
the states $(x,v)$ and $(y,w)$ of two interacting vehicles to the states $(x',v')$ and $(y',w')$.
During the interactions, the car locations do not change significantly, only 
the velocities. Therefore, we have
\begin{eqnarray}
& & \hspace{-5mm} \hat{w}_2(x',v',y',w'|x,v,y,w) \nonumber \\
&=& \lim_{\Delta t \rightarrow 0} \frac{1}{\Delta t}
[1-\hat{p}(\rho)] \, |v-w| 
\nonumber \\ 
& & \times \delta\bbox(v' - [v + f(x,v,y,w)\, \Delta t] \bbox) \delta (x'-x) \nonumber \\[1mm]
& & \times \delta\bbox(w' -[w + f(y,w,x,v)\, \Delta t] \bbox)\delta(y' - y)
\end{eqnarray}
(Helbing, 1997a).
Herein, the force $f$ is specified in agreement with Eq.~(\ref{agreement}).
The location and velocity of the interaction partner are
represented by $y$ and $w$, respectively. Quantities without and with
a prime ($'$) represent the respective values before and after the
interaction. 
To simplify the equations, we will assume hard-core
interactions corresponding to abrupt braking maneuvers, when neighboring
particles approach each other and have a distance $d^*$, i.e. $y = (x\pm d^*)$.
Moreover, let us assume the relations
\(
 (v'-w') = - \varepsilon (1-\mu) (v - w)  
\)
and
\(
 (v'+w') = (1-\mu) v + (1+\mu) w
\)
corresponding to
\begin{eqnarray}
 v' &=& \frac{1 - \mu - \varepsilon(1-\mu)}{2} v + \frac{1 + \mu +
\varepsilon (1-\mu)}{2} w \, ,\nonumber \\
 w' &=& \frac{1 - \mu + \varepsilon(1-\mu)}{2} v + \frac{1 + \mu -
 \varepsilon (1-\mu)}{2} w \,.
\end{eqnarray}
By variation of the single parameter $\mu$ reflecting the asymmetry of interactions,
this allows us to switch between {\em vehicle interactions} for $\mu = 1$ 
(velocity adaptation) and {\em granular interactions} 
for $\mu = 0$ ({\em conservation of momentum}, see
Foerster \etal, 1994).  In addition, the parameter $\varepsilon$ allows us to control the
degree of {\em energy dissipation} in granular interactions, where
$\varepsilon (1-\mu) = 1$ corresponds to the elastic case with {\em energy conservation}
relevant for idealized fluids. 
\par
Prigogine's Boltzmann-like interaction term results
for $\mu=1$, $d^* = 0$ (point-like vehicles), and the
so-called {\em vehicular chaos approximation}
\(
\tilde{\rho}_2(x,v,y,w,t) \approx
\tilde{\rho}(x,v,t)\tilde{\rho}(y,w,t) ,
\)
which assumes statistical independence of the vehicle velocities
$v$ and $w$. This is only justified in free traffic. 
At higher densities, we must take into account
the increase of the interaction rate by a density-dependent 
factor $\hat{\chi} \ge 1$ (Chapman and Cowling, 1939;
Lun \etal, 1984; Jenkins and Richman, 1985), since the proportion 
$1/\hat{\chi} \ge 0$ of free space is reduced by the vehicular space 
requirements (Helbing, 1995b, d, 1996a, b; Wagner \etal, 1996;
Klar and Wegener, 1997). For simplicity, we will assume that the
overtaking probability is just given by this proportion of free space, i.e.
\begin{equation}
 \hat{p} = 1/\hat{\chi} 
\end{equation}
(Treiber \etal, 1999).
At high vehicle densities, the velocities of successive cars will
additionally be correlated. This is particularly clear for vehicle
platoons (Islam and Consul, 1991), 
in which the variation of vehicle velocities is considerably
reduced (see Sec.~\ref{IIA4}). For the
treatment of vehicle platoons  there have been several suggestions 
(Andrews, 1970a, b, 1973a, b; Beylich, 1979, 1981). Related to an
approach by Lampis (1978) and others (Poethke, 1982, see Leutzbach, 1988;
Hoogendoorn and Bovy, 1998a, b; Hoogendoorn, 1999), one possibility is to distinguish a
fraction $\hat{c}(\rho)$ of free vehicles and a fraction $[1-\hat{c}(\rho)]$ of
bound (i.e. obstructed) vehicles (Helbing, 1997a, d). 
The most realistic description of vehicle platoons, however, is to set up
equations for vehicle clusters of different sizes 
(Beylich, 1979, 1981). Unfortunately, this ends up with a very complex
hierarchy of spatio-temporal equations, from which useful
macroscopic equations have not yet been successfully derived. 
Instead, we may simply assume a  {\em correlation} $r$ of successive vehicle velocities,
which increases from zero to positive values 
when the density changes from low to high densities (see Fig.~\ref{COR}). Taking also
into account the fact that vehicle velocities are more or less
Gaussian distributed (cf. Figs.~\ref{VELDIST} and
\ref{WAGNER}; Pampel, 1955; Gerlough and Huber, 1975; May, 1990;
see also Alvarez \etal, 1990), one may approximate the 
two-vehicle velocity-distribution function by a bivariate
Gaussian function
\begin{equation}
  \tilde{P}_2(v,w) = \frac{\sqrt{\mbox{det }R}}{2\pi} \; \mbox{e}^{-R(v,w)/2}
\label{GausS}
\end{equation}
with the quadratic form
\(
 R(v,w) = [ (v-V)^2/\theta
 - 2r (v-V)(w-V_{_+})/\sqrt{\theta \theta_{_+}} 
 + (w-V_{_+})^2/\theta_{_+} ] / (1 - r^2)
\)
and its determinant
$\mbox{det } R = 1/[\theta\theta'(1-r^2)]$ (Shvetsov and Helbing, 1999). 
Herein, $V$ denotes the average velocity and $\theta$ the velocity variance
of vehicles at location $x$ and time $t$, whereas the variables with the subscript ``+''
refer to the location $x_{_+}= [x+ d^*(V)]$ of the interaction point. 
Often, one chooses the interaction distance
\begin{equation}
 d^*(V) = \gamma \left( \frac{1}{\rho_{\rm max}} + T V \right) \, ,
\end{equation}
where $\gamma \ge 1$ is an anticipation factor.
I should mention that an analytical solution of the gas-kinetic equations is
not available, even for the stationary case, but the above formula  seems to
be a good approximation of numerical results (Wagner, 1997a). The bivariate Gaussian distribution
is favourable for the evaluation of the interaction integrals, and it makes sense
for $3\sqrt{\theta} \le V$, so that negative velocities are negligible.
\par\begin{figure}[htbp]
\caption[]{Stationary velocity distributions at different vehicle densities 
obtained by numerical solution of a gas-kinetic traffic model 
(after Wagner, 1997a).\label{WAGNER}}
\end{figure}
Summarizing the above considerations, a good approximation for the
{\em pair-distribution function} $\tilde{\rho}_2$ at high vehicle densities is given by
\begin{equation}
 \tilde{\rho}_2(x,v,x_{_+},w,t) = \hat{\chi}_{_+} \rho(x,t) \rho(x_{_+},t) 
 \tilde{P}_2(v,w;x,x_{_+},t) \, ,
\end{equation}
and the resulting Enskog-like gas-kinetic traffic 
equation reads
\begin{eqnarray}
  \frac{\partial \tilde{\rho}}{\partial t} &=& 
 -\frac{\partial ( \tilde{\rho}  v)}{\partial x} 
 -\frac{\partial}{\partial v} \left( \tilde{\rho} \frac{V_0-v}{\tau} \right) 
 + \frac{\partial^2}{\partial v^2} (\tilde{\rho} \tilde{D}) \nonumber \\
 &+& (1-\hat{p}_{_+}) \hat{\chi}_{_+} \rho \rho_{_+}
 \bigg[ \int\limits_{w>v} \!\!\! dw \, (w-v) \tilde{P}_2(w,v) \nonumber \\
 & & -  \int\limits_{w<v} \!\!\! dw \, (v-w) \tilde{P}_2(v,w) \bigg] \, .
\label{fcontin}
\end{eqnarray}

\subsubsection{The non-local, gas-kinetic-based traffic 
model (GKT model)}\label{IIB5d}

The probably most advanced macroscopic traffic models presently available is obtained
by multiplying equation (\ref{kineq}) with $v^k$ ($k\in \{0,1,2\}$) and subsequent integration
over $v$. The corresponding calculations are more or less straightforward, but very 
cumbersome, so the results were only eventually gained (Helbing, 1995d, 1996a, b, 1997a, 1998a;
Helbing and Treiber, 1998a; Shvetsov and Helbing, 1999). 
They can be cast into the general form of the 
continuity equation (\ref{common1}) and the velocity equation
(\ref{common2}), but instead of $V_{e}$ we have a non-local expression 
\begin{equation}
 V^{e}  = V_0 - \underbrace{\tau [1-\hat{p}(\rho_{_+})]\hat{\chi}(\rho_{_+}) \rho_{_+} 
 B(\Delta V,\theta_{_{\!\Delta}})}_{\rm Braking\ Interaction\ Term} \, .
\label{eqVderiv}
\end{equation}
This non-locality has some effects which other macroscopic
models produce by their pressure and viscosity (or diffusion) terms.
Related to this is the higher numerical efficiency of the non-local GKT model used
in the simulation package MASTER (Helbing and Treiber, 1999).
The dependence of the braking interaction on
the effective dimensionless velocity difference 
\(
 \Delta V = (V-V_{_+})/\sqrt{\theta_{_{\!\Delta}}} 
\)
between the actual position $x$ and the interaction point $x_{_+}$
takes into account effects of the velocity variances $\theta$, $\theta_{_+}$
and the correlation $r$ among successive cars (cf. Fig.~\ref{COR}):
\(
 \theta_{_{\!\Delta}} = (\theta - 2 r \sqrt{\theta \theta_{_+}} + \theta_{_+})  .
\)
The {\em ``Boltzmann factor''} 
\begin{equation}  
\label{B}
B(\Delta V,\theta_{_{\!\Delta}}) =  \theta_{_{\!\Delta}} \left\{ 
    \Delta V \Phi(\Delta V) 
           + [1+(\Delta V)^2] E(\Delta V) \right\} 
\label{boltzfact}
\end{equation}
in the braking term is monotonically increasing with $\Delta V$ and
contains the normal distribution
\(
 \Phi(z) = \mbox{e}^{-z^2/2}/\sqrt{2\pi}
\)
and the Gaussian error function 
\(
 E(z) = \int\limits_{-\infty}^{z} dz' \, \Phi(z') . 
\)
\par
To close the system of equations, we specify the velocity 
correlation $r$ as a function of the density according to empirical observations
(see Fig.~\ref{COR}). Moreover,  
for a description of the presently known properties
of traffic flows 
it seems sufficient to describe the variance by the equilibrium relation
\(
 \theta = \tilde{D} \tau
\)
of the dynamical variance equation
(Helbing, 1996b, 1997a), which also determines the traffic pressure via
$P = \rho \theta$ or, more exactly, Eq.~(\ref{presscorr}). Empirical data 
(Helbing, 1996b, 1997a, c; Treiber \etal, 1999) suggest that the variance
is a density-dependent fraction $A(\rho)$ of the squared average velocity:
\begin{equation}
 \theta = A(\rho) V^2 \, . 
\label{th1}
\end{equation}
This guarantees that the velocity variance will vanish if the average
velocity goes to zero, as required, but it will be finite otherwise. 
The variance prefactor $A$ is higher in congested traffic than in
free traffic (see Fig.~\ref{ARHO}). 
\par
Finally, the average time clearance of vehicles should correspond to the safe time clearance $T$,
if the vehicle density is high. Hence, in dense and homogeneous traffic
situations with $V_{_+}= V$ and $\theta_{_+} = \theta$, 
we should have $V = (1/\rho - 1/\rho_{\rm max})/T$. This is to be identified with the
equilibrium velocity
\begin{equation}
 V_{e}(\rho) = \frac{\tilde{V}^2}{2V_0}
       \left( - 1 + \sqrt{1 + \frac{4 V_0^2}{\tilde{V}^2}}
       \right) \, ,
\label{Ve}
\end{equation}
where
\(
\tilde{V}^2 = V_0 / [\tau \rho A(\rho) (1-\hat{p})\hat{\chi}]  .
\)
For the {\em ``effective cross section''} we, therefore, obtain
\begin{equation}
 [1-\hat{p}(\rho)]\hat{\chi}(\rho) = \frac{V_0 \rho T^2}{\tau A(\rho_{\rm max}) 
 (1-\rho/\rho_{\rm max})^2} \, ,
\label{bleibt}
\end{equation}
which makes also sense in the low-density limit $\rho \to 0$, since it
implies $\hat{\chi} \to 1$ and $\hat{p}\to 1$. 

\subsubsection{Navier-Stokes-like traffic equations}\label{IIB5e}\label{NAVI}

The above derivation of macroscopic traffic equations was carried out
up to {\em Euler} order only, based on the (zeroth order)
approximation of local equilibrium. This assumes that the {\em form} 
$\tilde{\rho}_0(x,v,t)$ of the phase-space density in equilibrium
remains unchanged in dynamic situations, 
but it is given by the {\em local} values of the density $\rho(x,t)$ and possibly
other macroscopic variables, see e.g. Eq.~(\ref{FORM}). 
For fluids and granular media, however, it is known
that the phase-space density depends also on their gradients, which is a consequence
of  the finite adaptation time $T_0$ needed to
reach local equilibrium (Huang, 1987). The related
correction terms lead to so-called Navier-Stokes equations, which have been
also derived for gas-kinetic traffic models (Helbing, 1996b, 1998d; 
Wagner, 1997b; Klar and Wegener, 1999a, b). 
For ordinary gases, the Navier-Stokes terms (transport terms)
are calculated from the kinetic equation by means of the Chapman-Enskog method  
(Enskog, 1917; Chapman and Cowling, 1939; Liboff, 1990). 
Here, I will sketch the more intuitive
relaxation-time approximation (Huang, 1987; 
for details see Helbing and Treiber, 1998c).
This assumes that\\
(i) the deviation $\delta \tilde{\rho}(x,v,t) = [\tilde{\rho}(x,v,t)
- \tilde{\rho}_0(x,v,t)]$ is small compared to $\tilde{\rho}_0(x,v,t)$,
so that
\begin{eqnarray}
 & & \frac{d_v \tilde{\rho}}{d t} + 
 \frac{\partial}{\partial v} \!\left( \!\tilde{\rho}\frac{v_0 - v}{\tau} \!\right)\!
 - \frac{1}{2} \frac{\partial^2 (\tilde{\rho} \tilde{D})}{\partial v^2}
 - \!\left( \frac{\partial \tilde{\rho}}{\partial t} \right)_{\rm int}
 \nonumber \\
 &\approx & \frac{d_v \tilde{\rho}_0}{d t} \!+\! 
 \frac{\partial}{\partial v} \!\left(\! \tilde{\rho}_0 \frac{v_0 - v}{\tau} \!\right)\!
 \!-\! \frac{\theta}{2\tau} \frac{\partial^2 \tilde{\rho}_0}{\partial v^2}
 \!-\! \left( \frac{\partial \tilde{\rho}_0}{\partial t} \right)_{\rm int}
 \!\!\!\!\!-\! {\cal L}(\delta \tilde{\rho}) \nonumber \\
& & 
\label{inn}
\end{eqnarray}
with 
\(
 d_v/dt \equiv \partial/\partial t
  + v \, \partial/\partial x = \partial/\partial t
  + V_{e}  \partial/\partial x + \delta v \, \partial/\partial x 
\)
and $\delta v(x,t) = [v - V_{e}(x,t)]$,\\
(ii) the effect of the linear interaction operator ${\cal L}$ can be characterized by 
its slowest eigenvalue $-1/T_0 < 0$ (which depends on the density and velocity variance):
\begin{equation}
 {\cal L}(\delta \tilde{\rho}) \approx - \frac{\delta
   \tilde{\rho}(x,v,t)}{T_0} \, .
\label{rela}
\end{equation}
\par
In order to reach a time-scale separation,
the equilibrium distribution is usually expressed in terms of the macroscopic variables
with the slowest dynamics related 
to the conserved quantities. In fluids, these are the particle
number, momentum, and kinetic energy, while in traffic, this is the particle number only.
Between the dynamics of the
velocity moments $\langle v^k \rangle$ and $\langle v^{k+1}\rangle$ with $k\ge 1$, there is no
clear separation of time scales.
Therefore, it is consequential to express the equilibrium 
phase-space density as a function of the vehicle density $\rho(x,t)$ only:
\begin{equation}
 \tilde{\rho}_0(x,v,t) = \frac{\rho}{[2\pi\theta_{e}(\rho)]^{1/2}}
 \exp \left\{ - \frac{[v-V_{e}(\rho)]^2}{2\theta_{e}(\rho )} \right\} \, ,
\label{FORM}
\end{equation}
where $\theta_{e}(\rho) = A(\rho) [V_{e}(\rho)]^2$. 
The same is done with the pair distribution function $\tilde{P}_2(v,w)$. This allows us to determine
\begin{equation}
\frac{d_v\tilde{\rho}_0}{dt} 
 = \frac{d\tilde{\rho}_0}{d \rho} \frac{d_v\rho}{dt} 
 = \frac{d\tilde{\rho}_0}{d\rho} \left( \frac{\partial \rho}{\partial t}
 + V_{e} \frac{\partial \rho}{\partial x} + \delta v \frac{\partial \rho}{\partial x} \right) \, .
\label{xxx}
\end{equation}
The relation $d \tilde{\rho}_0/d \rho$ can be simply obtained by differentiation of
$\tilde{\rho}_0$, whereas 
\(
\partial \rho/\partial t
 + V_{e} \, \partial \rho/\partial x = - \rho \, \partial V_{e}/\partial x
\)
follows from  the continuity equation (\ref{cont}) in Euler approximation.
Inserting (\ref{rela}) to (\ref{xxx}) into (\ref{inn}) finally yields
\begin{eqnarray*}
  \delta \tilde{\rho}(x,v,t) &=& - T_0 \bigg\{\! \tilde{\rho}_0 \!\bigg[\!
 \frac{1}{\rho} \!+\! \frac{\delta v}{\theta_{e}}\frac{d V_{e}}{d \rho}
 \!+\! \frac{1}{2\theta_{e}}\!\left(\! \frac{\delta v^2}{\theta}\! -\! 1\!\right)\! \frac{d \theta_{e}}
 {d \rho} \!\bigg]\! \nonumber \\
 & & \times \!\left(\! - \!\rho \frac{\partial V_{e}}{\partial x} 
\!+\! \delta v \frac{\partial \rho}{\partial x} \!\right)\!  
\! +\! \frac{\tilde{\rho}_0}{\tau} \!\bigg(\! -\! \frac{v_0 - V_{e}}{\theta_{e}} \delta v 
 \nonumber \\
 & &  \!+ \frac{\delta v^2}{\theta_{e}} \!-\! 1\! \bigg) 
 \!-\! \frac{\tilde{\rho}_0}{2\tau} \!\bigg(\! \frac{\delta v^2}{\theta_{e}}\! -\! 1\!\bigg)\!
 - \!\left( \!\frac{\partial \tilde{\rho}_0}{\partial t} \!\right)_{\rm int}\! \bigg\} \, . 
\end{eqnarray*}
Integrating this over $v$ gives exactly zero, i.e. there are no corrections to the
vehicle density. However, multiplying this with $\delta v$ and 
afterwards integrating over $\delta v$ 
yields corrections $\delta V(x,t) = [V(x,t) - V_{e}\bbox(\rho(x,t)\bbox)]$ 
to the equilibrium velocity according to 
\begin{eqnarray}
 \rho \,\delta V &=& - T_0 \rho \bigg[ \frac{\theta_{e}}{\rho} \frac{\partial \rho}{\partial x}
 + \frac{d\theta_{e}}{d\rho} \frac{\partial \rho}{\partial x} - \frac{dV_{e}}{d\rho}
 \rho \frac{\partial V_{e}}{\partial x} \nonumber \\
 & & \quad + \frac{V_{e}-v_0}{\tau} + (1-p_{_+})\chi_{_+} \rho_{_+} B \bigg] \label{navier1} \\
 &=& T_0 \bigg[ - \frac{\partial (\rho\theta_{e})}{\partial x}
 + \left( \rho \frac{dV_{e}}{d\rho} \right)^2 \frac{\partial \rho}{\partial x} 
  + \frac{\rho}{\tau} ( V^{e} - V_{e} ) \bigg] \, , \nonumber 
\end{eqnarray}
where $B$ is again the Boltzmann factor (\ref{boltzfact}).
Interestingly enough, this Navier-Stokes correction contains all terms and, therefore, 
captures all effects of the dynamic velocity equation (\ref{common2}) in Euler approximation,
which makes the success of the Euler-like GKT model understandable. Considering 
\(
V_{_+} = V_{e}(\rho_{_+}) = V_{e}\bbox(\rho(x+d,t)\bbox)
 \approx  [ V_{e}\bbox(\rho(x,t)\bbox) + d (dV_{e}/d\rho) \partial \rho/\partial x ]  
\)
and $\theta_{_+} \approx  [\theta_{e}\bbox(\rho(x,t)\bbox) 
+ d (d\theta_{e}/d\rho) \partial \rho/\partial x]$, 
we can carry out the linear Taylor approximation
\begin{eqnarray}
 \frac{\rho}{\tau} (V^{e} - V_{e}) &\approx & \frac{\rho d}{\tau}
 \left( \frac{\partial V^{e}}{\partial \rho_{_+}} 
 + \frac{\partial V^{e}}{\partial V_{_+}}\frac{d V_{e}}{d \rho} 
 + \frac{\partial V^{e}}{\partial \theta_{_+}}\frac{d\theta_e}{d\rho} \right)
 \frac{\partial \rho}{\partial x}
 \nonumber \\
 &=& \left( \frac{\partial (\rho\theta_{e})}{\partial \rho}
- \frac{\partial P_{\rm eff}}{\partial \rho} \right) \frac{\partial \rho}{\partial x} 
\label{navier2}
 \end{eqnarray}
for the purpose of linear stability analysis. This 
defines the {\em effective} traffic pressure $P_{\rm eff}$ which, apart from 
the kinematic dispersion effect $\rho\theta$, contains
additional terms arising from vehicle interactions, as in Payne's model and in Sec.~\ref{IIB5f}.
It can be shown that, in contrast to $d(\rho\theta_{e})/d\rho$,
$d P_{\rm eff}/d \rho$ is non-negative for reasonably chosen parameter values
and thereby resolves the problem of vehicle acceleration into congested areas
(see Helbing, 1996b, 1997a; Helbing and Greiner, 1997). 
Inserting Eqs.~(\ref{navier1}) and (\ref{navier2})
into the continuity equation (\ref{cont}) results in an equation of type
(\ref{Burg2}) with the diffusion function
\begin{equation}
  D(\rho) = T_0 \left[ \frac{dP_{\rm eff}}{d\rho} - \left( \rho \frac{dV_{e}}{d\rho} \right)^2 \right] \, .
\end{equation}
This diffusion function in the Navier-Stokes-like traffic model
with {\em eliminated} dynamic velocity equation becomes negative exactly when
the Euler-like traffic model {\em with} dynamic velocity equation
becomes linearly unstable. 
Unfortunately, the resulting equation is even more difficult so solve numerically 
than the Lighthill-Whitham model. This problem can, however, be resolved by taking into
account higher-order terms suppressing high frequency oscillations.

\subsubsection{Simultaneous micro- and macrosimulation}\label{IIB5f}

We will now discuss a way of obtaining macroscopic from microscopic traffic
models which is different from the gas-kinetic approach (see Hennecke \etal,
2000; Helbing \etal, 2001b).
For this, we define an average velocity by, for example, linear interpolation between 
the velocities of neighboring vehicles: 
\begin{equation}
V(x,t)=\frac{v_{\alpha}(t)[x_{\alpha-1}(t)-x]
+v_{\alpha-1}(t)[x-x_{\alpha}(t)]}
{x_{\alpha-1}(t)-x_{\alpha}(t)} 
\end{equation}
($x_{\alpha-1}\ge x \ge x_{\alpha}$).
While the derivative with respect to $x$ gives us 
\(
 \partial V/\partial x 
 = [v_{\alpha-1}(t)-v_{\alpha}(t)]/[x_{\alpha-1}(t)-x_{\alpha}(t)] ,
\)
the derivative with respect to $t$ gives us the {\em exact} equation
\begin{equation}
  \left( \frac{\partial}{\partial t} 
 +V\frac{\partial}{\partial x}\right)V =
A(x,t) \, ,
\end{equation}
where
\begin{equation}
A(x,t)=
\frac{a'_{\alpha}(t)[x_{\alpha-1}(t)-x]+a'_{\alpha-1}(t)[x-x_{\alpha}(t)]}
{x_{\alpha-1}(t)-x_{\alpha}(t)}
\end{equation}
is the linear interpolation of the  
vehicle accelerations $a'_{\alpha}=dv_\alpha/dt$ characterizing the microscopic
model. For most car-following models, the acceleration function can 
be written in the form
$a'_{\alpha}=a'(v_{\alpha},\Delta v_{\alpha}, s_{\alpha})$, where
$\Delta v_{\alpha}=(v_{\alpha}-v_{\alpha-1})$ 
is the approaching rate, and $s_{\alpha}=(x_{\alpha-1} -
x_\alpha - l_{\alpha-1})$ the netto distance to the vehicle in front.
In order to obtain a macroscopic system of partial differential equations for
the average velocity and density, the arguments $v_{\alpha}$, 
$\Delta v_{\alpha}$, and $s_{\alpha}$
of the  vehicle acceleration have to be expressed 
in terms of macroscopic fields, in particular the local vehicle
density $\rho(x,t)$ defined by
\begin{equation}
 \frac{1}{\rho(x,t)}\!=\!\frac{1}{\rho_{\rm max}} 
\!+\! \frac
 {s_{\alpha}(t)[x_{\alpha-1}(t)\!-\!x]\!+\!s_{\alpha-1}(t)[x\!-\!x_{\alpha}(t)]}
 {x_{\alpha-1}(t)\!-\!x_{\alpha}(t)} .
\end{equation}
Specifically, we make the following approximations:
$A(x,t)\approx a'\bbox(V(x,t),\Delta V(x,t),S(x,t)\bbox)$, with
$\Delta V(x,t)=[V(x,t)-V_{_+}(x,t)]$, 
$S(x,t)=\frac{1}{2}[\rho^{-1}(x,t)+\rho^{-1}_{_+}(x,t)] 
 - \rho_{\rm max}^{-1}$, and
the non-locality given by $g_{_+}(x,t) = g\bbox(x+1/\rho(x,t),t\bbox)$
with $g\in \{\rho,V\}$.
In this way, we obtain a non-local macroscopic velocity equation,
which supplements the continuity equation (\ref{cont}) for the vehicle density
and defines, for a given microscopic traffic model, a complementary 
macroscopic model. Note that it does not contain a pressure term
$(1/\rho)\partial P/\partial x$ with $P=\rho\theta$, 
in contrast to the GKT model described above, since we have assumed
identical vehicles and a deterministic dynamics corresponding to
$\theta =0$. For heterogeneous traffic, we expect the additional
term $-(1/\rho)\partial (\rho \theta_{e})/\partial x$, where $\theta_{e}(\rho)$
reflects the resulting, density-dependent velocity variance of the vehicles.
For the intelligent driver model (see Sec.~\ref{IIB1c}), the above
procedure leads to a remarkable agreement, at least for identical
vehicles. It is even possible to carry out {\em simultaneous}
micro- and macrosimulations of neighboring freeway sections
without any significant changes in the 
density and velocity profiles or their propagation speed. This has been
shown for downstream propagating perturbations, for
traffic jams propagating upstream, and for complex spatio-temporal
traffic patterns (Hennecke \etal, 2000; Helbing \etal, 2001b).

\subsubsection{Mesoscopic traffic models}\label{IIB5g}

For the sake of completeness, I also mention some 
``mesoscopic'' traffic models, which describe the microscopic vehicle
dynamics as a function of macroscopic fields. Examples are the model
by Wiedemann and Schwerdtfeger (1987) 
applied in the simulation tool DYNEMO (Schwerdtfeger, 1987), and the model by
Kates \etal\ (1998) used in the simulation tool ANIMAL.

\section{Properties of traffic models}\label{ANA}

In order to see which model ingredients 
are required to reproduce certain observations, 
we will start with simple one-lane models for identical vehicles,
and add more and more aspects throughout the next sections.

\subsection{Identical vehicles on homogeneous freeways} \label{IIC}

\subsubsection{``Phantom traffic jams'' and stop-and-go traffic}\label{IIC1}

For a long time, traffic research has aimed at reproducing the fundamental diagram 
and the kind of phantom traffic jams observed by Treiterer and others
(see Sec.~\ref{IIA5a}). Beforehand, the fundamental diagram was normally fitted by calibration of the
parameters in the respective equilibrium flow-density relation of the model under consideration. 
This procedure, however, makes only sense in the density regime(s) of stable traffic,
as we will see that unstable traffic does not just oscillate around the
fundamental diagram. 
\par
Most traffic models have the following mechanism of traffic instability in common:\\
(i) {\em Overreaction} of drivers, which is either reflected by a positive
slow-down probability $p$ or a delayed reaction
(relaxation time $\tau$ or reaction time $\Delta t$).\\
(ii) {\em Chain reaction} of followers: Before a decelerated vehicle  manages to return to 
its previous speed, the next car approaches and has to brake as well
(Nagel and Paczuski, 1995).
Overreacting followers are getting slower and slower, until traffic comes to
a standstill, although everyone likes to drive fast.
\par
The instability condition for the cellular automaton by Nagel and Schreckenberg 
with small slowdown probability $p\approx 0$ follows
directly from (ii). Starting with equally distributed vehicles on a circular road,
a vehicle reaches the leader in one time step $\Delta t$, if the average
clearance $(L-N\,\Delta x)/N = (L/N - \Delta x)$ becomes smaller than the
average distance moved with the
free velocity  $(\hat{v}_{\rm max} - p)\Delta x/\Delta t$. 
Hence, the average vehicle density $\varrho
= N/L$ above which jams are formed, is
\(
  \varrho \approx 1/[(\hat{v}_{\rm max} + 1)\Delta x]  
\)
(Nagel and Herrmann, 1993; L\"ubeck \etal, 1998; Eisenbl\"atter \etal, 1998). 
A formula for finite slowdown probabilities $p> 0$,
which takes into account that fluctuations may add up and initiate jam formation, 
has recently been derived by Gerwinski and Krug (1999). They calculated the
resolution speed $C_0(p) = \hat{C}_0(p) \Delta x/\Delta t$ of the downstream front of
a ``megajam'' and obtained 
\begin{equation}
 \varrho = \frac{\hat{C}_0(p)}{[\hat{v}_{\rm max} -p+\hat{C}_0(p)]\Delta x} 
\end{equation}
with $\hat{C}_0(p) \le (1 - p)$. 
\par
For most traffic models, however, 
the instability of traffic flow is investigated by means of a 
{\em linear stability analysis.}
This will be sketched in the following.
\par
For {\em microscopic} models of the form
\[
 \frac{dv_\alpha(t+\Delta t)}{dt} =
 \frac{v^{e}\bbox(s_\alpha(t),v_\alpha(t),\Delta v_\alpha(t)\bbox) - v_\alpha(t)}{\tau}
\, ,
\]
often with the simple specification $v^{e}(s,v,\Delta v) = v_{e}(s)$, 
we start our analysis by assuming a circular road
of length $L$ with homogeneous traffic flow of density $\varrho$.
Homogeneity implies that all $N = \varrho L$ vehicles $\alpha$ have identical distances
$d =(s+l) = 1 / \varrho$ to the respective leading vehicle, and the
relative velocity $\Delta v_\alpha$ is zero. Hence, the situation is similar to
a lattice of coupled particles in solid state physics, with the difference that\\
(i) the particles {\em move} with the identical velocity $v = v_{e}(d)$
along  the trajectories $x_0(0) - \alpha (s+l) + v t= x_0(0) - \alpha d + vt$,\\
(ii) the interactions are repulsive and anisotropic, and\\
(iii) there is a driving force counteracting the repulsion.\\ 
The dynamics of this spatially 
homogeneous equilibrium state in the presence of disturbances 
is, therefore, not obvious, but as usual we can carry out 
a stability analysis of the linearized model equations 
\begin{eqnarray*}
 \frac{d\,\delta v_\alpha(t)}{dt} &+& \Delta t \frac{d^2\delta v_\alpha(t)}{dt^2}=
 \frac{1}{\tau} 
 \bigg\{ \frac{\partial v^{e}}{\partial s} [\delta x_{\alpha-1}(t) - \delta x_\alpha(t)]
\nonumber \\
& & \hspace*{-14mm} +\frac{\partial v^{e}}{\partial v} \delta v_\alpha(t)
 + \frac{\partial v^{e}}{\partial \Delta v}[\delta v_\alpha(t) - \delta v_{\alpha-1}(t)]
 - \delta v_\alpha(t) \bigg\}
\, . 
\end{eqnarray*}
Due to linearization, this equation is only valid
for small disturbances $\delta v_\alpha(t) = \delta x_\alpha(t)/\delta t
= [v_\alpha(t) - v] \ll v$ and
$\delta x_\alpha(t) = x_\alpha(t) - [x_0(0) - \alpha d + v t] \ll d$. Its 
general solution has the form of a Fourier series:
\[
 \delta x_\alpha(t) = \frac{1}{N} \sum_{k=0}^{N-1} c_{k} \exp
 \bigg( 2\pi {\rm i} \frac{\alpha k}{N} + (\lambda_k - {\rm i}\omega_k)t \bigg)
\, ,
\]
where $c_k$ are the amplitudes of oscillations with wave length
$L/k$, frequency $\omega_k$, and
exponential growth parameter $\lambda_k$ ($k\in \{1,2,\dots,N\}$). 
Inserting this into the linearized model equations gives consistency relations for
$(\lambda_k - {\rm i}\omega_k)$, which have the form of
so-called {\em characteristic polynomials.} These determine the
possible eigenvalues $(\lambda_k-{\rm i}\omega_k)$. The homogeneous
solution is stable with respect to small perturbations only if
$\lambda_k < 0$ for all $k\in \{1,2,\dots,N\}$. 
Otherwise it is potentially unstable with respect to fluctuations $\xi_\alpha(t)$. 
Specific analyses for different car-following models yield the
instability conditions presented in Sec.~\ref{IIB1}. For details see
Herman \etal \ (1959) or Bando \etal \ (1995a). Typically one finds that traffic becomes 
unstable if the reaction time $\Delta t$ or the relaxation time $\tau$
exceed a certain critical value. 
\par
For {\em macroscopic} traffic models, we proceed similarly. Let us assume
spatially homogeneous traffic of average density $\varrho$ and 
velocity $V_{e}(\varrho)$, which is slightly disturbed according to
$\delta \rho(x,t) = [\rho(x,t) - \varrho] \ll \varrho $ and $\delta V(x,t) = [V(x,t)
- V_{e}(\varrho)] \ll V_{e}(\varrho)$. The linearized density equation 
(\ref{common1}) reads, apart from fluctuations,
\begin{equation}
 \frac{\partial \delta \rho}{\partial t} + V_{e}(\varrho) 
\frac{\partial \delta \rho}{\partial x} = - \varrho
 \frac{\partial \delta V}{\partial x} + D(\varrho) \frac{\partial^2 \delta \rho}{\partial x^2} 
\, ,
\end{equation}
while the linearized velocity equation (\ref{common2})  is
\begin{eqnarray}
 \frac{\partial \delta V}{\partial t} &+& V_{e}(\varrho) \frac{\partial \delta V}{\partial x}
 = - \frac{1}{\varrho} \frac{dP(\varrho)}{d\rho} \frac{\partial \delta \rho}{\partial x}
 + \nu(\varrho) \frac{\partial^2 \delta V}{\partial x^2} \nonumber \\
 &+& \frac{1}{\tau(\varrho)} \left[ \frac{dV_{e}(\varrho)}{d\rho} \delta \rho(x,t)
 - \delta V(x,t) \right] \, .  \label{EQN}
\end{eqnarray}
\par
The general solution of Eq.~(\ref{EQN}) is given by Fourier series
of $\delta \rho(x,t)$ and $\delta V(x,t)$:
\begin{eqnarray*}
  \delta \rho(x,t) &=& \sum_l \!\int\! dk \, \rho_k^l \exp \bbox( {\rm i} k x 
  + [ \lambda_k^l(\varrho) - {\rm i}\omega_k^l(\varrho) ] t \bbox) \, , \nonumber \\
   \delta V(x,t) &=& \sum_l \int dk \, V_k^l \exp \bbox( {\rm i} k x 
  + [ \lambda_k^l(\varrho) - {\rm i}\omega_k^l(\varrho) ] t \bbox) \, ,
\end{eqnarray*}
where $\rho_k^l$ and $V_k^l$ are amplitudes.
The possible values of $(\lambda_k^l - {\rm i}\omega_k^l)$ are given
by the eigenvalues $\tilde{\lambda}_k^l = 
[\lambda_k^l - {\rm i}\bbox(\omega_k^l - k V_{e}(\varrho)\bbox)]$
of the matrix 
\[
 \left( \!\!
\begin{array}{ccc}
 -D(\varrho)k^2-\tilde{\lambda}_k^l \! &\! ,\! &\! - {\rm i} k \varrho \\
-\frac{{\rm i}k}{\varrho} \frac{dP(\varrho)}{d\rho}
+ \frac{1}{\tau(\varrho)} \frac{dV_{e}}{d \rho} \!&\! ,\! &\! 
 - \nu(\varrho) k^2  - \frac{1}{\tau(\varrho)} -\tilde{\lambda}_k^l
\end{array} \!\! \right) \, .
\]
Stability requires the real values $\lambda_k^l(\varrho)$ of the solutions of the corresponding
characteristic polynomial to be negative for all wave numbers $k$.
This leads to the stability condition
\begin{equation}
\varrho \sqrt{\tau(\varrho)} \left| \frac{dV_{e}(\varrho)}{d\rho} \right|
\le  \left[ \tau(\varrho) \frac{dP(\varrho)}{d\rho} + D(\varrho) 
\right]^{1/2} \, ,
\label{stabcond}
\end{equation}
where equality yields the critical densities $\rho_{\rm c2}$ and
$\rho_{\rm c3}$ determining the range of linear instability.
For details see K\"uhne and R\"odiger (1991), and Helbing (1997a). As a consequence, traffic flow
in the Lighthill-Whitham model with $D(\varrho)=0$ and $\tau \rightarrow 0$ is marginally stable, 
while Eq.~(\ref{burg1}) and the related Burgers equation (\ref{BURG}) 
are always stable, if only $D(\varrho) > 0$.
Payne's model with $D(\varrho)=0$ and $dP(\varrho)/d\rho 
= |dV_{e}(\varrho)/d\rho|/[2\tau(\varrho)]$
becomes unstable on the condition
\(
 \varrho \left| dV_{e}(\varrho)/d\rho \right| > 1/(2\varrho \tau) .
\)
This agrees with the instability condition (\ref{bandinst}) of the optimal velocity model,
if we consider that $dv'_{e}/dd = (dV_{e}/d\rho)/ (dd/d\rho) = -\rho^2 (dV_{e}/d\rho)$.
According to this, stability can be improved by decreasing the relaxation time $\tau$
(i.e. by higher acceleration).
The instability condition for the models by K\"uhne, Kerner, and Konh\"auser
with $D(\varrho)=0$ and $dP(\varrho)/d\rho = \theta_0$ is
\(
  \varrho \left| dV_{e}(\varrho)/d\rho \right| > \sqrt{\theta_0} ,
\)
given the road is long (K\"uhne and R\"odiger, 1991; Kerner and Konh\"auser, 1993).
Hence, traffic flow becomes unstable if the velocity-density relation $V_{e}(\rho)$
falls too rapidly with increasing density, so that drivers cannot 
compensate for changes in the traffic situation sufficiently fast. 

\subsubsection{(Auto-)Solitons,
Korteweg-de-Vries equation, and Ginzburg-Landau equation} \label{IIC2}

Instead of using the original equations, in non-linear
science it is common to investigate spatio-temporal
patterns appearing in the vicinity of the instability threshold by
means of approximate {\em model equations} based on series expansions
(Cross and Hohenberg, 1993; Haken, 1997). In this way, the original
equations can sometimes be replaced by the {\em Korteweg-de-Vries equation}
\begin{equation}
 \frac{\partial \tilde{V}}{\partial \tilde{t}} + (1 + \tilde{V})
  \frac{\partial \tilde{V}}{\partial \tilde{x}} + \tilde{\nu}
 \frac{\partial^3 \tilde{V}}{\partial \tilde{x}^3} = 0 \, ,
\end{equation}
where the tilde ($\tilde{\hphantom{n}}$) indicates a scaling of the
respective variables, 
or by the Ginzburg-Landau equation (Cross and Hohenberg, 1993). K\"uhne
(1984b) was the first to apply this approach to the study of traffic
dynamics (see also Sick, 1989). Moreover, Kurtze and Hong (1995) could
derive a perturbed
(modified) Korteweg-de-Vries equation and soliton-like solutions
from the Kerner-Konh\"auser model
for densities immediately above the critical density $\rho_{\rm cr}$.
A similar result was obtained by Komatsu and Sasa (1995) for the
microscopic optimal velocity model (see also Whitham, 1990;
Igarashi \etal, 1999a). 
\par
A valid derivation of such model equations requires that the
solutions of the original equations in the vicinity of the instability
threshold are of small amplitude and stable, otherwise the approximate
equations can be completely misleading (Cross
and Hohenberg, 1993). Therefore, the application of approximate
model equations to traffic is restricted to the practically irrelevant
case with
$\rho_{\rm c1} \approx \rho_{\rm c2} \approx \rho_{\rm c3}
\approx \rho_{\rm c4}$, where instability
barely exists. In the optimal velocity model, this condition would be 
fulfilled, if the inverse density and relaxation time were close to the
critical point $(d_{\rm cr},1/\tau_{\rm cr}) =
(d_{c},2\,dv'_{e}(d_{c})/d\,d)$, where $d_{c}$ is the turning point of
the velocity-distance relation (\ref{Band}) with
$d^2v'_{e}(d_{c})/d\,d^2 = 0$. For that case, Nagatani (1998b, c; see also
Muramatsu and Nagatani, 1999)
derived the time-dependent {\em Ginzburg-Landau equation}
\begin{equation}
  \frac{\partial S'}{\partial t}  = - \left( \frac{\partial}{\partial x'}
  - \frac{1}{2} \frac{\partial^2}{\partial x'{}^2} \right) \frac{\delta
\Phi'(S')}{\delta S'}
\end{equation}
with
$x' = \alpha + 2 t \tau [ dv'_{e}(d_{c})/d\,d ]^2$, 
$S'=(d-d_{c})$, 
\[
 \Phi'(S') = \!\int\! dx' \! \left[ \frac{1}{48} \frac{dv'_{e}(d_{c})}{d\,d} 
 \left( \frac{\partial S'}{\partial x'} \right)^2 \! + \phi'(S') \right] \, ,
\]
and the {\em thermodynamic potential}
\begin{eqnarray*}
  \phi'(S') &=& - \frac{dv'_{e}(d_{c})}{d\,d} \left( \tau
\frac{dv'_{e}(d_{c})}{d\,d} - \frac{1}{2} \right) S'{}^2 \nonumber \\
 &+& \frac{1}{24} \left| \frac{d^3v'_{e}(d_{c})}{d\,d^3} \right|  S'{}^4 .
\end{eqnarray*}
Moreover, Nagatani (1998b, c) calculated the uniform and kink solutions,
the coexistence
curve defined by $\partial \phi'/\partial S' = 0$, and the spinodal line
given by $\partial^2 \phi'/\partial S'{}^2 = 0$
(see also Komatsu and Sasa, 1995). 
In his formalism, the inverse relaxation time $1/\tau$ plays the role of
the temperature in a conventional
phase transition, and the headways $d$ corresponds to the order
parameter.
See also the study by Reiss \etal (1986) for an
earlier application of thermodynamic ideas to traffic. 
\par
Approximate equations for non-linear solutions of {\em large}
amplitudes have been developed by Kerner \etal\ (1997),
based on the theory of {\em autosolitons} (Kerner and Osipov, 1994).
Moreover, the features of
wide moving jams have been compared with autosolitons in physical systems
(Kerner, 1995). They remind of properties found in some
{\em reaction-diffusion systems} (Mikhailov, 1991a, b; Meron, 1992; Kapral and Showalter, 1995;
Woesler \etal, 1996).

\subsubsection{Jam formation: Local breakdown and cluster effects,
segregation, and self-organized criticality (SOC)} \label{IIC3}

Wave formation in traffic has some particular features. It turns out that the
traffic equations are so highly non-linear that the linear approximation
is only of very limited use. For example, the wave number $k^*$ associated with
the largest growth rate $\lambda_{k^*}^l(\varrho)$ at a given average density
$\varrho$ does not at all determine the finally resulting wave length via $2\pi/k^*$.
Moreover, the forming waves are typically not periodic. Even when starting with a
small sinusoidal perturbation, the wave amplitude will not just grow linearly.
Instead, the extended perturbation will drastically change its shape, until 
it eventually becomes localized. Kerner and Konh\"auser (1994) call this the
{\em ``local breakdown effect''}. The resulting perturbation has a characteristic
form which can be approximated by the function 
\begin{eqnarray}
 \rho(x,t_0) &=& \varrho + \Delta \rho
\bigg[ \mbox{cosh}^{-2}\bigg(\frac{x-x_0}{w_{_+}}\bigg) \nonumber \\
&-& \frac{w_{_+}}{w_{_-}} \mbox{cosh}^{-2}
\bigg(\frac{x-x_0-(w_{_+} +w_{_-})}{w_{_-}}\bigg)\bigg]
\label{pertform}
\end{eqnarray}
with suitable parameters $t_0$, $x_0$, $w_{_+}$ and $w_{_-}$ 
determining its location and width (see Fig.~\ref{KERNER}a).
When this shape has been reached, the perturbation grows more and more,
playing the role of a nucleus for jam formation (Kerner and Konh\"auser, 1994).
While small-amplitude perturbations flow with the traffic, the propagation speed 
becomes slower with increasing perturbation 
amplitude and eventually becomes negative. The final result is
a wide traffic jam of characteristic form, which propagates upstream.
This is, because vehicles are leaving the standing jam at the front, while
new ones are joining the traffic jam at its end. The traffic jam is
localized, which is known as
{\em ``local cluster effect''} (Kerner and Konh\"auser, 1994; Herrmann and Kerner, 1998).
Moreover, it is normally surrounded by free traffic flow. Hence, one could say that 
there is a {\em phase separation (segregation)} between free and congested traffic
(Kerner and Konh\"auser, 1994; Kerner and Rehborn, 1996a).
The result of this {\em noise-induced ordering} process
(Helbing and Platkowski, 2000) reminds of
an equilibrium between two different phases (a freely moving, ``gaseous'' state and a
jammed, ``condensed'' state). In other words, traffic jams absorb as much cars
as necessary to have free traffic in the rest of the system. The resulting state
of the system is {\em not} a partial congestion with all vehicles moving slow, 
as the velocity-density relation $V_{e}(\rho)$ would suggest.
\par
After the traffic jam has developed a stationary shape, it moves with
constant velocity $C<0$ upstream the road. Therefore, we have
the relations $\rho(x,t) = \rho(x -Ct,0)$ and  $Q(x,t) = Q(x-Ct,0)$.
Inserting this into the continuity equation (\ref{cont}) gives 
\( 
 -C \, \partial \rho(x_*,0)/\partial x_*
 + \partial Q(x_*,0)/\partial x_* = 0  
\)
with $x_* = (x - Ct)$, which is solved by
\begin{equation}
  Q(x_*,0) =   Q_0 + C \rho(x_*,0) = J\bbox(\rho(x_*,0)\bbox)
\end{equation}
with a suitable constant $Q_0$ (Kerner and Konh\"auser, 1994).
Hence, the flow-density relation of a fully developed traffic jam is
a linear curve with negative slope $C < 0$, the so-called {\em jam line} $J(\rho)$.
While first-order macroscopic traffic models like the Lighthill-Whitham model have to assume
this linear relation for congested traffic, in second-order models with a dynamic velocity
equation it is normally a self-organized relation which often differs
significantly from the fundamental diagram $Q_{e}(\rho)$ (Kerner and Konh\"auser, 1994; 
Kerner and Rehborn, 1996a). Therefore,
the fundamental diagram should be fitted only in the range of stable traffic flow,
while congested traffic flow does not necessarily relate to the fundamental
diagram.
\par
Finally, note that the above described mechanism of jam formation and the
existence of the jam line $J(\rho)$  seems to apply to all models with
a deterministic instability mechanism, i.e. to models with a linearly unstable
density regime (Bando \etal, 1995b; Herr\-mann and Kerner, 1998;
Helbing and Schreckenberg, 1999;
Treiber \etal, 1999, 2000). Many cellular automata
and other probabilistic traffic models have a different mechanism of
jamming, but show a {tendency} of phase segregation
between free and congested traffic as well 
(Nagel, 1994, 1996; see also Schreckenberg \etal, 1995; 
Chowdhury \etal, 1997a; L\"ubeck \etal, 1998;
Roters \etal, 1999). In a strict sense, phase
segregation is found in slow-to-start models, while it
is restricted in the Nagel-Schreckenberg model. Interestingly enough,
the cruise control variant of the latter and a few other cellular
automata display power-law behavior in the density variations
(Choi and Lee, 1995; Cs\'{a}nyi and Kert\'{e}sz, 1995;
Yukawa and Kikuchi, 1995; Zhang and Hu, 1995;
Nagatani, 1995b, 1998a). Therefore, Nagel and Herrmann (1993) as well as
Nagel and Paczuski (1995) have interpreted jam formation as {\em self-organized 
critical phenomenon} (Bak \etal, 1987), which is related
with {\em fractal self-similarity} (Nagel and Rasmussen, 1994).

\subsubsection{Instability diagram, stop-and-go waves, metastability, and hysteresis} \label{IIC4}

To characterize the parameter-dependence of the
possible states of a system resulting in the long run, {\em phase diagrams}
are a very powerful method. They are of great
importance in thermodynamics with various applications in 
metallurgy, chemistry, etc. Moreover, they allow us to compare very different
kinds of systems like equilibrium and nonequilibrium ones, or microscopic and
macroscopic ones, whose  equivalence 
cannot simply be shown 
by transformation to normal forms (Kuramoto, 1989;
Manneville, 1990; Eckmann \etal, 1993). 
Defining {\em universality classes} by mathematically equivalent
phase diagrams, one can even compare so different systems as
physical, chemical, biological and social ones, 
which is done in {\em general systems theory} (Buckley, 1967; von Bertalanffy, 1968; Rapoport, 1986).
\par
The phase diagram for the different traffic states on a homogeneous, circular one-lane road 
as a function of the density $\varrho$ is usually called the {\em instability diagram.}
Most traffic models predict a stable traffic flow at small vehicle densities
and unstable traffic flow above a certain critical density $\rho_{\rm cr}$. 
At very high densities, many models with a deterministic instability mechanism
behave stable again corresponding to creeping traffic, 
while the Nagel-Schreckenberg model and other cellular automata
predict unstable traffic. Based on the calculation of limiting cases, a 
more detailled picture has been suggested by K\"uhne (1991a). He distinguishes
supercritical and subcritical regimes, solitary waves, as well as shock fronts with either
increasing or decreasing densities in downstream direction.
\par
Simplifying matters a little, based on a numerical analysis of their macroscopic traffic model,
Kerner and Konh\"auser found the following picture (see Fig.~\ref{KERNER}):
Below some density $\rho_{\rm c1}$, any kind of disturbance eventually disappears.
Between the densities $\rho_{\rm c1}$ and $\rho_{\rm c2}$, 
one wide traffic jam builds up, given a large enough perturbation. A series of traffic jams
appears in a density range between $\rho_{\rm c2}$ and some density
$\rho_{\rm c3}$. One so-called ``anticluster'' or ``dipole layer'' can be triggered, if the density $\varrho$ is
between $\rho_{\rm c3}$ and $\rho_{\rm c4}$, while any disturbance disappears 
in stable traffic above some density $\rho_{\rm c4}$. The critical densities $\rho_{{\rm c}k}$ 
depend mainly on the choice of the pressure function, the relaxation time, 
and the velocity-density relation. 
\par
The interpretation of these findings
is as follows (Kerner and Konh\"auser, 1994): 
Traffic is linearly unstable between the densities $\rho_{\rm c2}$ and
$\rho_{\rm c3}$. Therefore, the slightest inhomogeneity can cause a traffic jam.
Such an inhomogeneity can even be the transition region between a traffic jam and
the surrounding traffic. Hence, an existing  traffic jam will produce a new traffic jam
(downstream of it), which generates another  traffic jam, and so on, until there are
not enough vehicles left to form another jam. Such an irregular series of traffic
jams is called {\em stop-and-go traffic}. K\"uhne's (1991a) interpretation of this irregularity was
based on {\em chaotic dynamics}, 
motivated by an approximate mapping of his traffic 
model on the {\em Lorenz equation} (K\"uhne and Beckschulte, 1993). 
However, numerical investigations indicate that
the largest {\em Lyapunov exponent} stays close to zero (Koch, 1996). 
\par
Stop-and-go waves have been compared with waves in
shallow water (K\"uhne, 1984b; Hwang and Chang, 1987), with the clogging of 
sand falling through a vertical pipe (Schick and Verveen, 1974; Baxter \etal, 1989; Lee, 1994;
P\"oschel, 1994; Peng and Herrmann, 1995) and of
lead spheres falling through a fluid column (Horikawa \etal, 1996; Nakahara and Isoda, 1997),
but the propagation direction is opposite.
\par
Kerner and Konh\"auser (1994) have shown that, in the density ranges
$[\rho_{\rm c1}, \rho_{\rm c2}]$ and $[\rho_{\rm c3},\rho_{\rm c4}]$,
an existing traffic jam does not trigger any further jams, 
because traffic flow is not linearly unstable anymore. 
In these density regimes, they found {\em metastable traffic},
which is characterized by a {\em critical amplitude} $\Delta \rho_{c}(\varrho)$
for the formation of traffic jams. This amplitude is zero for $\varrho =
\rho_{\rm c2}$ and $\varrho =\rho_{\rm c3}$, i.e. at the edge of unstable
traffic, while it grows towards the edge of stable traffic 
and is expected to diverge at
$\varrho=\rho_{\rm c1}$ and $\varrho = \rho_{\rm c4}$
(see Fig.~\ref{KERNER}). 
Perturbations with subcritical amplitudes $\Delta \rho < \Delta \rho_{c}$ 
are eventually damped out (analogous to the stable density ranges),
while perturbations with supercritical amplitudes $\Delta \rho > \Delta \rho_{c}$ 
grow and form traffic jams (similar to the linearly unstable density
ranges). 
The situation in metastable traffic is, therefore, similar to supersaturated vapor
(Kerner and Konh\"auser, 1994), 
where an overcritical nucleus is required for condensation {\em (``nucleation effect'')}.
\par
\begin{figure}[htbp]
\caption[]{Schematic illustration of (a) the chosen localized perturbation
and the density-dependent perturbation 
amplitudes $\Delta \rho_{c}$ required for jam formation in the metastable
density regime, (b) the instability diagram of homogeneous (and slightly inhomogeneous)
traffic flow, (c) the finally resulting traffic patterns as a function of the respective
density regime. (Simplified diagram
after Kerner \etal, 1995a, 1996; 
see also Kerner and Konh\"auser, 1994).
\label{KERNER}}
\end{figure}
Interestingly, there are also traffic models without the phenomenon of metastability.
Investigations by Krau{\ss} (1998a, b) for a Gipps-like family of traffic models
(Gipps, 1981; see also McDonald \etal, 1998) 
suggest the following general conclusions (see Fig.~\ref{KRAUSS1}):
\begin{itemize}
\item[(i)] Models with a high ratio between maximum acceleration $a$ and maximum
deceleration $b$ never show any structure formation, 
since traffic flow is always stable.
\item[(ii)] Models with a relatively high maximum deceleration $b$
display a {\em jamming transition}, which is not
hysteretic. As a consequence, there are no metastable high-flow states and the
outflow from jams is  maximal.
\item[(iii)] A {\em hysteretic} jamming transition with meta\-stable {\em high-flow 
states} and a charac\-teristic, reduced outflow $Q_{\rm out}$ from traffic jams (see Sec.~\ref{IIC5}) 
is found for relatively small accelerations and medium decelerations. 
\end{itemize}
\par\begin{figure}[htbp]
\caption[]{(a) Schematic phase diagram of model classes as a function of the
acceleration strength $a$ and the deceleration strength $b$.
(b) Models belonging to phase (i) have no critical density $\rho_{\rm cr}$ 
and display stable traffic in accordance with the
fundamental diagram (thick dashed and solid lines), while models belonging to phase
(ii) show a non-hysteretic jamming transition
with a resulting flow-density relation represented by solid lines.
(c) A hysteretic phase transition with metastable high-flow states is found for models
belonging to phase (iii). The model behavior is determined by the relative magnitude
of the outflow $Q_{\rm out}$ from congested traffic compared to the flow 
$Q_{\rm cr}= Q_{e}(\rho_{\rm cr})$
at which traffic becomes linearly unstable with respect to small perturbations.
(After Krau{\ss}, 1998a).
\label{KRAUSS1}\label{KRAUSS2}}
\end{figure}
The existence of high-flow states requires the flow $Q_{\rm out} = Q_{e}(\rho_{\rm out})$ to be
smaller than the flow $Q_{\rm cr} = Q_{e}(\rho_{\rm cr})$  at which traffic becomes linearly
unstable, so that there is a density range of metastable
traffic. By variation of model parameters, it is also possible to have the case
$Q_{\rm cr} < Q_{\rm out}$, where traffic breaks down before it can
reach high-flow states. This corresponds to  case (ii) of the above classification 
by Krau{\ss} (1998a, b), where traffic flow does never exceed the
jam line $J(\rho)$ (see Fig.~\ref{KRAUSS1}). 
The Nagel-Schreckenberg model belongs to class (ii). Based on an analysis of 
the correlation function (Eisenbl\"atter \etal, 1998; Schadschneider, 
1999; Neubert \etal, 1999a) and other methods, 
it has been shown that the jamming transition is not of hysteretic nature, but continuous 
for $p=0$. For $0 < p < 1$, there is a strong evidence for a 
crossover (Sasvari and Kert\'{e}sz, 1997;
Eisenbl\"atter \etal, 1998; Schadschneider, 
1999; Neubert \etal, 1999a; Chowdhury \etal, 2000a),
although some people believe in critical behavior (Roters \etal, 1999, 2000).
\par
Note that cellular automata with metastable
high-flow states do exist, basically all slow-to-start models 
(Takayasu and Takayasu, 1993; Benjamin \etal, 1996; Barlovic \etal, 1998).
However, it remains 
to be answered whether this metastability is of the same or a different type as described
above, since in probabilistic models, the transition from high-flow states to jamming
is due to the unfortunate adding up of fluctuations. One would have to check whether
perturbations of the form (\ref{pertform}) tend to fade away, when they are smaller than
a certain critical amplitude $\Delta \rho_{c}$, but always grow, when they are overcritical.
Alternatively, one could determine the phase diagram of traffic states in the presence
of bottlenecks (see Sec.~\ref{IID1}).

\subsubsection{Self-organized, ``natural'' constants of traffic flow}\label{IIC5}

If the fundamental diagram $Q_{e}(\rho)$ is not linear in the congested regime,
the jam line $J(\rho)$ could, in principle, be a function of the average density $\varrho$
throughout the system. The same applies to the propagation velocity $C$ of the
jam, which is given by the slope of the jam line. 
However, in some traffic models, the
jam line $J(\rho)$ and the propagation velocity $C$ are independent of the average 
density $\varrho$, the initial conditions, and other factors (Kerner, 1999b), 
which is very surprising. This was first shown by analytical investigations for wide jams
in the limit of small viscosity (Kerner and Konh\"auser, 1994; Kerner \etal, 1997), 
based on the theory of {\em autosolitons} (Kerner and Osipov, 1994). 
Meanwhile, there is also empirical evidence 
for this (Kerner and Rehborn, 1996a). 
\par
Note that the characteristic jam line $J(\rho)$ defines some other constants as well:
The intersection point $(\rho_{\rm out},Q_{\rm out})$ with the free branch of the fundamental
diagram $Q_{e}(\rho)$ determines, because of the phase segregation 
between jammed and free traffic, the density $\rho_{\rm out}$
downstream of traffic jams. The quantity $\rho_{\rm out}$ is a characteristic density between
$\rho_{\rm c1}$ and $\rho_{\rm c2}$, which depends on the choice of the traffic model
and the specification of the model parameters. Moreover, the value $Q_{\rm out}$ 
characterizes the outflow from traffic jams. Finally, the density $\rho_{\rm jam}$ at
which the jam line $J(\rho)$ becomes zero, agrees with the density inside of 
standing traffic jams. 
\par
In summary, we have the following characteristic constants of wide traffic jams:
the density $\rho_{\rm out}$ downstream of jams
(if these are not travelling through ``synchronized'' congested traffic, see Kerner, 1998b), 
the outflow $Q_{\rm out} = Q_{e}(\rho_{\rm out})$ from jams, the
density $\rho_{\rm jam}$ inside of traffic jams, for which $J(\rho_{\rm jam}) = 0$,
and the propagation velocity $C_0 = Q_{e}(\rho_{\rm out})/( \rho_{\rm out} - \rho_{\rm jam})$
of traffic jams. Note that these ``natural constants'' of traffic 
are not at all a trivial consequence of the
construction of the corresponding models. They are rather a result of
self-organization within traffic, i.e. dynamical invariants. These are related with
an attractor of the system dynamics, determining the form of the wave fronts. 
In non-linear systems, however, the size and shape of an attractor usually changes with
the control parameter, which is the density, here. The attractor 
in the congested traffic regime can be only independent of the density due
to the hysteretic nature of the transition and the associated
phase segregation between free and jammed traffic. 
\par
Not all traffic models display this independence. For example, although the microscopic
optimal velocity model produces a similar dynamics as the KK model (\ref{KK})
(Berg \etal, 2000), for certain velocity-distance functions $v'_{e}(d)$
its propagation velocity of traffic jams depends on the average density $\varrho$
(Herr\-mann and Kerner, 1998). This happens, if the resulting
jam density $\rho_{\rm jam}$ varies with the
vehicle velocity when approaching a jam and, therefore,  with the
average density $\varrho$ and the initial conditions. 
The density-dependence can, however, be suppressed by
suitable discretizations of the optimal velocity model (Helbing and Schreckenberg, 1999) or
by models in which the approaching process depends on the relative velocity, for example
the GKT model or the IDM (Treiber \etal, 1999, 2000). 
It is also suppressed by particular velocity-distance
functions such as the simplified relation
\(
 v_{e}(d) = v_0 \Theta' (d - d_0) ,
\)
where $\Theta'(z)$ denotes the Heaviside function, which is 1 for $z> 0$ and
otherwise 0. For this choice, the optimal velocity model 
can be exactly solved.  Sugiyama and Yamada (1997) find that the dynamics of fully developed
jams can be represented by a hysteresis loop in the speed-over-distance plane
rather than the above Heaviside function (see Fig.~\ref{HYST}). Moreover, the characteristic
constants of traffic flow are related through the equation  
\(
 1/\rho_{\rm out}  = 1/\rho_{\rm jam} + v_0 T' \, , 
\)
where $T' = (t_\alpha - t_{\alpha - 1})$ is the characteristic time interval 
between the moments $t_\alpha$, when successive cars $\alpha$ have a distance
$d_0$ to the respective leading car $(\alpha - 1)$ and start accelerating in order
to leave the downstream jam front. The densities $\rho_{\rm out}$ and
$\rho_{\rm jam}$ of free and jammed traffic, respectively, follow from
$1/\rho_{\rm out} = d_0 + v_0 T'/2$. Moreover, the resolution velocity of jam fronts
can be determined as
\(
  C_0 = [x_\alpha(t_\alpha) - x_{\alpha - 1}(t_{\alpha - 1})]/[t_{\alpha} - t_{\alpha - 1}]
  = - 1/\rho_{\rm jam} T' = - 1/\rho_{\rm jam} T.
\)
Finally, the time interval $T'=T$ between accelerating vehicles is given by  
\(
 T'/\tau = 2 ( 1 - {e}^{-T'/\tau}) .
\)
\par
\begin{figure}[htbp]
\caption[]{Hysteresis loop of fully developed
traffic jams  (solid lines with arrows), emerging in the optimal velocity model
(\ref{BANDO}) with the equilibrium speed-distance relation $v_{e}(d) = v_0 \Theta' (d - d_0)$, see the
dashed line. (After Sugiyama and Yamada, 1997.) \label{HYST}}
\end{figure}
Summarizing the above results, one finds characteristic traffic constants, if the
jam density $\rho_{\rm jam}$ developing at the upstream end of traffic jams is
independent of the surrounding traffic. In contrast to $\rho_{\rm jam}$, 
the time interval $T'$ between accelerating 
vehicles is determined by the dynamics at the {\em downstream} jam front. It results
from the fact that the initial conditions of accelerating vehicles at jam fronts are 
more or less identical:
The cars have equal headways $d_{\rm jam} \approx 1/\rho_{\rm jam}$ to the respective leading
vehicle and start with the same velocity $v \approx 0$. The calculation of the characteristic
constants is nevertheless a difficult task. However, for different models,
analytical results have been
obtained by Kerner \etal\ (1997), Helbing and Schreckenberg (1999) as well
as Gerwinski and Krug  (1999).

\subsubsection{Kerner's hypotheses}\label{IIC6}

It appears that the above theoretical results cover many of the observations 
summarized in Sec.~\ref{IIA5}. However, Kerner (1998a, b, 1999a, b,
2000a, b, c) was surprised by the following aspects and has, therefore,
questioned previously developed traffic models:\\
(i) Synchronized flows of kind (i) and (ii) (see Sec.~\ref{IIA5b})
can exist for a long time (Kerner and Rehborn, 1996b), i.e., they can be
stable, at least with respect to fluctuations of
small amplitude (see also the discussion of stable congested
flow by Westland, 1998).\\
(ii) Wide jams and stop-and-go traffic are rarely formed spontaneously
(Kerner and Rehborn, 1997; Kerner, 1999c, 2000c), but occur reproducible
at the same freeway sections during rush hours. Instead of a transition from free traffic
to stop-and-go traffic, one normally observes two successive transitions: 
a hysteretic one from free flow
to synchronized flow (Kerner and Rehborn, 1997)
and another one from synchronized flow 
to stop-and-go waves (Kerner, 1998a; see Sec.~\ref{IIA5c}). 
\par
In conclusion, the transition to synchronized flow appears more frequently
than the transition to stop-and-go traffic. 
For an explanation, 
Kerner (1998a, b, 1999a, c, 2000a, b) has, therefore, developed some
hypotheses in the spirit of his three-phase traffic theory (see
Secs.~\ref{IIA2} and \ref{IIA5}), which I try to sketch in the following:\\ 
(i) The whole multitude of hypothetical homogeneous and stationary
(steady) states of synchronized flow is related to a two-dimensional
region in the flow-density plane. This region is the same for a multi-lane
road and for a one-lane road.\\
(ii) All hypothetical homogeneous and stationary (steady) states of free
flow and synchronized flow are stable with respect to infinitesimal 
perturbations.\\
(iii) There are two qualitatively different kinds of nucleation effects
and the related first-order phase transitions in the hypothetical
homogeneous states of traffic flow:\\
-- the nucleation effect responsible for jam formation,\\
-- the nucleation effect responsible for the phase transition from free
to synchronized flow.\\
(iv) At each given density in homogeneous states of free flow, the
critical amplitude of a local perturbation of the density and/or
velocity, which is needed for a phase transition from free flow
to synchronized flow, is considerably lower than the critical
amplitude of a local perturbation needed for jam formation in free flow.
In other words, the probability of the phase transition from free flow
to synchronized flow is considerably higher than the probability of
jam emergence in free flow.\\
(v) The jam line $J(\rho)$ (see Sec.~\ref{IIC3}) determines the threshold
of jam formation. All
(i.e. an infinite number of) states of traffic flow related to the line
$J$ are threshold states with respect to jam formation: Traffic states
below the jam line $J(\rho)$ would always be stable, while states above it
would be metastable. At the same distance above the line $J(\rho)$,
the critical amplitude of local perturbations triggering jam formation
is higher in free flow than in synchronized flow. Therefore, a
transition to a wide moving jam occurs much more frequently
from synchronized than from free flow.\\ 
(vi) The transition from free to synchronized
flow is due to an avalanche self-decrease in the mean
probability
of overtaking, since it is associated with a synchronization among
lanes.\\
(vii) The hysteretic transition to wide jams is related to an
avalanche-like growth
of a critical perturbation of the density, average velocity, or traffic flow, see 
the nucleation effect described in Sec.~\ref{IIC4}.
\par 
Simulation models and further empirical evidence for these hypotheses are still to be found.

\subsection{Transition to congested traffic at bottlenecks and ramps}\label{IID}

Effects of changes in the number of lanes
and ramps have been simulated since the early days of
applied traffic simulations (see, for example, Munjal \etal, 1971; Phillips, 1977;
Cremer, 1979; Makigami \etal, 1983), but a systematic study of the observed
phenomena, their properties, mechanisms, and preconditions has been missing.
Later, there have been various theoretically motivated studies by physicists as well.
These assume, for example,
local ``defects'',  ``impurities'', or bottlenecks with
reduced ``permeability'', which are mostly associated with a
locally decreased velocity. Typical results for one single defect are\\ 
(i) a drop of the freeway capacity to a value $Q_{\rm bot}$ at the impurity,\\
(ii) a segregation into one area of free and and one area of congested traffic,\\
(iii) a localization of the downstream front of congested traffic at the
bottleneck,\\
(iv) a shock-like propagation of the upstream congestion front in agreement
with the Lighthill-Whitham theory, particularly formula (\ref{shockprop}) with $Q_+ = Q_{\rm bot}$,\\ 
(v) a stationary length of the congested area on circular roads, since the
inflow $Q_-$ is determined by the outflow $\tilde{Q}_{\rm out}$ from the bottleneck.
\par
Investigations of this kind have been carried out for particle hopping models
like the TASEP with random sequential update (Janowky and Lebowitz, 1992, 1994)
and for traffic models with parallel update (Chung and Hui, 1994;
Csahok and Vicsek, 1994; Emmerich and Rank, 1995). One is certainly tempted to
compare the congested traffic appearing in these models with the observed ``synchronized'' 
congested flow discussed before, but several elements seem to be missing:\\
(i) the hysteretic nature of the transition, characterized by the nucleation effect, 
i.e. the requirement of an overcritical perturbation,\\
(ii) the typical dynamics leading to the formation of ``synchronized'' congested
flow (see Sec.~\ref{IIA5b}), and\\
(iii) the wide scattering of flow-density data. 
\par
In other simulations with inhomogeneities, researchers have recognized 
wave-like forms of congested traffic
behind bottlenecks
(Csahok and Vicsek, 1994; Hilliges, 1995;
Klar \etal, 1996; Nagatani, 1997c; Lee \etal, 1998).
One can even find {\em period doubling scenarios} upstream of stationary
bottlenecks (Nagatani, 1997c) or due to time-dependent local perturbations
(Koch, 1996; Nicolas \etal, 1994, 1996), which are analytically understandable. Moreover,
Kerner \etal\ (1995a) have simulated slightly inhomogeneous traffic on
a circular freeway with on- 
and off-ramps, for which in- and outflows were chosen identically. 
They found the interesting phenomenon of a localized stationary 
cluster along the on-ramp, which triggered a traffic jam, when the on-ramp flow
was reduced. However, these and other simulations with the KK model (\ref{KK})
were not fully consistent with the above
outlined phenomena, so that Kerner developed the hypotheses sketched 
in Sec.~\ref{IIC6}. I believe, the reason for the discrepancy
between observations and simulations may have been the assumptions of
periodic boundary conditions and uniform traffic, i.e. identical driver-vehicle units
(see below and Sec.~\ref{IIE3}). This is, however, still a controversial topic 
which certainly requires more detailed investigations in the future.
\par
In 1998, Lee \etal\ have tried to reproduce the hysteretic phase transition
to ``synchronized'' congested flow by simulation of a circular 
road with the KK model including on- and off-ramps, but 
with different parameters and a modified velocity-density function $V_{e}(\rho)$.
By a temporary peak in the on-ramp flow, they managed to trigger a
form of stop-and-go traffic that was propagating upstream, but
its downstream front was pinned at the location of the ramp.
They called it the ``recurring hump'' state (RH) and compared it to {\em autocatalytic
oscillators}. Free traffic would correspond to a
point attractor and the oscillating traffic state to a stable {\em limit cycle}. 
In terms of non-linear dynamics, the transition corresponds to a 
{\em Hopf bifurcation}, but a subcritical one, since the critical
ramp flow depends on the size of the perturbation. (After Helbing and Treiber, 1998b). 
\par
Lee {\em et al.} have pointed out that free traffic survives a 
pulse-type perturbation of finite amplitude, if the ramp flow is 
below a certain critical value. However, once a RH state has formed,
it is self-maintained until
the ramp flow falls below another critical value which is smaller
than the one for the transition from FT to the RH state. This proves
the hysteretic nature of the transition. Moreover, Lee {\em et al.}
have shown a gradual spatial transition from the RH state to
free flow downstream of the ramp. They have also demonstrated a
synchronization among neighboring freeway
lanes as a result of lane changes. Therefore, they have suggested 
that their model can describe the empirically observed first-order phase 
transition to synchronized flow, where the two-dimensional scattering
of ``synchronized'' congested traffic is understood as a result of 
the fact that the amplitude of the oscillating traffic state
depends on the ramp flow. The similarity with empirical data, however, was
rather loose. Also the central question regarding an explanation of
homogenous and stationary congested traffic (see Secs.~\ref{IIC6}
and \ref{IIA5b}) remained unsolved.
\par
Independently of Lee \etal, Helbing and Treiber (1998a) had submitted another 
study two weeks later. Based on the non-local, gas-kinetic-based traffic model,
they simulated a freeway section with open boundary
conditions and one on-ramp, since the transition to ``synchronized'' congested flow 
had been mostly observed close to ramps. Ramps were modelled by means of a
source term in the continuity equation:
\begin{equation}
 \frac{\partial \rho(x,t)}{\partial t} + \frac{\partial}{\partial x}[\rho(x,t)V(x,t)]
 = \frac{Q_{\rm rmp}(t)}{IL} \, .
\label{sourceterm}
\end{equation}
Herein, $Q_{\rm rmp}(t)$ represents the ramp flow per freeway lane entering 
the $I$-lane freeway over a ramp section of effective length $L$. 
The simulation scenario assumed that the sum 
\begin{equation}
 Q_{\rm down}= Q_{\rm up} + Q_{\rm rmp}(t)/I 
\end{equation}
of the flow $Q_{\rm up}$
upstream of the ramp and the ramp flow $Q_{\rm rmp}(t)/I$ per lane had reached 
the regime of metastable traffic flow 
with $Q_{e}(\rho_{\rm c1}) < Q_{\rm down} < Q_{e}(\rho_{\rm c2})$ downstream of the
ramp. Hence, without any perturbation, there was free traffic flow. However,
a short overcritical peak in the ramp flow $Q_{\rm rmp}(t)$
triggered a growing perturbation, which travelled downstream in the beginning,
but changed its propagation speed and direction as it grew larger. Consequently,
it returned to the ramp {\em (``boomerang effect'')}
and initiated a continuously growing region of
extended congested traffic when it arrived there (see Fig.~\ref{TR}). 
The reason for this is basically the 
{\em dynamical reduction of the freeway capacity} to the outflow 
$\tilde{Q}_{\rm out}$ 
from congested traffic (cf. Secs.~\ref{IIA5} and \ref{IIC5}).
This allows us to understand what Daganzo \etal\ (1999) call 
the ``activation'' of a bottleneck. Note that the outflow $\tilde{Q}_{\rm out}$ depends on the
ramp length $L$ (Helbing and Treiber, 1998a; Treiber \etal, 2000), 
but in some traffic models and in reality it is also a function of the 
ramp flow $Q_{\rm rmp}$, the average speed $V_{\rm bot}$ in the congested area, and possibly
other variables (such as the delay time by congestion, since drivers may change their behavior).
We expect the relations
$\partial \tilde{Q}_{\rm out}/\partial L \ge  0$, $\partial \tilde{Q}_{\rm out}/\partial 
Q_{\rm rmp} \le  0$,  $\partial \tilde{Q}_{\rm out}/\partial V_{\rm bot} \ge  0$, and
\begin{equation}
\lim_{Q_{\rm rmp} \rightarrow 0} \lim_{V_{\rm bot}\rightarrow 0} \lim_{L\rightarrow \infty}
\tilde{Q}_{\rm out}(Q_{\rm rmp}/I,V_{\rm bot},L) = Q_{\rm out} \, .
\end{equation}
Note that both, $\tilde{Q}_{\rm out}$ and $Q_{\rm out}$ are also functions of the
model parameters $V_0$, $T$, etc. (and of their spatial changes along the road).
\par\unitlength=0.5mm 
\begin{figure}[htbp]
\caption[]{Spatio-temporal evolution of the density 
after a small peak in the inflow from the on-ramp. The on-ramp merges with
the main road at $x=0$\,km. Traffic flows from left to right.
In (a), the parabolically shaped region of high density
corresponds to the resulting ``synchronized'' congested flow. 
Plot (b) illustrates in more detail that the congestion starts to build up 
downstream of the bottleneck (cf. Sec.~\ref{IIA5b}).
The time-dependent inflows $Q_{\rm up}$ at the upstream boundary 
and $Q_{\rm rmp}/I$ at the on-ramp are displayed in (c).
(d) The simulated traffic flows at two cross sections 
show the temporary drop below the finally resulting bottleneck flow (--~--) and discharge 
flow (---) observed in real traffic, cf. Fig.~\ref{FUNDAM}(b).
(After Helbing and Treiber, 1998a; Helbing \etal, 2001a, b.)\label{TR}} 
\end{figure}
The (congested) bottleneck flow 
\begin{equation}
Q_{\rm bot} = \tilde{Q}_{\rm out} - Q_{\rm rmp}/I 
 = \tilde{Q}_{\rm out} - \Delta Q
\label{botflow}
\end{equation}
immediately upstream of the ramp is given by the outflow $\tilde{Q}_{\rm out}$ minus the ramp flow
$Q_{\rm rmp}/I$ per lane defining the {\em bottleneck strength} $\Delta Q$. 
Let $\rho_{\rm bot}>\rho_{\rm c2}$ be the associated density
according to the congested branch of the fundamental diagram, i.e. 
$Q_{\rm bot} = Q_{e}(\rho_{\rm bot})$. Then, the congested traffic flow upstream of the
bottleneck is homogeneous, if $\rho_{\rm bot}$ falls in the stable or metastable
range, otherwise one observes different forms of congestion (see Sec.~\ref{IID1}).
Hence, the above hysteretic transition to
{\em homogeneous congested traffic} (HCT) can
explain the occurence of homogeneous and stationary (steady) 
congested traffic states. The scattering of ``synchronized''
congested flow can be accounted for by a mixture of different vehicles types such as
cars and trucks (Treiber and Helbing, 1999a, see Sec.~\ref{IIE3}). However, apart from
a heterogeneity in the time headways (Banks, 1999), there are also 
other plausible proposals for this scattering, which are more difficult to verify 
or falsify empirically:  complex dynamics with forward and backward propagating shock waves
(Kerner and Rehborn, 1996a; Kerner, 1998b, 2000a, b), changes in the behavior of ``frustrated'' drivers
(Krau{\ss}, 1998a), anticipation effects
(Wagner, 1998a; Knospe \etal, 2000a, b), non-unique equilibrium solutions
(Nelson and Sopasakis, 1998; Nelson, 2000),  
or multiple metastable oscillating states (Tomer \etal, 2000).
Non-unique solutions are also expected for the non-integer car-following model
(\ref{einfach}), where the vehicle acceleration $dv/dt$ is always zero when the relative velocity
$\Delta v$ becomes zero, no matter how small the distance is.  However, most of these
states are not stable with respect to fluctuations.
\par
Let us now explain why the downstream front of extended congested traffic
is located at the bottleneck and stationary: If it would be located downstream 
of the ramp for some reason, its downstream front would travel
upstream as in traffic jams. So the question is why it does not 
continue to move upstream, when the downstream front has reached the bottleneck.
When it would pass the on-ramp, 
it would continue to emit the flow $\tilde{Q}_{\rm out}$, leading 
to an increased flow  $(\tilde{Q}_{\rm out} + Q_{\rm rmp}/I) >
(Q_{\rm up} + Q_{\rm rmp}/I)$ at the ramp, which would
again queue up. The spatial extension of the congested area is only reduced, when the
traffic flow $Q_{\rm down}$ falls below the outflow $\tilde{Q}_{\rm out}$ of congested
traffic. This is usually the case after the rush hour.
\par
Finally, I should mention that 
congested traffic can be also triggered by a perturbation in 
the traffic flow $Q_{\rm up}(t)$ entering the upstream
freeway section. It can be even caused by a supercritical ``negative''
perturbation with a {\em reduction} of the 
vehicle density and flow as, for example, at an off-ramp (Helbing, 2001).
This is, because vehicles will accelerate into the region of reduced 
density, followed by a successive deceleration. Therefore, a negative perturbation of the 
density is rapidly transformed into the characteristic perturbation 
which triggers congested traffic (see Sec.~\ref{IIC3}).

\subsubsection{Phase diagram of traffic states at inhomogeneities}\label{IID1}

Theoretical investigations of traffic dynamics have, for a long time, mainly been carried out
for circular roads. The corresponding results were, in fact, misleading, since it is
usually not possible to enter the linear unstable
region $\varrho > \rho_{\rm c2}$ of congested traffic on a homogeneous
road. High densities were produced by means of the initial condition, but they 
cannot be reached by any natural 
inflow into a homogeneous freeway with open boundaries, no matter how high the inflow is.
Consequently, spontaneously forming traffic jams 
are rather unlikely as well, apart from
temporary ones caused by sufficiently large perturbations. However, the situation
changes drastically for inhomogeneous roads with ramps or other bottlenecks.
\par
Helbing \etal\ (1999) have  systematically explored 
the resulting traffic states at inhomogeneities triggered by a fully developed perturbation
(jam) as a function of the ramp flow $Q_{\rm rmp}/I$ per freeway lane and the
upstream traffic flow $Q_{\rm up}$.  In this way, they found a variety
of congested traffic states and their relation to each other
(see Fig.~\ref{PHASEDIAG}). Free traffic (FT)
was, of course, observed, if the total downstream traffic flow $Q_{\rm down}
= ( Q_{\rm up} + Q_{\rm rmp}/I)$ was stable with respect to large
perturbations, i.e. if 
\(
 Q_{\rm down} < Q_{e}(\rho_{\rm c1}) . 
\)
As expected, extended congested traffic states were found when 
the traffic flow $Q_{\rm down}$ exceeded the maximum 
flow $Q_{\rm max} = \max_\rho Q_{e}(\rho)$,
i.e. the theoretical freeway capacity for homogeneous traffic. However, this
classical kind and many other kinds of congestion occurred 
even for smaller traffic flows $Q_{\rm down} < Q_{\rm max}$, triggered by perturbations.
Consequently, all these kinds of congestions could be avoided by suppressing perturbations
with suitable technical measures (see Sec.~\ref{CONT}).
\par\begin{figure}[htbp]
\caption[]{Numerically determined phase diagram of the traffic states forming in the vicinity of
an on-ramp as a function of the 
inflows $Q_{\rm up}$ and $Q_{\rm rmp}/I$ per freeway lane
on the main road and the 
on-ramp. Displayed are the phases of homogeneous congested traffic (HCT),
oscillatory congested traffic (OCT),
triggered stop-and-go traffic (TSG),
moving localized clusters (MLC),
pinned localized clusters (PLC),
and free traffic (FT).
The states are triggered by a fully developed localized cluster
travelling upstream and passing the ramp (cf. 
Fig.~\protect\ref{STATES}). Dashed 
lines indicate the theoretical phase boundaries. The lowest diagonal line
$Q_{\rm down} = Q_{e}(\rho_{\rm c1})$ separates free traffic from wide
jams, while the central diagonal line $Q_{\rm down} = \tilde{Q}_{\rm out}
\approx Q_{e}(\rho_{\rm c2})$ separates localized from extended forms of congested traffic.
The localized cluster states between these two diagonals require metastable downstream traffic.
Finally, the highest (solid) diagonal line represents the condition
$Q_{\rm down} = Q_{\rm max}$ limiting
the region, in which a breakdown of traffic flow is caused by
exceeding the theoretically possible freeway capacity $Q_{\rm max}$
(upper right corner). All congested traffic states below this line are caused
by perturbations and, therefore, avoidable by technical control measures.
(After Helbing \etal, 1999, 2001a.)\label{PHASEDIAG}}
\end{figure}
The ``classical'' form of extended congested traffic is HCT
(see Fig.~\ref{STATES}a), which is formed if the density
$\rho_{\rm bot}$ in the congested region falls into the \mbox{(meta-)}stable regime. 
Decreasing the ramp flow $Q_{\rm rmp}$ will increase the flow $Q_{\rm bot}
= (\tilde{Q}_{\rm out} - Q_{\rm rmp}/I)$ in the congested region and decrease the
related density $\rho_{\rm bot}$. When it falls into the unstable
regime, i.e. when we have $\rho_{\rm bot} < \rho_{\rm c3}$ or 
\(
Q_{\rm bot} > Q_{e}(\rho_{\rm c3}) ,
\)
we can expect {\em oscillating congested traffic} (OCT, see Fig.~\ref{STATES}b). 
However, in deterministic models and
without any additional perturbations, homogeneous congested traffic extends 
up to  $Q_{\rm bot} > Q_{e}(\rho_{\rm cv})$ with a smaller 
density $\rho_{\rm cv} < \rho_{\rm c3}$. The reason is that, above a density of
$\rho_{\rm cv}$, traffic flow is {\em convectively} stable
(Helbing \etal, 1999a), i.e. perturbations are convected
away (Manneville, 1990; Cross and Hohenberg, 1993). 
While the transition from free traffic to congested traffic
is  hysteretic (see Sec.~\ref{IID}), the transition from the HCT to the OCT state is
continuous. A suitable order parameter to characterize this transition is the oscillation
amplitude. 
\par
\begin{figure}[htbp]
\caption[]{\label{STATES}
Spatio-temporal dynamics of typical representatives of congested states
which are triggered at an on-ramp by a fully developed perturbation 
travelling upstream. The middle of the on-ramp is located at $x=8.0$\ km. 
The displayed states are
(a) homogeneous congested traffic (HCT), 
(b) oscillatory congested traffic (OCT), 
(c) triggered stop-and-go traffic (TSG), 
(d) a pinned localized cluster (PLC).
(From Helbing \etal, 2001a, b; after Helbing \etal, 1999a.)} 
\end{figure}
Let us assume we further reduce the ramp flow, so that congestion is expected to
become less serious. Then, at some characteristic ramp flow $Q_{\rm rmp}^0$, 
the oscillation amplitude becomes so large
that the time-dependent flow-density curve
reaches the free branch of the fundamental diagram
$Q_{e}(\rho)$. For 
\(
Q_{\rm rmp} < Q_{\rm rmp}^0
\)
we, therefore, speak of a 
{\em triggered stop-and-go state} (TSG). A suitable order parameter would be the average
distance between locations of maximum congestion. Drivers far upstream of
a bottleneck will be very much puzzled by the alternation of jams and free traffic,
since these jams seem to appear without any plausible reason, analogous to phantom traffic jams.  
The triggering mechanism of repeated jam formation is as follows: When the upstream travelling
perturbation passes the inhomogeneity, it triggers another 
small perturbation, which travels downstream.
As described in Secs.~\ref{IIA5} and \ref{IIC3}, 
this pertubation changes its propagation speed and direction while
growing. As a consequence, the emerging jam returns like a boomerang and triggers
another small perturbation while passing the inhomogeneity. This process repeats time and
again, thereby causing the triggered stop-and-go state. 
\par
If the traffic flow $Q_{\rm down}$ downstream of the  inhomogeneity is 
not linearly unstable, the triggered small perturbation cannot grow, and we have only
a single {\em localized cluster} (LC), which normally passes the inhomogeneity and moves upstream. 
The condition for this {\em moving localized cluster} (MLC, see Fig.~\ref{STATES}c) or wide
moving jam is obviously 
\(
{Q}_{\rm down} < Q_{e}(\rho_{\rm c2}) \, . 
\)
However, if 
\(
Q_{\rm up} < Q_{e}(\rho_{\rm c1}) , 
\)
i.e. when the flow upstream of the 
bottleneck is stable, a congested state cannot survive upstream of the inhomogeneity.
In that case, we find a standing or {\em pinned localized cluster} (SLC/PLC) at the location
of the bottleneck (see Fig.~\ref{STATES}d). 
Oscillatory forms of pinned localized clusters (OPLC) are possible
as well  (Lee \etal, 1999; Treiber \etal, 2000). 
The transition between localized clusters and extended forms
of congested traffic is given by the condition 
\(
Q_{\rm down} > \tilde{Q}_{\rm out} ,
\)
requiring
that the traffic flow $Q_{\rm down}$ exceeds the outflow $\tilde{Q}_{\rm out}$ from
congested traffic. 
\par
Finally, I would like to repeat that the transition between free and congested traffic is of
first order (i.e. hysteretic), while the transitions between extended forms of 
congested traffic (HCT, OCT, and TSG) seem to be continuous. As a consequence of the former,
one can have direct transitions from free traffic to homogeneous or oscillating congested traffic
(HCT or OCT) without crossing localized cluster (LC) states or triggered stop-and-go
waves (TSG). Normally, the traffic flows $Q_{\rm up}$ and $Q_{\rm rmp}$ during the
rush hour increase so drastically that a phase point belonging to the area of
extended congested traffic (CT) is reached before free traffic flow breaks down. 
The theory predicts this to happen at weekdays characterized by high traffic flows, 
while at the same freeway section
LC or TSG states should rather appear at weekdays with less pronounced rush
hours.
\par
It should be stressed that the above analytical 
relations for the transitions between free traffic and the
five different forms of congested traffic (PLC, MLC, TSG, OCT, and HCT) are in
good agreement with numerical results (see Fig.~\ref{PHASEDIAG}). 
Moreover, according to the quantities
appearing in these analytical relations, the transitions are 
basically a consequence of the instability diagram, combined with  
the fundamental diagram and the jam line. Therefore, we expect that the above 
phase diagram is universal for all microscopic and macroscopic traffic models
having the same instability diagram. This prediction has been meanwhile
confirmed for several models.  Lee \etal\ (1999) 
have found analogous transitions for
the KK model (\ref{KK}), but recognized a region of tristability, 
where it is a matter of history whether one ends up with free traffic, a pinned localized
cluster, or oscillating congested traffic. 
This tristability is probably a result of a dynamical variation of
$\tilde{Q}_{\rm out}$, as pointed out by 
Treiber \etal\  (2000), who have also reproduced the phase diagram for the 
IDM (see Sec.~\ref{IIB1c}). Moreover, they
have presented empirical data confirming the existence of all five 
congested traffic states. These have been successfully reproduced in simulation scenarios
with the IDM, using the measured boundary conditions. Only one single parameter, the safe time
clearance $T$, has been varied to calibrate both, the capacity of the different freeway
sections and their bottlenecks. It has also been demonstrated that different kinds of bottlenecks
have qualitatively the same effects. This requires a generalized definition
of bottleneck strengths $\Delta Q$ according to 
\begin{equation}
  \Delta Q = Q_{\rm rmp}/I + \tilde{Q}_{\rm out} - \tilde{Q}'_{\rm out} \, ,
\end{equation}
where $\tilde{Q}_{\rm out}$ denotes the outflow from congested 
traffic at the freeway section with the
largest capacity and $\tilde{Q}'_{\rm out}$ the outflow at the respective bottleneck. Spatial
changes in the number $I(x)$ of 
lanes can be reflected by virtual ramp flows (Shvetsov and Helbing, 1999):
\begin{equation}
 \frac{Q_{\rm rmp}}{IL} = - \frac{\rho V}{I} \frac{\partial I(x)}{\partial x} \, .
\end{equation}
This follows from the continuity equation $\partial (I\rho)/\partial t
+ \partial (I\rho V)/\partial x = 0$. 
\par
Recently, Lee \etal\ (2000) have presented an empirical phase diagram, but the
corresponding freeway stretch contains {\em several} capacity changes, which complicates
matters. Such freeway sections show often congestion with a stationary (pinned)
upstream end. It can also happen that a jam travels through a short section of
stable traffic, because there is a finite {\em penetration depth} (Treiber \etal, 2000).
Finally, I would like to mention a number of other simulation studies with open boundaries,
which have recently been carried out (Bengrine \etal, 1999; 
Cheybani \etal, 2000a, b; Mitarai and Nakanishi, 1999, 2000;
Popkov and Sch\"utz, 2000; Popkov \etal, 2000).
\par
Summarizing the results, a
comparison between numerical simulations and empirical
data suggest the following identifications:
Homogeneous congested traffic (HCT) seems to be the same as ``sychronized'' traffic flow (ST) of
type (i) (see Sec.~\ref{IIA5b}), while  oscillating congested traffic (OCT) seems to be related to
synchronized flow of type (iii). Homogeneous-in-speed states, which are also called
synchronized flow of type (ii), are found where congested
traffic relaxes towards free traffic (see Figs.~\ref{SCATTER}b and \ref{SPEED}; 
Tilch and Helbing, 2000, Fig.~3, right). Moreover,
the term ``recurring hump state'' (RH state, see Lee \etal, 1998, 1999)
summarizes all oscillating forms of congested traffic (OCT, TSG, and OPLC), while 
``non-homogeneous congested traffic'' was used for all forms of unstable traffic 
(Tomer \etal, 2000). 
\par
Let us now ask, what the above theory predicts for models with another
instability diagram. If the model is stable in the whole density range such as
the Burgers equation or the TASEP, we just expect two states: free traffic
(FT) and homogeneous congested traffic (HCT), which is in agreement with numerical
(Janowsky and Lebowitz, 1992, 1994) and exact analytical results (Sch\"utz, 1993).
If the model displays stable traffic at small densities and unstable
traffic at high densities, one expects free traffic (FT) and 
oscillating traffic (OCT or TSG). This was, for example, observed by Emmerich and Rank (1995)
for the Nagel-Schreckenberg model (see also Diedrich \etal, 2000; Cheybani \etal, 2000a). 
For the case of stable traffic at small and high densities, but unstable traffic at
medium densities, one should find free traffic (FT), oscillating traffic states
(OCT or TSG), and homogeneous congested traffic (HCT). This could possibly be tested with
the Weidlich-Hilliges model. The existence of moving and pinned localized
clusters (MLC/PLC) would normally require a metastable state between free traffic and
linearly unstable traffic. Finally, assume there were models with stable traffic 
at low densities and metastable traffic above a certain critical density $\rho_{\rm c1}$,
but no range of linearly unstable traffic, as suggested by Kerner
(see Sec.~\ref{IIC6}). Then,
oscillatory congested traffic (OCT) and triggered stop-and-go waves (TSG) should not
exist.
Identifying localized cluster states (MLC/PLC) with wide moving jams,
but homogeneous congested traffic (HCT) with synchronized flow
would give a similar picture to the three-phase traffic theory proposed by
Kerner and Rehborn (see Sec.~\ref{IIA2}): The possible traffic states
would be free traffic, wide jams, and synchronized flow. 
Transitions to synchronized flow would
occur, when the traffic flow $Q_{\rm down}$ would exceed the maximum capacity
$Q_{\rm max}$. This would tend to happen at bottlenecks, where
the metastable density regime is entered first. The transition
would cause a queueing of vehicles and a capacity drop due to
the increased time headways of accelerating vehicles at downstream congestion fronts.
In contrast, transitions to wide moving jams would always have to be triggered by supercritical
perturbations. This would most frequently happen in congested areas
upstream of bottlenecks (``pinch effect'', see Sec.~\ref{IIA5c}), as
metastable traffic is rather unlikely on homogeneous stretches of a
freeway. Supplementary, the scattering of flow-density
data in ``synchronized''  congested traffic could 
be explained by some source of fluctuations, e.g. heterogeneous driver-vehicle
behavior (see Sec.~\ref{IIE3}).
\par
The above considerations demonstrate how powerful the  method of the phase 
diagram is. Based on a few characteristic quantities (the critical flows), one can 
predict the dynamical states of the respective model as well as the phase boundaries. 
Moreover, one can classify traffic models
into a few universality classes. It does not matter whether the model is of microscopic or
macroscopic, of deterministic or of probabilistic nature.

\subsubsection{Coexistence of states and ``pinch effect''} \label{IID2}

In simulations with the IDM, Treiber and Helbing (1999b) have
simulated a scenario which reminds of the pinch effect reported by Kerner (1998a) 
and similar observations by Koshi \etal\ (1983).
This scenario is characterized by the spatial coexistence of different forms of congested
traffic, which can occur upstream of bottlenecks even on rampless road sections,
as the transitions between the concerned 
congested states are {\em continuous} (see Sec.~\ref{IID1}).
Starting with homogeneous congested traffic (HCT) at a bottleneck,
Treiber and Helbing found oscillating congested traffic (OCT) upstream of it, but stop-and-go
traffic even further upstream (see also Helbing \etal, 2001b). 
For comparison, upstream of a bottleneck
Kerner has observed transitions from synchronized flow to a ``pinch region'' 
characterized by narrow clusters, which eventually merged to form wide moving jams
(see Fig.~\ref{COEXIST} and
Sec.~\ref{IIA5c}). Personally, I valuate this phenomenon as strong evidence for the existence of
linearly unstable traffic.
\par\begin{figure}[htbp]
\caption[]{Illustration of the spatial coexistence of HCT, OCT and TSG states in a simulation
with the IDM for an inhomogeneous freeway without ramps. Traffic flows in positive $x$-direction.
The spatio-temporal density plot shows 
the breakdown to homogeneous congested traffic near the inhomogeneity and stop-and-go waves
emanating from this region. 
The inhomogeneity corresponds to an increased safe time clearance $T'> T$
between $x=0$ km and $x=0.3$ km, reflecting more careful driving. 
Downstream of the inhomogeneity, vehicles accelerate into free traffic.
(From Treiber and Helbing, 1999b; Helbing \etal, 2001b.)\label{COEXIST}}
\end{figure}
According to the model by Treiber and Helbing (1999b), the conditions for this 
spatial coexistence of congested traffic states are the following. 
The density $\rho_{\rm bot}$ in the congested region immediately 
upstream of the bottleneck should be in the linearly unstable, but convectively stable
range $[\rho_{\rm cv},\rho_{\rm c3}]$, where perturbations are convected away in 
upstream direction (Manneville, 1990; Cross and Hohenberg, 1993).
In this case, traffic flow will appear stationary and homogeneous close to the bottleneck,
but small perturbations will grow as they propagate upstream in the congested
region starting at the bottleneck. If the perturbations propagate faster than
the congested region expands, they will reach the area of free traffic upstream of the
bottleneck. During rush hours, it is quite likely that this free flow is in the metastable
range between $\rho_{\rm c1}$ and $\rho_{\rm c2}$. Consequently, sufficiently large perturbations
can trigger the formation of jams, which continue travelling upstream, while
small perturbations are absorbed. Note that this phenomenon should
be widespread, since the range $[\rho_{\rm cv},\rho_{\rm c3}]$ of convectively stable traffic 
can be quite large, so it is likely to appear when the freeway flow $Q_{\rm up}$ 
exceeds the critical flow $Q_{e}(\rho_{\rm c1})$.

\subsection{Heterogeneous traffic}\label{IIE}

In reality, driver-vehicle units are not identical, but behave differently, which may be
reflected by individual parameter sets. This is easy to do in microscopic traffic models,
while gas-kinetic and macroscopic models require generalizations. Such generalized
models have been developed for a continuous distribution of desired velocities 
(Paveri-Fontana, 1975; Helbing, 1995c, d, 1996a, 1997a; Wagner \etal,
1996; Wagner, 1997a, b, c). However, if several driver-vehicle parameters are varied,
it is more practical to distinguish several vehicle classes only, 
such as aggressive drivers (``rabbits'') and timid ones (``slugs'') (Daganzo, 1995c, 1999a)
or cars and  trucks (long vehicles; see, for example, 
Helbing, 1996d, 1997a, d; Hoogendoorn and Bovy, 1998a, b, 2000b; Hoogendoorn, 1999;
Shvetsov and Helbing, 1999).
New phenomena arising from the heterogeneity of driver-vehicle units are
platoon formation and scattering of flow-density data, as is outlined in the following.

\subsubsection{Power-law platoon formation and quenched disorder}\label{IIE1}

The simplest models for heterogeneity are particle hopping models with {\em quenched
disorder}. For example, Evans (1996), Krug and Ferrari (1996), 
Karimipour (1999a, b), as well as Sepp\"al\"ainen and Krug (1999)
study a simplified version of a model
by Benjamini \etal\ (1996). It corresponds to the one-dimensional driven
lattice gas known as TASEP, but with particle-specific, constant jump rates $q_\alpha$.
Since overtaking was not allowed, 
Krug and Ferrari have found a sharp phase transition between a low-density regime, where
all particles are queueing behind the slowest particle, and a high-density regime, where
the particles are equally distributed. While, at low densities,
the slow particles ``feel free traffic'' until the critical density is reached
(cf. the truck curve in Figs.~\ref{suppl}b, c, \ref{NAT}a), 
the  growth of particle clusters (``platoons'') is characterized by a power-law coarsening.
If particles move ballistically with individual velocities $v_\alpha$ and
form a platoon when a faster particle reaches a slower one,
the platoon size $n_{\rm pl}(t)$ grows according to
\begin{equation}
 n_{\rm pl}(t) \sim t^{(m'+1)/(m'+2)} \, ,
\end{equation} 
where the exponent $m'$ characterizes the distribution $P_0(v)
\sim (v - v_{\rm min})^{m'}$ of free velocities
in the neighborhood of the minimal desired velocity $v_{\rm min}$
(Ben-Naim \etal, 1994; Ben-Naim and Krapivsky, 1997, 1998, 1999;
Nagatani, 1996c; cf. also Nagatani, 1995a; Nagatani \etal, 1998; for spatial inhomogeneities
see the results by Krug, 2000). 
Above the critical density, the differences among fast and slow particles become irrelevant, 
because there is so little space that {\em all} particles have to move slower than preferred. 
The formation of platoons has been compared
with {\em Bose-Einstein condensation}, where the steady-state velocity of the particles is
analogous to the fugacity of the ideal Bose gas (Evans, 1996; Krug and Ferrari, 1996;
Chowdhury \etal, 2000b).
\par
Platoon formation and power-law coarsening has also been found in microscopic models
with parallel update (Ben-Naim \etal, 1994; Nagatani, 1996c; Ben-Naim and Krapivsky, 1997,
1998, 1999; Nagatani \etal, 1998; see also Fukui and Ishibashi, 1996a; Nagatani, 2000).
An example is the
Nagel-Schreckenberg model with quenched disorder, i.e. vehicle-specific slow-down
probabilities $p_\alpha$ (Ktitarev \etal, 1997; Knospe \etal, 1999).
\par
In real traffic, platoons remain limited in size (see Sec.~\ref{IIA4}). This is probably because of
occasional possibilities for lane changes on multi-lane roads. A model for 
platoon size distributions has been developed by Islam and Consul (1991).

\subsubsection{Scattering}\label{IIE3}

In order to model the observed scattering of congested flow-density data,
Helbing \etal \ have simulated traffic with different driver-vehicle units.
In contrast to the models discussed in Sec.~\ref{IIE1}, they used traffic models
with a linearly unstable density regime, namely the GKT 
model (Treiber and Helbing, 1999a),
the discrete optimal velocity model (Helbing and Schreckenberg, 1999), 
and the IDM (Treiber \etal, 2000).
Since they could determine the time-dependent variation of the percentage of
long vehicles in their empirical data, they distinguished two vehicle classes,
cars and trucks (long vehicles). These were characterized by two different sets of parameters, i.e.
two different fundamental diagrams. Consequently, the effective fundamental diagram
was a function of the measured proportion of long vehicles. Performing simulations
with the empirically obtained boundary conditions and truck proportions (see Fig.~\ref{TRUCKFRAC}), 
they managed to reproduce the
observed transition from free to ``synchronized'' congested flow semi-quantitatively,
see Fig.~\ref{SCATTER} (Treiber and Helbing, 1999a). 
\par\begin{figure}[htbp]
\caption{\label{SCATTER}
The displayed points in flow-density space correspond
to empirical 1-minute data 
(dark crosses) and related simulation results (grey boxes)
at two different cross sections of the freeway. At the section upstream
of a bottleneck (left), we find  ``synchronized'' congested 
flow, while the close downstream section (right) displays homogeneous-in-speed states.
The simulations manage to reproduce both, the quasi-linear 
flow-density relation at small densities and the scattering over 
a two-dimensional region in the congested regime.
For comparison, we have displayed the equilibrium flow-density relations
for traffic consisting of 100\% cars (---),
and  100\% trucks (-~-~-). The respective model parameters were weighted with the
measured, time-dependent truck fraction, see Fig.~\ref{TRUCKFRAC}.  
(After Treiber and Helbing, 1999a; Helbing \etal, 2001a, b.)}
\end{figure}
In particular, the spatial dependence of the flow-density
data on the respective freeway cross section was well reproduced, including the
relaxation to free traffic downstream of the bottleneck, which leads to
homogeneous-in-speed states, i.e. synchronized flow of type (ii).
Moreover, the flow-density data had a linear structure in the regime of free traffic,
while they were widely scattered in the congested regime. The most important
requirement for this was a considerable difference in the safe time clearance $T_\alpha$
of cars and trucks (cf. Fig.~\ref{PARAMETERS2}). This is certainly in agreement with facts, and it
is also compatible with the explanation of randomly sloped flow-density changes
suggested by Banks (see Sec.~\ref{IIA2}). If this interpretation of the scattering of flow-density
data is correct, one should observe a smaller variation of flow-density data on days when 
truck traffic is reduced or prohibited. However, some scattering is always expected due
to the wide distribution of time headways (see Figs.~\ref{TGAPS} and \ref{PARAMETERS2}).

\subsection{Multi-lane traffic and synchronization} \label{IIF}

\subsubsection{Gas-kinetic and macroscopic models}\label{IIFa}

Traffic models have been also developed for multi-lane 
traffic (Gazis \etal, 1962; Munjal \etal, 1971;
Munjal and Pipes, 1971; Makigami \etal, 1983; 
Holland and Woods, 1997; Daganzo, 1997b; Daganzo \etal, 1997). 
In gas-kinetic and macroscopic traffic models, this is reflected by additional
lane-changing terms (R{\o}rbech, 1976; Michalopoulos \etal, 1984;
Helbing, 1996d, 1997a, d; Helbing and Greiner, 1997;
Hoogendoorn and Bovy, 1998a, b; Hoogendoorn, 1999;
Shvetsov and Helbing, 1999; Klar and Wegener, 1999a, b).
Representing the vehicle density in lane $i$ by
$\rho_i(x,t)$ and the lane-specific average velocity by $V_i(x,t)$, the lane-specific 
density equations of the form (\ref{sourceterm}) are complemented by the additional contributions
\begin{equation}
 + \frac{\rho_{i-1}}{\tau_{i-1}^+} - \frac{\rho_i}{\tau_i^+}
 + \frac{\rho_{i+1}}{\tau_{i+1}^-} - \frac{\rho_i}{\tau_i^-} \, .
\end{equation}
Herein, $1/\tau_i^+$ is the density-dependent lane-changing rate from lane $i$ to 
lane $(i+1)$, while $1/\tau_i^-$ describes analogous changes to lane $(i-1)$.
Empirical  measurements of lane-specific data and
lane-changing rates are rare (Sparman, 1978; Hall and Lam, 1988;
Chang and Kao, 1991; McDonald \etal, 1994;
Brackstone and McDonald, 1996; Brackstone \etal, 1998).
Based on his data, Sparman (1978) proposed the relation 
\(
  1/\tau_i^{\pm} = c_i^{\pm} \rho_i (\rho_{\rm max} - \rho_{i\pm 1})
\)
with suitable lane-dependent constants $c_i^\pm$. 
\par
The additional terms in the velocity equations include
\begin{equation}
  + \frac{\rho_{i-1}}{\tau_{i-1}^+} (V_{i-1} - V_i) + \frac{\rho_{i+1}}{\tau_{i+1}^-}
  (V_{i+1} - V_i) \, ,
\end{equation}
which lead to a {\em velocity adaptation} in neighboring lanes. This is responsible for the
{\em synchronization} among lanes in congested traffic (Lee \etal, 1998) and allows
the treatment of multi-lane traffic by effective one-lane models (Shvetsov and Helbing, 1999).
Similar adaptation terms are obtained for the variances. More detailed calculations
yield further terms reflecting that overtaking vehicles transfer their
higher speed to the neighboring lane (Helbing, 1997a). 
\par
The lane-specific traffic equations allow us to derive simplified
equations for the whole freeway cross section. It turns out that, compared to the
effective one-lane model used before, one has 
to correct the expression for the traffic pressure at small densities,
where the average velocities in the neighboring lanes are not synchronized
(Helbing, 1997a, d):
\begin{equation}
 P(x,t) \approx \rho(x,t)[ \theta(x,t) +  \langle (V_i - V)^2 \rangle ] \, .
\label{presscorr}
\end{equation}
The density-dependent contribution $ \langle (V_i - V)^2 \rangle$
reflects the difference in the lane-specific
speeds $V_i$, which also contributes to the overall velocity 
variance of vehicles (see Fig.~\ref{suppl}a). In contrast,
the interaction frequency among vehicles, 
which enters the braking interaction term in (\ref{eqVderiv})
and $V^{e}$, depends on the {\em lane}-specific variance $\theta$. 

\subsubsection{Microscopic models and cellular automata}\label{IIFb}

A microscopic modelling of lane changes 
(Fox and Lehmann, 1967; Levin, 1976; Daganzo, 1981; Mahmassani and Sheffi, 1981;
Ahmed \etal, 1996)
turns out to be an intricate problem for several reasons:\\
(i) Because of legal regulations, lane-changing is not symmetric, at least
in Europe.\\
(ii) Drivers accept smaller gaps than normal in cases of lane closures, i.e.
they behave differently.\\
(iii) The minimum gaps required for lane-changing depend on the own velocity and the 
speed differences to the neighboring vehicles in the adjacent lane.\\ 
(iv) Without assuming fluctuations or heterogeneous vehicles, it is hard to reproduce
the high rates of lane changes, which drivers can
reach by means of tricky strategies.\\
As a consequence, classical gap acceptance models 
do not work very well (Gipps, 1986; Yang and Koutsopoulos, 1996).
Benz (1993)  and Helbing (1997a) have, therefore, suggested a
flexible lane-changing criterion. 
Lane-changing rules for cellular automata have been developed
as well (Nagatani, 1993, 1994a, b, 1996a; Zhang and Hu, 1995; Rickert \etal, 1996; 
Chowdhury \etal, 1997a; Wagner \etal, 1997; Awazu, 1998; Nagel \etal, 1998;
Brilon and Wu, 1999). 
\par
In modern lane-changing models, the lane is changed if both, an {\em incentive
criterion} and a {\em safety criterion} are fulfilled (Nagel \etal, 1998). The incentive
criterion checks whether the considered driver $\alpha$ could go faster on
(one of) the neighboring lane(s). The lane, for which the absolute value
$|f_{\alpha\beta}|$ of the repulsive interaction force with the respective
leading vehicle $\beta$ is smallest, is the preferred one. It is, however,
reasonable to introduce a certain threshold value for lane changing.
The safety criterion is fulfilled, if the anticipated repulsive
interaction forces to neighboring vehicles on the adjacent lane
stay below a certain critical value which depends on the friendliness
of the driver $\alpha$. The IDM and other models considering velocity differences
take into account that the gap required for lane changing must be
greater if the new following vehicle approaches fast, i.e. if 
its relative velocity with respect to the lane-changing vehicle is
large. 
\par
To obtain reasonable lane changing rates, one has to either work with a probabilistic
model or distinguish different vehicle types like cars and trucks. In a small density
range around 25 vehicles per kilometer and lane, simulations by Helbing and Huberman (1998) 
indicated a transition to a {\em coherent movement} of cars and trucks, which they could confirm
empirically (cf. Figs.~\ref{suppl}b, c and \ref{NAT}a). 
In this state, fast and slow vehicles drive with the same speed, i.e. they are
moving like a solid block. This is caused by a breakdown of the lane-changing rate
(see Fig.~\ref{NAT}b),
when the freeway becomes crowded, so that drivers do not anymore find sufficiently large
gaps for overtaking (see Fig.~\ref{NAT}c).
However, the coherent state is destroyed when traffic flow
becomes unstable. Then, vehicle gaps are again widely scattered (cf. Fig.~\ref{NAT}d), 
and lane changing is sometimes possible. 
\par\unitlength10mm
\begin{figure}[htbp]
\caption[]{The transition to coherent vehicle traffic, see (a), is related to the
breakdown of the lane-changing rate at densities above 20 vehicles per kilometer, see (b).
This is related to the rapid decay of the proportion of successful
lane changes (i.e. the quotient between actual and desired lane changes), see (c). 
(d) Opportunities for lane-changing are rapidly diminished when gaps corresponding to 
about twice the safe headway required for lane-changing cease to
exist, i.e. when the quotient of the standard deviation of vehicle headways to
their mean value is small. (After Helbing and Huberman, 1998; see also Helbing, 2001).
The breakdown of the lane-changing rate seems to imply a decoupling of the
lanes, i.e. an effective one-lane behavior. However, this 
is already the {\em result} of a self-organization process based on
two-lane interactions, since
any significant perturbation of the solid-block state (like different 
velocities in the neighbouring lanes) will cause frequent
lane changes (Shvetsov and Helbing, 1999). By filling large gaps, the gap distribution is considerably
modified. This eventually reduces possibilities for lane changes, 
so that the solid-block state is restored. 
\label{NAT}}
\end{figure}
Properties of multi-lane traffic have recently found a considerable interest, as is
documented by further studies (Chowdhury \etal, 1997b;
Belitsky \etal, 2001; see also Goldbach \etal, 2000;
Lubashevsky and Mahnke, 2000).

\subsection{Bi-directional and city traffic} \label{IIG}

Traffic models have also been developed for bi-directional traffic
(Dutkiewicz \etal, 1995; Lee \etal, 1997;
Simon and Gutowitz, 1998; Fouladvand and Lee, 1999) and city traffic 
(see, for example, TRANSYT by Robertson, 1969a, b; Schmidt-Schmiedebach, 1973; 
Herman and Prigogine, 1979; SCOOT by Hunt \etal, 1982;
Cremer and Ludwig, 1986; Williams \etal, 1987;
Mahmassani \etal, 1990;
MAKSIMOS by Putensen, 1994; MIXSIM by Hoque and McDonald, 1995;
PADSIM by Peytchev and Bargiela, 1995). 
Interested readers should consult
the reviews by Papageorgiou (1999),  Chowdhury \etal\ (2000b),
and the corresponding chapters in
{\em Traffic and Granular Flow '97,'99} (Schreckenberg and Wolf, 1998;
Helbing \etal, 2000c).  Here, I will be very short.
\par
Traffic dynamics in cities is quite different from freeways, since it is to a large extent
determined by the intersections.
The first model for city traffic in the physics literature, the ``BML model'', goes back to
Biham \etal\ (1992). It is based on a two-dimensional square lattice, 
in which each site represents a crossing of north-bound and east-bound traffic.
The sites can be either empty, or occupied by a north-bound or east-bound
vehicle. East-bound vehicles are synchronously updated every odd time step, 
while north-bound vehicles are updated every even time step, reflecting 
synchronized and periodic traffic signals at the crossings. During the parallel updates,
a vehicle moves ahead by one site, if this is empty, otherwise it has to wait.
Biham \etal\ (1992) observed a first-order transition from free traffic
to jammed traffic with zero velocity throughout the system, 
which appears at a finite density due to a gridlock of north-bound
and east-bound traffic along diagonal lines (see also Fukui and Ishibashi, 1996b). 
This first-order transition is also 
characteristic for most other models of city traffic. For a detailled discussion of the
related literature see the review by Chowdhury \etal\ (2000b).

\subsection{Effects beyond physics} \label{IIH}

There are some effects which are not at all or only partly represented
by the traffic models described in Sec.~\ref{IIB}:\\
(i) the anticipation behavior of drivers (Ozaki, 1993; Wagner, 1998a; 
Lenz \etal, 1999; Knospe \etal, 2000a, b),\\
(ii) the reaction to winkers,\\
(iii) the avoidance of driving side by side,\\
(iv) the velocity adaptation to surrounding traffic (Herman \etal, 1973),\\
(v) the tolerance of small time clearances for a limited time (Treiterer and Myers, 1974),\\
(vi) the reduction of gaps to avoid vehicles cutting in (Daganzo, 1999a),\\
(vii) the avoidance of the truck lane, even if it is underutilized 
(Brackstone and McDonald, 1996; Helbing, 1997a, b; ``Los Gatos effect'': Daganzo, 1997c, 1999a; 
Helbing and Huberman, 1998),\\
(viii) motivations for lane changing (Redelmeier and Tibshirani, 1999),\\
(ix) effects of attention (Michaels and Cozan, 1962; Todosiev, 1963;
Todosiev and Barbosa, 1963/64; Michaels, 1965),\\
(x) changes of driver strategies (Migowsky \etal, 1994),\\
(xi) effects of road conditions and visibility, etc.
\par
Certainly, such kinds of effects can be also modelled in a mathematical way,
and this has partly been done in 
so-called sub-microscopic models of vehicle dynamics (Wiedemann, 1974)
and vehicle simulators such as VISSIM (Fellendorf, 1996) or PELOPS
(Ludmann \etal, 1997).
However, most of the above effects are hard to test empirically and difficult to verify or falsify.
Therefore, we are leaving the realm of physics, here, and eventually entering the realm
of driver psychology (Daganzo, 1999a).

\subsection{Traffic control and optimization} \label{CONT}\label{IIi0}

Traffic control and optimization is an old and large field
(see, e.g., Cremer, 1978; Cremer and May, 1985;
Smulders, 1989; Dougherty, 1995; K\"uhne, 1993; 
K\"uhne \etal, 1995, 1997; Taale and Middelham, 1995; Schweitzer \etal, 1998a), 
which can be certainly not discussed in detail, here. 
Overviews of optimization approaches by physicists can be found in the
Proceedings of the Workshops on {\em Traffic and Granular Flow}
(Wolf \etal, 1996; Schreckenberg and Wolf, 1998; 
Helbing \etal, 2000c; see also Claus \etal, 1999), while the
readers interested in the engineering literature are referred to the 
books and review articles by Pagageorgiou (1995, 1999) as well as 
Lapierre and Steierwald (1987).
These also discuss the available software packages for the simulation and optimization of
street networks and/or signalling of traffic lights such as 
TRANSYT (Robertson, 1969a, b), SCOOT (Hunt \etal, 1982),
OPAC (Gartner, 1983), PRODYN (Farges \etal, 1984), 
CRONOS (Boillot \etal, 1992), and many others. I should also mention the 
cellular-automata-based projects OLSIM (Kaumann \etal, 2000; 
see {\tt http://www.traffic.uni-duisburg.de/OLSIM/}) and
TRANSIMS (Nagel and Barrett, 1997; Rickert and Nagel, 1997; Simon and Nagel, 1998; see 
{\tt http://www-transims. tsasa.lanl. gov}).
A complementary list of microsimulation tools can be found at {\tt 
http:// www.its.leeds.ac.uk/smartest/links.html}.
\par
While optimization of city traffic is based on a synchronization
of traffic lights, most optimization approaches for freeways are
based on a homogenization of vehicle traffic, as stop-and-go traffic, jams, and
congested traffic are associated with reduced efficiency due to the capacity drop
(see, e.g. Huberman and Helbing, 1999).
Remember that this capacity drop is often triggered by perturbations in the
flow (cf. Sec.~\ref{IID}), which should, therefore, be suppressed by 
technical measures such as intelligent speed limits, adaptive on-ramp controls,
traffic-dependent re-routing, and driver assistance systems (see, e.g., Treiber and Helbing, 2001).

\section{Pedestrian traffic}\label{III}

Pedestrian crowds have been empirically
studied for more than four decades now
(Hankin and Wright, 1958; Oeding, 1963; Hoel, 1968; Older, 1968;
Navin and Wheeler, 1969;  Carstens and Ring, 1970; 
O'Flaherty and Parkinson, 1972; Weidmann, 1993).
The evaluation methods applied were based on direct observation,
photographs, and time-lapse films. Apart from behavioral
investigations (Hill, 1984; Batty, 1997),
the main goal of these studies was to develop 
a {\em level-of-service concept}
(Fruin, 1971; Polus \etal, 1983; M\={o}ri, and Tsukaguchi, 1987),
{\em design elements} of pedestrian facilities
(Schubert, 1967; Boeminghaus, 1982; Pauls, 1984; Whyte, 1988; Helbing, 1997a;
Helbing \etal, 2001c), or {\em planning guidelines} 
(Kirsch, 1964;
Predtetschenski and Milinski, 1971;
Transportation Research Board, 1985; 
Davis and Braaksma, 1988;
Brilon \etal, 1993). The latter have usually
the form of {\em regression relations}, 
which are, however, not very
well suited for the prediction of pedestrian flows in pedestrian zones
and buildings with an exceptional architecture, or in extreme conditions
such as evacuation (see Sec.~\ref{IIIC1}). Therefore, a number of
simulation models have been proposed, e.g. {\em queueing models} 
(Yuhaski and Macgregor Smith, 1989; Roy, 1992; L{\o}v{\aa}s, 1994),
{\em transition matrix models} (Garbrecht, 1973),
and {\em stochastic models} (Mayne, 1954; Ashford et al., 1976),
which are partly related to each other. In addition, there are
models for the {\em route choice behavior} of pedestrians
(Borgers and Timmermans, 1986a, 1986b; Timmermans et al., 1992; Helbing, 1992a).
\par
None of these concepts adequately takes into account the
self-organization effects occuring in pedestrian crowds. 
These may, however, lead to unexpected obstructions due to 
mutual disturbances of pedestrian flows. 
More promising with regard to this is the approach by Henderson. 
He conjectured that pedestrian crowds behave 
similar to gases or fluids (Henderson, 1971, 1974;  Henderson and
Lyons, 1972; Henderson and Jenkins, 1973; see also Hughes, 2000, 2001). 
This could be partly confirmed (see Sec.~\ref{IIIA}). 
However, a realistic gas-kinetic or fluid-dynamic
theory for pedestrians must contain corrections due to their 
particular interactions
(i.e. avoidance and deceleration maneuvers) which, of course, do not
conserve momentum and energy. Although such a theory can be actually
formulated (Helbing, 1992a, b; Hoogendoorn and Bovy, 2000a), for practical applications 
a direct simulation of {\em individual} pedestrian motion is favourable,
since this is more flexible.  As a consequence, current research focusses on the
{\em microsimulation} of pedestrian crowds, which also allows one to consider 
incoordination by {\em excluded volume effects} 
related to the discrete, ``granular'' structure of pedestrian flows. In this connection, 
a {\em behavioral force model} of individual
pedestrian dynamics has been developed
(Helbing, 1991, 1996d, 1997a, 1998c; Helbing \etal, 1994; 
Helbing and Moln\'{a}r, 1995, 1997; Moln\'{a}r, 1996a, b;
Helbing and Vicsek, 1999; Helbing \etal, 2000a, b) (see Secs.~\ref{ID1} and \ref{IIIB}).
A discrete and simple forerunner of this model was proposed by {Gipps} and 
{Marksj\"o} (1985). I also like to mention recent
{\em cellular automata} of pedestrian dynamics (Bolay, 1998; Blue and Adler, 1998, 1999;
Fukui and Ishibashi, 1999a, b; Muramatsu \etal, 1999; Muramatsu and
Nagatani, 2000a, b;  Kl\"upfel \etal, 2000; Burstedde \etal, 2001), 
{\em emergency and evacuation models} (Drager \etal, 1992; 
Ebihara \etal, 1992; Ketchell \etal, 1993;
Okazaki and Matsushita, 1993; Still, 1993, 2000; Thomson and Marchant, 1993; 
L{\o}v{\aa}s, 1998; Hamacher and Tjandra, 2000; Kl\"upfel \etal, 2000), and
{\em AI-based models} (Gopal and Smith, 1990; Reynolds, 1994, 1999; Schelhorn 
\etal 1999). 

\subsection{Observations} \label{Obs}\label{IIIA}

Together with my collaborators
I have investigated pedestrian motion for several years and evaluated 
a number of video films. 
Despite the sometimes more or less
``chaotic'' appearance of individual pedestrian behavior, one can find
regularities, some of which become best visible in 
time-lapse films like the ones produced by Arns (1993). While
describing these, I also summarize results of other pedestrian
studies and observations (after Helbing, 1997a, 1998c; Helbing \etal, 2001c):\\ 
(i) Pedestrians feel a strong aversion to taking detours or moving
opposite to their desired {\em walking direction}, even if the direct way is crowded. However, 
there is also some evidence that pedestrians normally choose the {\em fastest} route to their
next destination, but not the {\em shortest one} (Ganem, 1998). In general, pedestrians
take into account detours as well as the comfort of walking, thereby minimizing
the effort to reach their destination (Helbing \etal, 1997a, b). 
Their ways can be approximated by polygons.\\ 
(ii) Pedestrians prefer to walk with an individual desired speed, which
corresponds to the most comfortable (i.e. least energy-consuming)
{\em walking speed} (see Weidmann, 1993) as long as it is not necessary to
go faster in order to reach the 
destination in time. The desired speeds within
pedestrian crowds are Gaussian distributed with a mean value of approximately 
1.34\,m/s and a standard deviation of about 0.26\,m/s (Henderson, 1971).
However, the average speed depends on the situation (Predtetschenski and Milinski, 1971),
sex and age, the time of the day, the purpose of the trip, the surrounding,
etc. (see Weidmann, 1993).\\
(iii) Pedestrians keep a certain {\em distance} to other pedestrians and borders
(of streets, walls, and obstacles; see Transportation Research Board, 1985; 
Brilon \etal, 1993). 
This distance is smaller the more a
pedestrian is in a hurry, and it decreases with growing pedestrian density. 
Resting individuals (waiting on a railway platform for a train, sitting in 
a dining hall, or lying at a beach) are uniformly distributed
over the available area if there are no 
acquaintances among the individuals. Pedestrian density increases (i.e. 
interpersonal distances lessen) around particularly attractive places.
It decreases with growing velocity variance (e.g. on a dance floor, see
Helbing, 1992b, 1997a). Individuals knowing each other may form {\em groups} which are
entities behaving similar to single pedestrians. Group
sizes are {Poisson} distributed (Coleman, 1964; Coleman and James, 1961; Goodman, 1964).\\
(iv) In situations of {\em escape panic} (Helbing \etal, 2000b), 
individuals are often nervous, and sometimes they develop blind actionism. Moreover
people move or try to move considerably faster than normal
(Predtetschenski and Milinski, 1971).
Individuals start pushing, and interactions among people 
become physical in nature.
Moving and, in particular, passing of a bottleneck 
becomes incoordinated (Mintz, 1951). 
At exits, arching and clogging are observed (Predtetschenski and Milinski, 1971),
and jams are building up (Mintz, 1951).
The physical interactions in the jammed crowd 
add up and cause dangerous pressures up to
4,500 Newtons per meter (Elliott and Smith, 1993; Smith and Dickie, 1993),
which can bend steel barriers or tear down brick walls.
Escape is further slowed down by fallen or injured people turning
into ``obstacles''.
Finally, people tend to show herding behavior, i.e., 
to do what other people do (Quarantelli, 1957; Keating, 1982). 
As a consequence, alternative exits are often overlooked or not efficiently used
in escape situations (Keating, 1982; Elliott and Smith, 1993).\\
(v) At medium and high pedestrian densities,
the motion of pedestrian crowds shows some striking {\em analogies
with the motion of gases, fluids, and granular flows}. For example, I found that
footprints of pedestrians in snow look similar to {\em streamlines} of fluids
(Helbing, 1992a). 
At borderlines between opposite directions of walking one can observe
{\em ``viscous fingering''} (Kadanoff, 1985; Stanley and Ostrowsky, 1986). 
The emergence of pedestrian streams through 
standing crowds (Arns, 1993; Helbing 1997a, Fig.~2.6; 1998c, 
Fig.~1; Helbing \etal, 2001c, Fig.~1) appear analogous to the
{\em formation of river beds} 
(St{\o}lum, 1996; Rodr\'{\i}guez-Iturbe and Rinaldo, 1997; Caldarelli, 2000).
Moreover, similar to {\em segregation or stratification phenomena} in granular media
(Santra \etal, 1996; Makse \etal, 1997),
pedestrians spontaneously organize in lanes of uniform walking direction,
if the pedestrian density is high enough 
(Oeding, 1963; Helbing, 1991, 1997a, Fig.~2.5; 1998c, Fig.~2; Helbing \etal, 2001c, Fig.~2).
At bottlenecks (e.g. corridors, staircases, or doors), 
the passing direction of pedestrians oscillates 
(Helbing \etal, 1994; Helbing and Moln\'{a}r, 1995). 
This may be compared to the {\em ``saline oscillator''} (Yoshikawa \etal, 1991) or the
granular {\em ``ticking hour glass''} (Wu \etal, 1993; Pennec \etal, 1996).
Finally, one can find the propagation of {\em shock waves} 
in dense pedestrian crowds pushing forward (see also Virkler and Elayadath, 1994).
The {\em arching and clogging} in panicking crowds (Helbing \etal, 2000b) 
is similar to the outflow of rough granular media through small openings 
(Ristow and Herrmann, 1994; Wolf and Grassberger, 1997).

\subsection{Behavioral force model of pedestrian dynamics} \label{Model} \label{IIIB}

For reliable
simulations of pedestrian crowds we do not
need to know whether a certain pedestrian, say, turns to the right at 
the next intersection. It is sufficient to have a good estimate 
what percentage of pedestrians turns to the right. This can be
either empirically measured or calculated by means of route choice 
model like the one by Borgers and Timmermans (1986a, b). 
In some sense, the uncertainty about the individual behaviors is
averaged out at the macroscopic level of description, as in fluid
dynamics. 
\par
Nevertheless, I will now focus on a more flexible microscopic
simulation approach based on the behavioral force concept outlined
in Sec.~\ref{ID1}. Particular advantages of this approach are the
consideration of
(i) excluded volume effects due to the granular structure of a pedestrian crowd and
(ii) the flexible usage of space by pedestrians, requiring a (quasi-)continous
treatment of motion.
It turns out that these points are essential to reproduce the above mentioned
phenomena in a natural way.
\par
As outlined in Sec.~\ref{ID1}, we describe the different motivations of and influences on
a pedestrian $\alpha$ by separate force terms. First of all, the decisions and
intentions where to go and how fast are reflected by the driving term
$v_\alpha^0(t)\vec{e}_\alpha^0(t)/\tau_\alpha$, 
in which $v_\alpha^0(t)$ is the desired velocity and $\vec{e}_\alpha^0(t)$
the desired direction of motion. The interactions
among pedestrians are described by forces 
\begin{equation}
 \vec{f}_{\alpha\beta}(t) =\vec{f}_{\alpha\beta}^{\rm soc}(t) 
+  \vec{f}_{\alpha\beta}^{\rm ph}(t) \, ,
\end{equation}
which consist of
a {\em socio-psychological contribution} $\vec{f}_{\alpha\beta}^{\rm soc}(t)$ and
physical interactions $\vec{f}_{\alpha\beta}^{\rm ph}(t)$, if pedestrians come too close
to each other. The socio-psychological component reflects the tendency of pedestrians
to keep a certain distance to other pedestrians and may be described by a force of the
form
\begin{eqnarray}
  \vec{f}_{\alpha\beta}^{\rm soc}(t) &=& 
 A_\alpha\exp[(r_{\alpha\beta}-d_{\alpha\beta})/B_\alpha] \vec{n}_{\alpha\beta}
\nonumber \\
&\times & 
\left(\lambda_\alpha + (1-\lambda_\alpha)\frac{1+\cos(\varphi_{\alpha\beta})}{2}\right) \, .
\label{psychol}
\end{eqnarray}
Herein, $A_\alpha$ denotes the interactions strength and $B_\alpha$ the range of the
repulsive interaction, which are individual parameters and partly dependent on
cultural conventions. $d_{\alpha\beta}(t) = \|\vec{x}_\alpha(t) - \vec{x}_\beta(t)\|$ is the
distance between the centers of mass of pedestrians $\alpha$ and $\beta$, 
$r_{\alpha\beta} = (r_\alpha
+ r_\beta)$ the sum of their radii $r_\alpha$ and $r_\beta$, and 
\[
\vec{n}_{\alpha\beta}(t)= \bbox(n_{\alpha\beta}^1(t),n_{\alpha\beta}^2(t)\bbox)
= \frac{\vec{x}_\alpha(t)-\vec{x}_\beta(t)}{d_{\alpha\beta}(t)}
\]
the normalized vector pointing
from pedestrian $\beta$ to $\alpha$. Finally, with the choice $\lambda_\alpha < 1$,
we can take into account the anisotropic character of pedestrian interactions, as
the situation in front of a pedestrian has a larger impact on his or her behavior
than things happening behind. 
The angle $\varphi_{\alpha\beta}(t)$ denotes the angle between the 
direction $\vec{e}_\alpha(t) = \vec{v}_\alpha(t)/\|\vec{v}_\alpha(t)\|$ of
motion and the
direction $-\vec{n}_{\alpha\beta}(t)$ of the object exerting the repulsive force,
i.e. $\cos \varphi_{\alpha\beta}(t) = - \vec{n}_{\alpha\beta}(t)
\cdot \vec{e}_\alpha(t)$. One may, of course, take into account
other details such as a velocity-dependence of the forces and non-circular shaped
pedestrian bodies, but this does not have {\em qualitative} effects on the dynamical
phenomena resulting in the simulations. In fact,  most observed self-organization
phenomena are quite insensitive to the specification of the interaction forces,
as different studies have shown (Helbing and Moln\'{a}r, 1995; Moln\'{a}r, 1996a;
Helbing and Vicsek, 1999; Helbing \etal, 2000a, b). 
\par
The {\em physical interaction} $\vec{f}_{\alpha\beta}^{\rm ph}$
plays a role only when pedestrians have physical contact 
with each other, i.e. if  $r_{\alpha\beta} \ge d_{\alpha\beta}$.
In this case, we assume two additional forces
inspired by granular interactions (Ristow and Herrmann, 1994; Wolf and Grassberger, 1997):
First, a {\em ``body force''} $k(r_{\alpha\beta}-d_{\alpha\beta})
\, \vec{n}_{\alpha\beta}$ counteracting body compression
and a {\em ``sliding friction force''} $\kappa (r_{\alpha\beta} - d_{\alpha\beta})
\, \Delta v_{\beta\alpha}^t \,
\vec{t}_{\alpha\beta}$ impeding {\em relative} tangential motion:
\begin{eqnarray}
\vec{f}_{\alpha\beta}^{\rm ph}(t) &=& k \Theta(r_{\alpha\beta}-d_{\alpha\beta}) 
\vec{n}_{\alpha\beta} \nonumber \\
&+& \kappa \Theta(r_{\alpha\beta}-d_{\alpha\beta}) 
\Delta v_{\beta\alpha}^t \, \vec{t}_{\alpha\beta} \, ,
\label{FORMEL}
\end{eqnarray}
where the function $\Theta(z)$ is equal to its argument $z$, if $z> 0$, otherwise 0. 
Moreover, $\vec{t}_{\alpha\beta} = (-n_{\alpha\beta}^2, n_{\alpha\beta}^1)$ means the tangential
direction and $\Delta v_{\beta\alpha}^t = (\vec{v}_\beta -\vec{v}_\alpha) 
\cdot \vec{t}_{\alpha\beta}$
the tangential velocity difference, while $k$ and $\kappa$ represent large
constants. 
Strictly speaking, friction effects already set in before pedestrians
touch each other, because of the socio-psychological tendency not to pass other individuals
with a high relative velocity, when the distance is small. This is, however, not important
for the effects we are going to discuss later on.
\par
The {\em interaction with the boundaries} of walls and other obstacles 
is treated analogously to pedestrian interactions, i.e., if
$d_{\alpha b}(t)$ means the distance to boundary $b$,
$\vec{n}_{\alpha b}(t)$ denotes the direction perpendicular to it, and
$\vec{t}_{\alpha b}(t)$ the direction tangential to it,
the corresponding interaction force with the boundary reads
\begin{eqnarray}
 \vec{f}_{\alpha b} &=& \left\{ A_\alpha \exp[(r_{\alpha}-d_{\alpha b})/B_\alpha]
 + k \Theta(r_\alpha-d_{\alpha b}) \right\} \vec{n}_{\alpha b} \nonumber \\
 &-& \kappa \Theta(r_\alpha-d_{\alpha b}) (\vec{v}_\alpha\cdot\vec{t}_{\alpha b}) 
 \, \vec{t}_{\alpha b} \, .
\end{eqnarray}
\par
Moreover, we may also take into account time-dependent
attractive interactions towards window displays,
sights, or special attractions $i$ by social forces of the type
(\ref{psychol}). However, in comparison with repulsive interactions,
the corresponding interaction range $B_{\alpha i}$ is usually
larger and  the strength parameter $A_{\alpha i}(t)$ typically smaller, negative, and
time-dependent. Additionally, the joining behavior (Dewdney, 1987)
of families, friends, or tourist 
groups can be reflected by forces of the type $ \vec{f}_{\alpha \beta}^{\rm att}(t) = 
 - C_{\alpha\beta}  \vec{n}_{\alpha\beta}(t)$, which guarantee that acquainted individuals
join again, after they have accidentally been separated by other pedestrians.
Finally, we can take into account
unsystematic variations of individual behavior
and voluntary deviations from the assumed behavioral laws by a fluctuation
term $\vec{\xi}_\alpha(t)$. Therefore, the resulting pedestrian model corresponds to
equation (\ref{three}), but the sum of interaction forces $\sum_{\beta(\ne \alpha)}
\vec{f}_{\alpha\beta}(t)$ is replaced by the sum of all the above mentioned
attractive and repulsive interaction and boundary forces. 
The resulting expression looks rather complicated, 
but in the following, we will drop attraction effects
and assume $\lambda_\alpha= 0$ for simplicity, so that the interaction forces become
isotropic and conform with Newton's third law.  Moreover, the physical interactions
are mostly relevant in panic situations, so that we normally just have repulsive social 
and boundary interactions:
\(
 \sum_{\beta (\ne \alpha)} \vec{f}_{\alpha\beta}^{\rm soc}(t)
 + \sum_{b} \vec{f}_{\alpha b}(t) .
\)

\subsection{Self-organization phenomena and noise-induced
ordering}\label{SELF}\label{IIIC}

Despite its simplifications, the behavioral force model of pedestrian dynamics
describes a lot of observed phenomena quite realistically. Especially, it allows one to
explain various self-organized spatio-temporal patterns that are not externally
planned, prescribed, or organized, e.g. by traffic signs, laws, or behavioral
conventions. Instead, the spatio-temporal patterns discussed below
emerge due to the non-linear
interactions of pedestrians even without assuming strategical
considerations, communication, or imitative behavior of pedestrians.
All these collective patterns of motion are {\em symmetry-breaking phenomena}, 
although the model was formulated 
completely symmetric with respect to the right-hand
and the left-hand side (Helbing and Moln\'{a}r, 1997; Helbing, 1997a, 1998c;
Helbing \etal, 2001c).

\subsubsection{Segregation}\label{IIICa}

The microsimulations reproduce the
empirically observed {\em formation of lanes} consisting of pedestrians with the
same desired walking direction (Helbing \etal, 1994; Helbing and
Moln\'{a}r, 1995, 1997; Helbing, 1996d, 1997a; Moln\'{a}r, 1996a, b;
Helbing and Vicsek, 1999; Helbing \etal, 2001c; see Fig.~\ref{Ped3}). 
For open boundary conditions, these lanes are dynamically varying. Their number
depends on the width of the street (Helbing and Moln\'{a}r, 1995; Helbing, 1997a), 
on pedestrian density, and on the noise level. Interestingly, one finds
a {\em noise-induced ordering} (Helbing and Vicsek, 1999; Helbing and Platkowski, 2000):
Compared to small noise amplitudes, medium ones result in a more pronounced segregation
(i.e., a smaller number of lanes), while large noise amplitudes lead to
a ``freezing by heating'' effect (see Sec.~\ref{IIIC1a}).
\par\begin{figure}[htbp]
\caption[]{Formation of lanes in initially disordered pedestrian crowds 
with opposite walking directions (after Helbing \etal, 
2000a; Helbing, 2001; cf. also Helbing {\em et al.}, 1994;
Helbing and Moln\'{a}r, 1995). White disks represent pedestrians moving
from left to right, black ones move the other way round. Lane formation does not
require the periodic boundary conditions applied above, see the Java applet
{\tt http://www.helbing.org/Pedestrians/Corridor.html}.
\label{Ped3} }
\end{figure}
The conventional interpretation of lane formation is: Pedestrians
tend to walk on the side prescribed in vehicular traffic. However,
the above model can explain lane formation even without assuming a preference
for {\em any} side (Helbing and Vicsek, 1999; Helbing \etal,
2000a). The most relevant point is the higher relative velocity of
pedestrians walking in opposite directions. As a consequence, they
have more frequent interactions until they have segregated into
separate lanes. The resulting collective pattern of motion 
minimizes the frequency and strength of avoidance maneuvers,
if fluctuations are weak. 
Assuming identical desired velocities $v_\alpha^0 = v_0$,
the most stable configuration corresponds to a state with a
{\em minimization of interactions}
\[
 - \frac{1}{N} \sum_{\alpha \ne \beta} \tau \vec{f}_{\alpha\beta} \cdot \vec{e}_\alpha^0
 \approx  \frac{1}{N} \sum_\alpha (v_0 -  \vec{v}_\alpha\cdot\vec{e}_\alpha^0 )
 = v_0 ( 1 - E) \, .
\]
It is related with a maximum {\em efficiency}
\begin{equation}
 E = \frac{1}{N} \sum_{\alpha} 
 \frac{\vec{v}_\alpha \cdot \vec{e}_\alpha}
  {v_0} \label{EFF}
\end{equation}
{\em of motion} (Helbing and Vicsek, 1999). The efficiency $E$ with $0 \le E \le 1$ 
(where $N = \sum_\alpha 1$ is the respective number of pedestrians $\alpha$)
describes the average fraction of the desired speed $v_0$ with which 
pedestrians actually approach their destinations. That is, lane
formation 
``globally'' maximizes the average velocity into the respectively desired
direction of motion, although the model does not even assume that
pedestrians would try to optimize their behavior {\em locally}. This is
a consequence of the symmetrical interactions among pedestrians with
opposite walking directions. One can even show that a large class 
of driven many-particle systems, if they self-organize at all,
tend to globally optimize their state (Helbing and Vicsek, 1999).
This {\em optimal self-organization} was also observed in the formation of
coherent motion in a system of
cars and trucks, which was related with a minimization of the
lane changing rate (see Sec.~\ref{IIFb}). Another example is the
emergence of optimal way systems by bundling of trails (see
Fig.~\ref{tr}).

\subsubsection{Oscillations}\label{IIICb}

In simulations of bottlenecks like doors, 
oscillatory changes of the passing direction are observed, if people do not panic, see Fig.~\ref{Ped4} 
(Helbing \etal, 1994; Helbing and
Moln\'{a}r, 1995, 1997; Helbing, 1996d, 1997a; Moln\'{a}r, 1996a, b; Helbing \etal, 2001c). 
\par\begin{figure}[htbp]
\caption[]{Oscillations of the passing direction at a
bottleneck (after Helbing, 2001; cf. also Helbing \etal, 1994;
Helbing and Moln\'{a}r, 1995). Dynamic simulations are available at {\tt
http://www.helbing.org/Pedestrians/Door.html}. \label{Ped4} }
\end{figure}
The mechanism leading to alternating flows is the following: Once a
pedestrian is able to pass the narrowing, pedestrians
with the same walking direction can easily follow.
Hence, the number and ``pressure''
of waiting and pushing pedestrians becomes less than
on the other side of the narrowing where, consequently, 
the chance to occupy the passage
grows. This leads to a deadlock situation which is followed
by a change in the passing direction. Capturing the bottleneck is easier if it
is broad and short so that the passing direction changes more
frequently. Therefore, two separate doors close to the walls are more efficient than one single
door with double width. By self-organization, 
each door is, then, used by one walking direction for a long time, see Fig.~\ref{F3}
(Moln\'{a}r, 1996a, b; Helbing and Moln\'{a}r, 1997; Helbing, 1997a, 1998c; Helbing \etal, 2001c).
The reason is that pedestrians, who pass a door, clear the
way for their successors, similar to lane formation.
\begin{figure}[htbp]
\caption[]{\label{F3}If two alternative passageways are available, pedestrians
with opposite walking direction use different doors
as a result of self-organization (after Moln\'{a}r, 1996a, b; Helbing and Moln\'{a}r, 1997;
Helbing, 1997a, 1998c; Helbing \etal, 2001c).}
\end{figure}

\subsubsection{Rotation}\label{IIICc}

At intersections one is confronted with
various alternating collective patterns of motion which are very 
short-lived and unstable. For example, phases during which
the intersection is crossed in ``vertical'' or ``horizontal'' direction
alternate with phases of temporary
roundabout traffic, see Fig.~\ref{Ped5} (Helbing \etal, 1994;
Helbing, 1996d, 1997a;  Moln\'{a}r, 1996a, b;
Helbing and Moln\'{a}r, 1997; Helbing \etal, 2001c). This self-organized
round-about traffic is similar to the emergent rotation 
found for self-driven particles (Duparcmeur \etal, 1995).
It is connected with small detours, but decreases the
frequency of necessary deceleration, stopping, and avoidance maneuvers
considerably, so that pedestrian motion becomes more efficient on average.
The efficiency of pedestrian flow can be considerably increased by 
putting an obstacle in the center of the intersection, since this
favours the smooth roundabout traffic compared with the competing, inefficient
patterns of motion.
\par
\begin{figure}[htbp]
\caption[]{Self-organized, short-lived roundabout traffic in intersecting pedestrian
streams (from Helbing, 1996d, 1997a;  
Helbing and Moln\'{a}r, 1997; Helbing \etal, 2001c; 
see also Helbing \etal, 1994; Moln\'{a}r, 1996a, b).\label{Ped5}}
\end{figure}

\subsection{Collective phenomena in panic situations} \label{IIIC1}

The behavior of pedestrians in panic situations (e.g. in certain cases of emergency
evacuation) displays some characteristic features:\\
(i) People are getting nervous, resulting in a higher level of fluctuations.\\
(ii) They are trying to escape from the cause of panic, which can be reflected 
by a significantly higher desired velocity.\\
(iii) Individuals in complex situations, who do not know what is the right thing to do, 
orient at the actions of their neighbours, i.e. they tend to do what other people do.
We will describe this by an additional herding interaction, but attractive interactions
have probably a similar effect.
\par
We will now discuss the fundamental collective 
effects which fluctuations, increased desired velocities, and herding behavior can have.
In contrast to other approaches, 
we do not assume or imply that individuals in
panic or emergency situations would behave relentless and asocial, although they sometimes do.

\subsubsection{``Freezing by heating''}\label{IIIC1a}

The effect of getting nervous has been investigated by Helbing \etal\ 
(2000a). Let us assume 
that the individual level of fluctuations is given by
\begin{equation}
  \eta_\alpha = (1 - n'_\alpha) \eta_0 + n'_\alpha \eta_{\rm max} \, ,
\end{equation}
where $n'_\alpha$ with $0 \le n'_\alpha \le 1$ measures the nervousness of pedestrian
$\alpha$. The parameter $\eta_0$ means 
the normal and $\eta_{\rm max}$ the maximum fluctuation strength.
It turned out that, at sufficiently high pedestrian densities, 
lanes were destroyed by increasing the fluctuation strength (which is
analogous to the temperature). However, instead of the expected transition from the ``fluid'' 
lane state to a disordered, ``gaseous''
state, Helbing \etal\   
found the formation of a solid state. It was characterized by
a blocked situation with a regular (i.e. ``crystallized'' 
or ``frozen'') structure so that
this paradoxial transition was called {\em ``freezing
by heating''} (see Fig.~\ref{FREEZ}). Notably enough, the blocked state had
a {\em higher} degree of order, although the internal
energy was {\em increased} and the resulting state was {\em metastable}
with respect to structural perturbations such as the exchange of oppositely moving particles. 
Therefore, ``freezing by heating'' is just opposite to what one would
expect for equilibrium systems, and different from fluctuation-driven orderding
phenomena in metallic glasses and some
granular systems (Gallas \etal, 1992;
Umbanhowar \etal, 1996;
Rosato \etal, 1987), where fluctuations lead from a disordered
{\em metastable} to an ordered {\em stable} state. A model 
for related {\em noise-induced ordering} processes has been developed by
Helbing and Platkowski (2000). 
\par
The precondition for the unusual freezing-by-heating
transition are the additional driving term and the dissipative friction. Inhomogeneities in the 
channel diameter or other impurities which temporarily slow down pedestrians
can further this transition at the respective places. Finally note that a transition from
fluid to blocked pedestrian counter flows is also observed, when a critical particle density
is exceeded (Muramatsu \etal, 1999; Helbing \etal, 2000a).\\
\begin{figure}[htbp]
\caption[]{Noise-induced formation of a crystallized, ``frozen'' state in a periodic
corridor used by oppositely moving pedestrians 
(after Helbing \etal, 2000a; Helbing, 2001).\label{FREEZ} }
\end{figure}

\subsubsection{``Faster-is-slower effect''}\label{IIIC1b}

The effect of increasing the desired velocity was studied by
Helbing \etal\ (2000b). They found that
the simulated outflow from a room is well-coordinated and regular, if the
desired velocities $v_\alpha^0=v_0$ are normal. However, for desired
velocities above 1.5~m/s, i.e. for people in a rush (as in many panic situations), 
one observes an irregular
succession of arch-like blockings of the exit and avalanche-like
bunches of leaving pedestrians, when the arches break up (see Fig.~\ref{fig2}). 
This phenomenon is compatible with the empirical
observations mentioned above and comparable to intermittent clogging
found in granular flows through funnels or hoppers 
(Ristow and Herrmann, 1994; Wolf and Grassberger, 1997). 
\par\begin{figure}[htbp]
\caption[]{Panicking pedestrians often come so close to each other, that their
physical contacts lead to the build up of pressure and obstructing friction effects.
This results in temporary arching and clogging related with
inefficient and irregular outflows. (After Helbing \etal, 
2000b; Helbing, 2001.)\label{fig2} }
\end{figure}
As clogging is connected with delays,
trying to move faster (i.e., increasing $v_\alpha^0$) can cause a smaller
average speed of leaving, if the friction
parameter $\kappa$ is large (Helbing \etal, 2000b).
This {\em ``faster-is-slower effect''} is particularly tragic in the
presence of fires, where fleeing people sometimes reduce their own chances
of survival. As a consequence, models for everyday
pedestrian streams are not very suitable for realistic simulations of
emergency situations, which require at least modified
parameter sets corresponding to less efficient pedestrian behavior.
Related fatalities can be estimated by the number
of pedestrians reached by the fire front (see {\tt
http://angel.elte.hu/$\tilde{\hphantom{n}}$panic/}).
\par
Since the interpersonal friction was assumed to have, on average, no deceleration effect
in the crowd, if the boundaries are sufficiently remote,
the arching underlying the clogging effect requires a {\em combination} of
several effects:
First, slowing down due to a bottleneck such as a door, and second,
strong inter-personal friction, which becomes dominant when pedestrians get
too close to each other. It is, however, noteworthy that the faster-is-slower effect also occurs
when the sliding friction force changes continuously with the distance rather than being 
``switched on'' at a certain distance $r_\beta$ as in the model above.
\par
The danger of clogging can
be minimized by avoiding bottlenecks in the construction of
stadia and public buildings. Notice, however, that jamming can also
occur at widenings of escape routes! This comes from disturbances due to pedestrians, 
who expand in the wide area because of their repulsive
interactions or try to overtake each other.
These squeeze into the main stream again at the end of the 
widening, which acts like a bottleneck and leads to jamming. Significantly 
improved outflows in panic situations
can be reached by {\em columns} placed asymmetrically in front of the exits,
which reduce the pressure at the door and, thereby, also reduce
injuries (see {\tt http://angel.elte.hu/$\tilde{\hphantom{n}}$panic/}).

\subsubsection{``Phantom panics''}\label{IIIC1c}

Sometimes, panics have occured {\em without} any comprehensible reasons such as
a fire or another threatening event. 
Due to the ``faster-is-slower effect'',
panics can be triggered by small pedestrian counterflows (Elliott and Smith, 1993), 
which cause delays to the crowd
intending to leave. Consequently, stopped pedestrians in the back, who do not see the reason
for the temporary slowdown, are
getting impatient and pushy. In accordance with observations
(Helbing, 1991, 1997a), one may describe this by increasing the desired velocity, for example,
by the formula
\begin{equation}
 v_\alpha^0(t) = [1-n'_\alpha(t)]v_\alpha^0(0) + n'_\alpha(t) v_\alpha^{\rm max}  \, .
\end{equation}
Herein, $v_\alpha^{\rm max}$ is
the maximum desired velocity and $v_\alpha^0(0)$ the initial one, corresponding to the
expected velocity of leaving. The time-dependent parameter
\begin{equation}
 n'_\alpha(t) = 1 - \frac{\overline{v}_\alpha(t)}{v_\alpha^{\rm 0}(0)} 
\end{equation}
reflects the {\em nervousness},  where $\overline{v}_\alpha(t)$
denotes the average speed into the desired direction of motion. 
Altogether, long waiting times increase the desired velocity,
which can produce inefficient outflow. This further increases
the waiting times, and so on, so that this tragic feedback can
eventually trigger so high pressures that people are crushed 
or falling and trampled. It is, therefore, imperative, to have sufficiently
wide exits and to prevent counterflows, when big crowds want to 
leave (Helbing \etal, 2000b).

\subsubsection{Herding behavior}\label{IIIC1d}

Let us now discuss a situation in which pedestrians are
trying to leave a room with heavy smoke, but first have to find one of the
hidden exits (see Fig.~\ref{fig4}). Each pedestrian $\alpha$ may
either select an individual direction $\vec{e}_\alpha^*$ or 
follow the average direction $\langle \vec{e}_\beta(t) \rangle_\alpha$
of his neighbours $\beta$ in a certain radius $R_\alpha$ (Vicsek \etal, 1995), or
try a mixture of
both. We assume that both options are again weighted with 
the parameter $n'_\alpha$ of nervousness:
\begin{equation}
 \vec{e}_\alpha^0(t)
= {\rm Norm}\left[ \left(1- n'_\alpha \right) \vec{e}_\alpha^*
+ n'_\alpha  \, \langle \vec{e}_\beta(t) \rangle_\alpha \right] \, ,
\end{equation}
where ${\rm Norm}(\vec{z}) = \vec{z} / \|\vec{z}\|$
denotes normalization of a vector
$\vec{z}$ to unit length. As a consequence, we have individualistic behavior
if $n'_\alpha$ is low, but herding behavior if $n'_\alpha$ is high. Therefore,
$n'_\alpha$ reflects the degree of panics of individual~$\alpha$.
\par\begin{figure}[htbp]
\caption[]{Herding behavior of panicking pedestrians in a smoky room (black), 
leading to an inefficient use of
available escape routes (from Helbing \etal, 2000b).\label{fig4} }
\end{figure}
Our model suggests that neither
individualistic nor herding behavior performs well.
Pure individualistic behavior means that a few
pedestrian will find an exit accidentally, while the others will not leave the
room in time, before being poisoned by smoke. Herding may, in some cases, guide the
mass to do the right things, but this requires that there is someone who knows a 
suitable solution to the problem, e.g. trained personnel. However, in new and
complex crises situations, herding behavior normally
implies that the complete crowd is eventually moving into
the same and probably wrong or jammed direction, 
so that available exits are not efficiently used (see Fig.~\ref{fig4}),
in agreement with observations. According to simulation results by 
Helbing \etal\ (2000b), optimal chances of survival are expected for a certain
mixture of individualistic and herding behavior, where 
individualism allows some people to detect the exits and herding guarantees
that successful solutions are imitated by small groups of pedestrians. 

\section{Flocking and spin systems}\label{IV}\label{IIIC2}

Models for herding behavior have been developed already before in order to understand the
{\em flocking of birds} (Reynolds, 1987), the formation of fish schools, or the herding of sheep based on
a non-equilibrium analogue of ferromagnetic models (Vicsek \etal, 1995). More or less related
models of {\em swarm formation} are by Stevens (1992), Miramontes \etal\ (1993),
Hemmingsson (1995), Rauch \etal\ (1995),
Toner and Tu (1995), Albano (1996), 
Bussemaker \etal\ (1997),  Schweitzer (1997b),
Stevens and Schweitzer (1997), and Mikhailov and Zanette (1999) 
(see also Sec.~\ref{ID4}). 
\par
The possibly simplest model for the
behavior of birds looks as follows. Let us assume all birds $\alpha$ are moving with
a certain, finite speed $\|\vec{v}_\alpha(t)\| = v_\alpha^0 = v_0$. To make swarm formation
less trivial, we set  $\vec{f}_{\alpha\beta} = \vec{0}$ and 
do without the assumption of long-range attractive
and short-range repulsive interactions among birds.
Equation (\ref{four}) can, then, be simplified to 
\(
  \vec{e}_\alpha(t) = [\vec{e}_\alpha^0(t) + \vec{\chi}_\alpha(t)] ,
\)
which is solved together with
\begin{equation}
  \frac{d\vec{x}_\alpha}{dt} \approx \frac{[\vec{x}_\alpha(t+\Delta t) - \vec{x}_\alpha(t)]}{\Delta t}
 = v_0 \vec{e}_\alpha(t) \, .
\label{anteil}
\end{equation}
\par
We are now assuming that a bird $\alpha$ orients itself at the mean direction of
motion $\mbox{Norm}\bbox(\langle \vec{e}_\beta(t-\Delta t)\rangle_\alpha\bbox)
= \mbox{Norm}\bbox(\langle \vec{v}_\beta(t-\Delta t)\rangle_\alpha\bbox)$ of those birds $\beta$ 
which are in a neighborhood of radius $R_\alpha$ at time $(t-\Delta t)$. As a consequence,
the direction of motion $\vec{e}_\alpha^0$ is adapted to the average one in the immediate
environment with a reaction time of $\Delta t$:
\(
  \vec{e}_\alpha^0(t) = {\rm Norm}\bbox( \langle \vec{e}_\beta(t-\Delta t) \rangle_\alpha \bbox) \, .
\)
In order not to have velocity changes, we assume that the fluctuations $\vec{\chi}_\alpha(t)$
influence only the {\em directions}  $\vec{e}_\alpha(t)$, 
but stronger, the greater the {\em fluctuation level} $\eta$ is. The
vector $[\vec{e}_\alpha^0(t) + \vec{\chi}_\alpha(t)]$ is, therefore, also normalized.
Consequently, the above {\em self-propelled particle (SPP) model} 
can be summarized by Eq.~(\ref{anteil}) and 
\begin{equation}
  \vec{e}_\alpha(t+\Delta t) = {\rm Norm}\Big( {\rm Norm} 
\langle \vec{e}_\beta(t) \rangle_\alpha + \vec{\chi}'_\alpha(t) \Big) \, ,
\label{vicsek}
\end{equation}
where the fluctuations $\vec{\chi}'_\alpha(t)$ are uniformly distributed in a sphere of radius $\eta$.
\begin{figure}[htbp] 
\caption[]{Typical simulation results for the two-dimensional, self-propelled particle
model given by equations (\ref{anteil}) and (\ref{vicsek}). (a) Ferromagnetic
ordering for $v_0 = 0$. (b) For $v_0 > 0$ one observes a flocking of particles moving into
the same direction. (After Vicsek \etal, 1995.)\label{VT2}}
\end{figure}
\par
The computer simulations start with a uniform distribution of particles and mostly
use periodic boundary conditions. Figure~\ref{VT2} shows two representative snapshots of
the two-dimensional model variant for $v_0 = 0$ and $v_0 \ne 0$. An  exploration
of the model behavior in three dimensions yields a phase diagram with an ordered
phase and a disordered one (see Fig.~\ref{VT3}). For $v_0 \ne 0$, no matter how small
the average particle density $\varrho$ is, there is always a {\em critical
fluctuation strength} $\eta_{c}(\varrho)$, below which the directions $\vec{e}_\alpha(t)$
of motion are aligned and particle clusters are formed,
even without attractive interactions (Czir\'{o}k \etal, 1999b). 
In other words, birds are flocking, 
if the fluctuation strength is only small enough. Otherwise, they are flying around
in a disordered way. At a given fluctuation strength $\eta$, 
flocking is more likely, if their density $\varrho$ is high, 
which is in agreement with observations. Moreover,
the model reproduces that birds are keeping a certain ($R_\alpha$-dependent) distance
to each other, although no repulsive interactions have been assumed.
\par\begin{figure}[htbp]
\caption[]{
Phase diagram of the three-dimensional self-propelled particle (SPP) model
and the corresponding ferromagnetic system.  
The nonequilibrium SPP system becomes ordered in the whole region below the curve $\eta_c(\varrho)$
connecting the diamonds. In the static equilibrium case with $v_0 = 0$, the ordered region
extends only up to a finite ``percolation'' density, see the beginning of the area given by the
curve connecting the plus signs. (After Czir\'{o}k \etal, 1999b.)
}
\label{VT3}
\end{figure}
While the critical fluctuation strength $\eta_{c}(\varrho)$ separating ordered and disordered
behavior seems to be independent of the 
concrete velocity $v_0 >0$, the behavior of the related
equilibrium system with $v_0 = 0$ is considerably different. It corresponds to the behavior
of a {\em Heisenberg spin system} with homogeneous particle density $\varrho$, in which the
spins $\vec{e}_\alpha$ align parallel as in a ferromagnet, if the temperature (which is 
proportional to the fluctuation strength $\eta$) is below a critical temperature. However,
this {\em ferromagnetic ordering} only takes place if the particle density $\varrho$ exceeds 
a certain, so-called {\em percolation value} $\varrho_{c}\approx 0.75$ 
(see Fig.~\ref{VT3}). Surprisingly enough, the
behavior for $v_0= 0$ and $v_0 \ne 0$ is, therefore, fundamentally different. 
\par
In conclusion, non-equilibrium systems can considerably differ from equilibrium ones.
The motion of the particles in  a self-driven non-equlibrium system seems to have a
similar effect as long-range interactions, since it is mainly a matter of time, until
the particles meet and interact with each other. As a consequence, one can find a transition to
an ordered phase even in an analogous one-dimensional 
non-equilibrium system (Czir\'ok \etal, 1999a), 
while the related equilibrium system with $v_0=0$ is always disordered 
in the presence of fluctuations (Mermin and Wagner, 1966).

\subsection{Stock markets: Herding and oscillations} \label{VB2}

As is reflected in {\em bubbles} and {\em crashes} of stock markets,
the reinforcement of buying and selling decisions has sometimes features of
herding behavior
(see, e.g., Youssefmir \etal, 1998; Farmer, 1998; Lux and Marchesi, 1999). 
Remember that herding behavior is frequently
found in complex situations, where individuals do not know what is the right thing to do. Everyone
counts on collective intelligence then, believing that a crowd is following
someone who has identified the right action. However, as has been shown for panicking 
pedestrians seeking for an exit, such mass behavior can have undesireable results,
since alternatives are not efficiently exploited. 
\par
The corresponding herding model in Sec.~\ref{IIIC1d} 
could be viewed as paradigm for problem-solving behavior
in science, economics, and politics, where new solutions to complex problems have to 
be found in a way comparable to finding the exit of a smoky room. From the simulation results, 
we may conclude that people will
not manage to cope with a sequence of challenging situations, if everyone
sticks to his own idea egocentrically, but they will also fail, if everyone
follows the same idea, so that the possible spectrum of solutions is not adequately
explored. Therefore, the best strategy appears to be pluralism with a reasonable degree of readiness
to follow good ideas of others, while a totalitarian regime would probably not survive
a series of crises. This fits well into experimental data on the efficiency 
of {\em group problem solving} (Anderson, 1961; Kelley and Thibaut, 1969; Laughlin \etal, 1975),
according to which groups normally perform better than
individuals, but masses are inefficient in finding new solutions to complex
problems. 
\par
Considering the phase diagram in Fig.~\ref{VT3}, we may also
conjecture that, in the presence of a given level $\eta$ of fluctuations,
there is a certain critical density above which people tend to show {\em mass behavior},
and below which they behave individualistically. In fact,
the tendency of mass behavior is higher in dense populations than in
dilute ones. In summary, conclusions from findings for self-driven many-particle systems
reach far into the realm of the social, economic, and psychological sciences.
\par
Another example is the analogy between 
irregular oscillations of prices at the stock market and of pedestrian flows
at bottlenecks (see Sec.~\ref{IIICb}). The idea is as follows:
Traffic is a prime example of individuals competing for limited resources (namely, space). 
At the stock exchange market, we have a competition of two different groups:
optimistic traders (``bulls'') and pessimistic ones (``bears''). The optimists count on growing
stock prices and increase the price by their orders. In contrast, pessimists speculate
on a decrease in the price and reduce it by their selling of stocks. Hence, traders belonging to
the same group enforce each others' (buying or selling) actions,
while optimists and pessimists push (the price) into opposite directions. Consequently,
there is an analogy with opposite pedestrian streams pushing at a bottleneck, which
is reflected in a roughly similar dynamics (Helbing, 2001). 

\section{Summary and outlook} \label{VI}

In this review, I have shown that many aspects of traffic flow and other living
systems can be reflected by self-driven many-particle systems, which implies
that the respective interactions dominate
socio-psychological effects. Although there are still many interesting open questions
and controversial debates, calling for intensified research activities, one can
already state that traffic theory is a prime example of a mathematically
advanced, semi-quantitative description of human behavior. Meanwhile, it appears 
feasible to reproduce the observed complex transitions between 
different traffic states (see Sec.~\ref{IIA5}) by simulation models  which
are adapted to the measurement site by variation of one parameter value only 
and work with the empirically measured boundary conditions (see Sec.~\ref{IID1}). One could
even say there are kind of {\em natural laws} behind the behavior of drivers and pedestrians. 
This includes the self-organized, characteristic constants of traffic flow 
(see Secs.~\ref{IIA5a} and \ref{IIC5}), the collective patterns of motion observed in pedestrian crowds
(see Secs.~\ref{IIIC} and \ref{IIIC1}), and the trail formation behavior of humans 
(see Sec.~\ref{IF5}). 
\par
Self-driven many-particle systems display
a surprisingly rich spectrum of spatio-temporal {\em pattern formation phenomena}.
They were described by Newton's equation of motion (where Newton's third law
{\em actio = reactio} was sometimes relaxed), complemented
by driving forces and frictional dissipation. Based on such small modifications,
we often found considerably different behavior compared to analogous systems
from classical or equilibrium statistical mechanics, but there were also various analogies with
driven fluids or granular media (see Secs.~\ref{IIB5c}, \ref{IIC4} and \ref{IIIA}). Important factors 
for the finally resulting patterns were
(i) the specification of the desired velocities and directions of motion
in the driving term,  (ii) inhomogeneities 
and other boundary effects, (iii) the heterogeneity among the particles, and 
(iv) the degree of fluctuations.
Many of the systems were characterized by non-linear feedback mechanisms, so that
small perturbations could have strong effects like ``phantom traffic jams''
(see Secs.~\ref{IIA5a}  and \ref{IIC1}) or ``phantom panics'' (see Sec.~\ref{IIIC1c}). 
These effects were often counterintuitive, such as ``freezing by heating'' (see Sec.~\ref{IIIC1a})
or the ``faster-is-slower effect'' (see Sec.~\ref{IIIC1b}). 
Summarizing the discoveries, in the 
considered self-driven dissipative many-particle systems with repulsive interactions,
one can find all sorts of counter-intuitive {\em breakdown and structure formation
phenomena}, when
the ``temperature'' (i.e. the fluctuation strength) is increased. 
In contrast, from classical many-particle systems
with attractive interactions, we are used to the idea that increasing temperature 
{\em breaks up} structures and {\em destroys patterns}, 
i.e. fluid structures are replaced by gaseous ones, not by
solid ones, etc. Nevertheless, the combination of repulsive interactions with a driving
term and dissipation gives also rise to phenomena typical for attractive
interactions, e.g. the segregation effects observed in the formation of traffic jams 
(see Sec.~\ref{IIC3}) and pedestrian lanes (see Sec.~\ref{IIICa}), 
or the emergence of the coherent moving state in
a system of cars and trucks (see Sec.~\ref{IIFb}).
\par
The description of the phenomena in self-driven many-particle systems uses and generalizes
almost the complete spectrum of
methods developed in non-equilibrium statistical physics and non-linear dynamics.
It has close relations with kinetic gas theory and fluid dynamics, soft matter, solid state physics, 
and even quantum mechanics (see Sec.~\ref{MPA}), while the corresponding non-equilibrium 
thermodynamics is still to be developed. However, there are aspects of
{\em universality} found in traffic, as is exemplified by the power-law behavior of traffic flows
(see Secs.~\ref{IIA5a} and \ref{IIC3}) and the phase diagrams 
of traffic dynamics and model characteristics (see Figs.~\ref{KERNER}, \ref{KRAUSS2}, and
\ref{PHASEDIAG}). In the spirit of general systems theory
(Buckley, 1967; von Bertalanffy, 1968; Rapoport, 1986), one could even go one step
further: Many phenomena found in traffic systems, where drivers or pedestrians 
compete for limited space, may have implications 
for other biological and socio-economic systems, in which
individuals compete for limited resources. In fact, jamming and herding phenomena are
found in various living systems, from 
animal colonies, markets, administrations, or societies, up to
science, economics, and politics. Therefore, traffic is a good and, in particular, concrete
example to study certain aspects of socio-psychological phenomena in an experimental setup.
Possibly, this can help to gain a better understanding of more complex human behavior in the future.
\par
Self-driven many-particle systems are also interesting from a practical point of view.
They are related to aspects of {\em collective intelligence} (see, e.g., the work carried out at the Santa
Fe Institute, {\tt http://www.santafe.edu/sfi/research/ indexResearchAreas.html},
by scientists like Botee and Bonabeau, 1998; Kube and Bonabeau, 1998; Wolpert and Tumer, 1999). 
Based on suitable interactions, many-particle
and many-agent systems can organize themselves and reach an 
optimal state (Helbing and Vicsek, 1999). Such 
a distributed strategy based on local optimization is more efficient and robust than
classical, centralized control strategies. In particular, the outcome on large scales
is relatively insensitive to the local failure of control.

\section*{Acknowledgments}

I like to thank the German Research Foundation (DFG) 
and the  Federal Ministry for Education and Science (BMBF)
for financial support through the grants He 2789/1-1, He 2789/2-1, and the project SANDY,
grant no. 13N7092. Many thanks also for the warm hospitality
experienced at the Tel Aviv University, the Weizmann Institute of Science,
the Xerox Palo Alto Research Center (Xerox PARC), the E\"otv\"os University,
and the Collegium Budapest---Institute for Advanced Study. Furthermore, I am
grateful for the fruitful collaboration with and/or
comments by James Banks, Eshel Ben-Jacob, Kai Bolay, Andr\'{a}s Czir\'{o}k,
Ill\'{e}s Farkas,  Isaac Goldhirsch, Ansgar Hennecke, Bernardo Huberman, Boris Kerner,
Joachim Krug,
P\'{e}ter Moln\'{a}r, David Mukamel, Joachim Peinke, 
Andreas Schadschneider, Michael Schreckenberg, Gunter Sch\"utz,
Rudolf Sollacher, Benno Tilch, Martin Treiber, Tam\'{a}s Vicsek, and others,
who have also supplied various figures. Finally, I am obliged to the
Dutch {\em Ministry of Transport, Public Works and Water Management},
to the {\em Bosch GmbH, Stuttgart}, to the  {\it Autobahndirektion S\"udbayern},
and the {\it Hessisches Landesamt f\"ur Stra{\ss}en und Verkehrswesen}
for providing traffic data. Tilo Grigat, Daniel Kern, Martina Seifert, Benno Tilch,
Martin Treiber, and Torsten Werner 
have helped a lot with the references and corrections.

\end{document}